\documentclass{appolb}
\usepackage{amssymb,amsmath,bm}

\usepackage{epsf}
\usepackage{epsfig,rotate}
\usepackage{afterpage}
\usepackage{longtable}
\usepackage{cite}
\usepackage{color}

\newcommand{\cH}{{\cal H}}

\newcommand{\vub}{|V_{ub}|}

\newcommand{\sla}{\! \! \! \!  /~}
\def\nnb{\nonumber}

\def\epe{\varepsilon'/\varepsilon}
\newcommand{\tev}{\, {\rm TeV}}
\newcommand{\gev}{\, {\rm GeV}}

\newcommand{\no}{\nonumber}

\newcommand{\f}{\frac}
\newcommand{\mt}{m_{\rm t}}

\newcommand{\ts}{\tilde s}
\newcommand{\tc}{\tilde c}
\newcommand{\Vt}{\widetilde V}
\newcommand{\Vtr}{\widetilde V_0}

\newcommand{\be}{\begin{equation}}
\newcommand{\ee}{\end{equation}}
\newcommand{\bea}{\begin{eqnarray}}
\newcommand{\eea}{\end{eqnarray}}

\newcommand{\ba}{\begin{array}}
\newcommand{\ea}{\end{array}}

\newcommand{\ord}{{\cal O}}

\def\kpn{K^+\rightarrow\pi^+\nu\bar\nu}

\def\klpn{K_{L}\rightarrow\pi^0\nu\bar\nu}

\begin{document}
\headtitle{Minimal Flavour Violation and Beyond: Towards a Flavour Code 
\dots}
\headauthor{Andrzej J.~Buras}
\title{MINIMAL FLAVOUR VIOLATION AND BEYOND:\\
TOWARDS  A FLAVOUR CODE FOR SHORT\\ DISTANCE 
DYNAMICS
\thanks{Lecture presented at the 50th Cracow School of Theoretical Physics 
``Particle Physics at the Dawn of the LHC'', Zakopane, Poland, June 9-19, 2010.}
}
\author{Andrzej J. Buras
\address{Technical University Munich, Physics Department, D-85748 Garching, Germany,\\
TUM-IAS, Lichtenbergstr. 2a, D-85748 Garching, Germany
 \\
   }}
\maketitle
\begin{abstract}
 This decade should provide the first definitive signals of  New Physics (NP)
 beyond the Standard Model (SM) and the goal of these lectures is a review
 of flavour physics in various extensions of the SM that have been popular 
 in the last ten years. After an overture, two pilot sections and a
 brief summary of the structure of 
flavour violation and  CP violation in the SM, we will present the 
theoretical framework for  weak decays that will allow us to distinguish 
between different NP scenarios. 
Subsequently we will present 
twelve concrete BSM models summarizing the patterns of flavour violation
 characteristic for each
model. 
In addition to models with minimal flavour violation (MFV) 
accompanied 
by flavour blind phases we will discuss a number of extensions containing 
non-MFV sources of flavour and CP violation and, in particular, new local 
operators originating in right-handed charged currents and scalar currents.
Next we will address various anomalies in the data as seen from 
the point of view of the SM that appear very natural in certain extensions 
of the SM. 
In this presentation selected superstars of this field will play a very 
important role. These are processes that are 
very sensitive to NP effects and which are theoretically clean. Particular 
emphasis will be put on correlations between various observables that
could allow us to distinguish between various NP scenarios.
Armed with this knowledge we will
propose a coding system in a form of a  $3\times 3$ matrix which helps to 
 distinguish between 
various extensions of the SM. 
Finding which {\it flavour code}
is chosen by nature would be an important step towards 
the fundamental theory of flavour. We give several examples of flavour codes 
representing specific models.
 We believe that such studies 
combined with new results from the Tevatron, the LHC, Belle II, 
Super-Flavour-Facility in Rome and dedicated Kaon and lepton flavour violation 
experiments should allow  to improve significantly our knowledge about the 
dynamics at the shortest distance scales.
\end{abstract}

\centerline{\bf Overture}

The year 1676 was a very important year for the humanity. In this year 
Antoni van Leeuvenhoek (1632-1723) discovered the empire of bacteria. 
He called these small creatures {\it animalcula} (small animals). This 
discovery was a mile stone in our civilization  for at least two reasons: 
\begin{itemize}
\item
 He discovered invisible to us creatures which over thousands of years 
 were systematcally killing the humans, often responsible for millions 
 of death in one year. While Antoni van Leeuvanhoek did not know that 
 bacteria could be dangerous for humans, his followers like L. Pasteur (1822-1895),
 Robert Koch (1843-1910) and other {\it microbe hunters} not only realized 
 the danger coming from this tiny creatures but also developed weapons against
 this empire. 
\item
 He was the first human who looked at short distance scales invisible to 
 us, discovering thereby
 a new {\it underground world}. At that time researchers 
 looked mainly at large distances, discovering new planets and finding 
 laws, like Kepler laws, that Izaak Newton was able to derive from his 
 mechanics.
 \end{itemize}
 
 While van Leeuvanhoek could reach the resolution down to roughly 
$10^{-6}$m, over the last 334 years this resolution could be improved 
by twelve orders of magnitude. On the way down to shortest distance 
scales scientists discovered {\it nanouniverse} ($10^{-9}$m), 
{\it femtouniverse}   ($10^{-15}$m) relevant for nuclear particle physics 
and low energy elementary particle physics and finally 
{\it attouniverse} ($10^{-18}$m)
that is the territory of contemporary high energy elementary particle physics.

In this decade we will be able to improve the resolution of 
 the short distance scales by at
least
 an order of magnitude, extending the picture of fundamental physics 
down to scales $5\cdot 10^{-20}$m with the help of the LHC. Further resolution 
down to scales as short as $10^{-21}$m or even shorter scales
 should be possible with the help of 
high precision experiments in which flavour violating processes will play a 
prominent role. These notes deal with the latter route to the short 
distance scales and with new animalcula which hopefully will be discovered both 
at the LHC and through high precision experiments in the coming years.

\begin{figure}[thb]
\centerline{\includegraphics[width=0.65\textwidth]{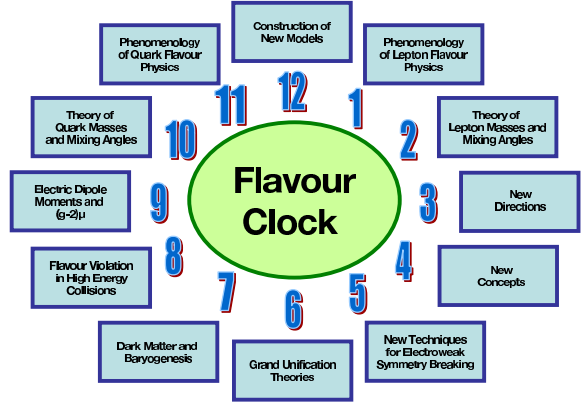}}
\caption{Working towards the Theory of Flavour around the Flavour Clock.}\label{Fig:1}
\end{figure}

\section{Introduction}
In our search for a fundamental theory of elementary particles we need to
improve our understanding of flavour 
in \cite{Buras:2009if,Isidori:2010kg,Fleischer:2010qb,Nir:2010jr,Antonelli:2009ws,Hurth:2010tk}. This is clearly a very ambitious goal that
requires the advances in different directions as well as continuos efforts of
many experts day and night, as depicted with the help of the  ''Flavour Clock''
in Figure~\ref{Fig:1}.

Despite  the impressive success of the CKM picture of flavour changing 
interactions \cite{Cabibbo:1963yz,Kobayashi:1973fv} in which also the GIM mechanism 
\cite{Glashow:1970gm} for the
suppression of flavour changing neutral currents (FCNC) 
plays a very important role, there are many open questions of 
theoretical and experimental nature that should be answered before we
can claim to have a theory of flavour.
Among the basic questions in flavour physics that could be answered in the 
present decade are the following ones:
\begin{enumerate}
\item
What is the fundamental dynamics behind the electroweak symmetry breaking 
that very likely plays also an important role in flavour physics?
\item
Are there any new flavour symmetries that could
help us to understand the existing hierarchies of fermion masses and the 
hierarchies in the quark and lepton flavour violating interactions?
\item
Are there any flavour violating interactions that are not governed by 
the SM Yukawa couplings? In other words, is  Minimal Flavour Violation 
(MFV)
the whole story?
\item
Are there any additional {\it flavour violating} CP-violating (CPV) phases that could 
explain certain anomalies present in the flavour data and 
simultaneously play
a role in the explanation of the observed baryon-antibaryon asymmetry 
in the universe (BAU)?
\item
Are there any {\it flavour conserving} CPV phases that could also help
in explaining the flavour anomalies in question and would be signalled 
in this decade  through  enhanced electric dipole moments (EDMs) of the
neutron, the electron and of other particles?
\item
Are there any new sequential heavy quarks and leptons of the 4th 
generation and/or new fermions with exotic quantum numbers like 
vectorial fermions?
\item
Are there any elementary neutral and charged scalar particles  
with masses below 1~TeV and having a significant impact on flavour physics?
\item
Are there any new heavy gauge bosons representing an enlarged gauge 
symmetry group?
\item
Are there any relevant right-handed (RH) weak currents that would help us to
make our fundamental theory parity conserving at short distance scales 
well below those explored by the LHC?
\item
How would one successfully address all these questions if the breakdown of 
the electroweak symmetry would turn out to be of a non-perturbative origin? 
\end{enumerate}

An important question is the following one:
will some of these questions be answered through the interplay of high
energy processes explored by the LHC with low energy precision experiments 
or are the relevant scales of fundamental flavour well beyond the energies
explored by the LHC and future colliders in this century? The existing 
tensions in some of the corners of the SM and still a rather big room for
NP contributions in rare decays of mesons and leptons and 
CP-violating observables, including in particular EDMs, give us hopes that indeed several phenomena required to answer at least
some of these questions could be discovered in this decade.

\section{Superstars of Flavour Physics in 2010-2015}
As far as high precision experiments are concerned a number of selected 
processes and observables will, in my opinion, play the leading role in 
learning about the NP in this new territory. This selection is based on 
the sensitivity to NP and theoretical cleanness. The former can be increased 
with the increased precision of experiments and the latter can improve with
the progress in theoretical calculations, in particular the non-perturbative
ones like the lattice simulations.

My superstars for the coming years are as follows:
\begin{itemize}
\item
The mixing induced CP-asymmetry $S_{\psi\phi}(B_s)$ that is
 tiny in the SM: $S_{\psi\phi}\approx 0.04$. The asymmetry
 $S_{\phi\phi}(B_s)$ is also important. It is also
 very strongly suppressed 
in the SM and is sensitive to NP similar to the one explored through 
the departure of $S_{\phi K_S}(B_d)$ from $S_{\psi K_S}(B_d)$ 
\cite{Fleischer:2007wg}.
\item
The rare decays $B_{s,d}\to\mu^+\mu^-$ that could be enhanced in certain 
NP scenarios by an order of magnitude with respect to the SM values.
\item
The angle $\gamma$ of the unitarity triangle (UT) that will be precisely 
measured
through tree level decays.
\item
$B^+\to\tau^+\nu_\tau$ that is sensitive to charged Higgs particles.
\item
The rare decays $K^+\to\pi^+\nu\bar\nu$ and $K_L\to\pi^0\nu\bar\nu$ that
belong to the theoretically cleanest decays in flavour physics.
\item
The decays $B\to X_s\nu\bar\nu$, $B\to K^*\nu\bar\nu$ and $B\to K\nu\bar\nu$ 
that are theoretically rather clean and are sensitive to RH currents.
\item
Numerous angular symmetries and asymmetries in $B\to K^*l^-l^-$.
\item
Lepton flavour violating decays like $\mu\to e\gamma$, $\tau\to e\gamma$, 
$\tau\to\mu\gamma$, decays with three leptons in the final state and 
$\mu-e$ conversion in nuclei.
\item
Electric dipole moments of the neutron, the electron, atoms and leptons.
\item
Anomalous magnetic moment of the muon $(g-2)_\mu$ that indeed seems to
be ''anomalous'' within the SM even after the inclusion of radiative corrections.
\item
The ratio $\varepsilon'/\varepsilon$ in $K_L\to\pi\pi$ decays 
which is known experimentally within 
$10\%$ and which should gain in importance in this decade due to improved 
lattice calculations.
\end{itemize}

Clearly, there are other stars in flavour physics but I believe that the 
ones above will play the crucial role in our search for the theory of 
flavour. Having experimental results on these decays and observables with
sufficient precision accompanied by improved theoretical calculations will
exclude several presently studied models reducing thereby our exploration
of short distance scales to a few avenues.

In the rest of the paper I will proceed as follows. In Section 3 I will 
recall those ingredients of the SM that are dominantly responsible for the 
pattern of flavour violation and CP violation in this model.   In Section 4
I will briefly recall the theoretical framework for weak decays that goes 
beyond the SM. 
In Section 5 we discuss several concrete BSM models. For each model we list the 
new particles and we recall the structure of their interactions with the 
ordinary quarks and leptons in particular paying attention to the 
Lorentz structure of these interactions. 
We summarize the patterns of flavour violation characteristic for each
model. In Section 6 we address a number of anomalies present in the 
data from the point of view of the models of Section 5. We also illustrate 
how the superstars listed in Section 2 when considered together can help 
in distinguishing between various NP scenarios.
In this context the  correlations 
between various observables will play a prominent role. 

Armed with all this knowledge we will propose in Section 7
a new classification of various NP effects by means of a
  coding system in a form of a  $3\times 3$ {\it flavour code matrix}. Each NP model
is characterized by a special code in which only some entries of 
the matrix in question are occupied. MFV, non-MFV sources and
{\it flavour blind} CP-violating phases on the one hand and LH-currents, 
RH-Currents
and scalar currents on the other hand are the fundamental coordinates 
in this 
code. They allow to classify transparently the models 
discussed in Section 5. We give several examples of flavour codes 
corresponding to specific models discussed in the text.
 As we 
will see some models, depending on the values of parameters involved, 
show their presence in a number of entries of this matrix.
Finally in Section 8 we will provide a brief summary. 
Recent reviews on flavour physics can be found 
in \cite{Buras:2009if,Isidori:2010kg,Fleischer:2010qb,Nir:2010jr,Antonelli:2009ws,Hurth:2010tk}.

\section{Patterns of flavour violation and CP violation in the SM}
Let us collect here  those ingredients of the SM which
are fundamental for the structure of flavour violating and CP-violating 
phenomena in this model.

\begin{itemize}
\item
The SM contains three generations of 
quarks and leptons.
\item
The gauge interactions are described by the group 
${\rm  SU(3)_C\times SU(2)_L\times U(1)_Y}$ spontaneously broken to
${\rm SU(3)_C\times U(1)_Q}$.
The strong interactions are mediated by eight gluons $G_a$, the
electroweak interactions by $W^{\pm}$, $Z^0$ and $\gamma$.
\item
Concerning {\it Electroweak Interactions}, the left-handed leptons and
quarks are put into ${\rm SU(2)_L}$ doublets:
\begin{equation}\label{2.31}
\left(\begin{array}{c}
\nu_e \\
e^-
\end{array}\right)_L\qquad
\left(\begin{array}{c}
\nu_\mu \\
\mu^-
\end{array}\right)_L\qquad
\left(\begin{array}{c}
\nu_\tau \\
\tau^-
\end{array}\right)_L~,
\end{equation}
\begin{equation}\label{2.66}
\left(\begin{array}{c}
u \\
d^\prime
\end{array}\right)_L\qquad
\left(\begin{array}{c}
c \\
s^\prime
\end{array}\right)_L\qquad
\left(\begin{array}{c}
t \\
b^\prime
\end{array}\right)_L~,     
\end{equation}
with the corresponding right-handed fields transforming as singlets
under $ SU(2)_L $. 

The {\it weak
eigenstates} $(d^\prime,s^\prime,b^\prime)$ and the corresponding {\it mass 
eigenstates} $d,s,b$ are connected through the CKM matrix
\begin{equation}\label{2.67}
\left(\begin{array}{c}
d^\prime \\ s^\prime \\ b^\prime
\end{array}\right)=
\left(\begin{array}{ccc}
V_{ud}&V_{us}&V_{ub}\\
V_{cd}&V_{cs}&V_{cb}\\
V_{td}&V_{ts}&V_{tb}
\end{array}\right)
\left(\begin{array}{c}
d \\ s \\ b
\end{array}\right)\equiv\hat V_{\rm CKM}\left(\begin{array}{c}
d \\ s \\ b
\end{array}\right).
\end{equation}
In the leptonic sector the analogous mixing matrix is a unit matrix
due to the masslessness of neutrinos in the SM. Otherwise we have the 
PMNS matrix.
\item
The charged current interactions mediated by $W^{\pm}$ are only 
between left-handed quarks in accordance with maximal parity 
breakdown observed in low-energy processes.
\item
The unitarity of the CKM matrix assures the absence of
flavour changing neutral current (FCNC) transitions at the tree level.
This means that the
elementary vertices involving neutral gauge bosons ($G_a$, $Z^0$,
$\gamma$) and the neutral Higgs are flavour conserving.
This property is known under the name of GIM mechanism \cite{Glashow:1970gm}.
\item
The fact that the $V_{ij}$s can a priori be complex
numbers allows  CP violation in the SM \cite{Kobayashi:1973fv}. 
\item
The CKM matrix can be parametrized by $s_{12}$, $s_{13}$, $s_{23}$ and 
a phase $\delta=\gamma$, with $\gamma$ being one of the angles of 
the Unitarity Triangle. While $\gamma\approx(70\pm 10)^\circ$, the 
$s_{ij}$ exhibit a hierarchical structure:
\begin{equation}\label{2.73}
s_{12}=| V_{us}|\approx 0.225, \quad s_{13}=| V_{ub}|\approx 4\cdot 10^{-3}, \quad s_{23}=|V_{cb}|\approx 4\cdot 10^{-2}~.
\end{equation}
\item
This pattern of $s_{ij}$ and the large phase $\gamma$ combined with the large 
top quark mass and GIM mechanism imply large CP-violating 
effects in the $B_d$ system ($S_{\psi K_s}\approx 0.7$), small CP-violating 
effects in the $B_s$ system ($S_{\psi \phi}\approx 0.04$) and tiny CP-violating 
 effects in the $K$ system ($|\varepsilon_K|\approx 0.002$).
\item
The EDMs predicted by the SM are basically unmeasurable in this decade. 
\item 
Lepton flavour violation in the SM is very strongly supressed.
\end{itemize}

Presently, this global structure of flavour violating interactions 
works rather well but as we will see 
below some deviations from the SM predictions are observed in the data 
although most of these {\it anomalies} being typically $2-3\sigma$ are 
certainly not conclusive.

\section{Theoretical Framework: Beyond the SM}
\subsection{Preliminaries}

The starting point of any serious analysis of weak decays in 
the framework of a 
given extension of the SM is the basic Lagrangian 
\be\label{basicL}
{\cal L} ={\cal L}_{\rm SM}(g_i,m_i,V^{i}_{\rm CKM})+
{\cal L}_{\rm NP}(g_i^{\rm NP},m_i^{\rm NP},
V^{i}_{\rm  NP}),
\ee
where $(g_i,m_i,V^{i}_{\rm CKM})$  denote the parameters of the SM and
$(g_i^{\rm NP},m_i^{\rm NP},V^{i}_{\rm  NP})\equiv\varrho_{\rm NP}$
 the additional 
parameters in a given NP scenario.

Our main goal then is to identify in weak decays the effects  decribed by  
${\cal L}_{\rm NP}$
 in the presence of 
the background from ${\cal L}_{\rm SM}$. 
In the first step one derives the Feynman rules following 
from (\ref{basicL}), which allows to calculate Feynman diagrams. But then we
 have to face two challenges:

\begin{itemize}
\item
our theory is formulated in terms of quarks, but experiments involve their 
bound states: $K_L$, $K^\pm$, $B_d^0$, $B_s^0$, $B^\pm$, $B_c$, $D$, $D_s$, 
etc. 
\item
NP takes place at very short distance scales $10^{-19}-10^{-18}~{\rm m}$, 
while $K_L$, $K^\pm$, $B_d^0$, $B_s^0$, $B^\pm$ and other mesons live at
$10^{-16}-10^{-15}~{\rm m}$.
\end{itemize}

The solution to these challenges is well known. 
One has to construct an effective 
theory relevant for experiments at low energy scales. 
Operator Product Expansion (OPE)
and Renormalization Group (RG) methods are involved here. 
They allow to separate 
the perturbative short distance (SD) effects, where NP is present, 
from long distance 
(LD) effects for which non-perturbative methods are necessary. Moreover RG 
methods allow an efficient summation of large logarithms 
$\log (\mu_{\rm SD}/\mu_{\rm LD})$. 
A detailed exposition of 
these techniques can be found in \cite{Buchalla:1995vs,Buras:1998raa}
 and fortunately we do not have to 
repeat them here. At the 
end of the day the formal expressions involving matrix elements of local operators 
and their Wilson coefficients can be cast into the following {\it Master Formula for 
Weak Decays} \cite{Buras:2001pn}.

\subsection{Master Formula for Weak Decays}

 The master formula in question reads:
\be\label{master}
{\rm A(Decay)}=\sum_i B_i \eta^i_{\rm QCD}V^i_{\rm CKM} 
F_i(m_t,{\rm \varrho_{NP}}),
\ee
where $B_i$ are non-perturbative parameters representing hadronic matrix
elements of the contributing operators, $\eta^i_{\rm QCD}$ stand symbolically 
for the renormalization group factors, $V^i_{\rm CKM}$ denote the relevant
combinations of the elements of the CKM matrix and finally 
$F_i(m_t,\varrho_{\rm NP})$
denote the loop functions resulting in most models from box and penguin
diagrams 
but in some
models  also representing tree level diagrams if such diagrams contribute.
The internal charm contributions have been suppressed in this formula but
they have to be included in particular in $K$ decays and $K^0-\bar K^0$ 
mixing.
$\varrho_{\rm NP}$ denotes symbolically all parameters beyond $m_t$, in 
particular the set $(g_i^{\rm NP},m_i^{\rm NP},V^{i}_{\rm  NP})$ in 
(\ref{basicL}).
It turns out to be useful to factor out $V^i_{\rm CKM}$ in all contributions
in order to see transparently the deviations from MFV
 that will play a prominent role in these lectures.

In the SM only a particular set of parameters $B_i$ is relevant as there are
no right-handed charged current interactions, the
functions $F_i$ are {\it real} and the flavour and CP-violating effects
 enter only through 
the CKM factors $V^i_{\rm CKM}$.
 This  implies that the
functions $F_i$ are universal with respect to flavour so that they are the
same in the $K$, $B_d$ and $B_s$ systems. Consequently a number of observables 
in these different systems are strongly correlated with each other 
within the SM.

The simplest class of extensions of the SM are  models with 
Constrained Minimal Flavour Violation (CMFV)
\cite{Buras:2000dm, Buras:2003jf,Blanke:2006ig}.
In these models
all flavour changing transitions are governed by the CKM matrix with the 
CKM phase being the only source of CP violation. Moreover, the $B_i$ factors 
in (\ref{master}) are only those that are also relevant in the SM.
This implies that relative to the SM  only the values of $F_i$ are 
modified but their universal character remains intact. In particular they
are real. This implies various correlations between different observables 
that we will discuss as we proceed.

In more general MFV models 
\cite{D'Ambrosio:2002ex,Chivukula:1987py,Hall:1990ac}
new parameters $B_i$ and 
$\eta^i_{\rm QCD}$, related to new operators, enter the game but if flavour 
blind CP-violating phases (FBPs) are absent or negligible the functions
$F_i$ still remain real quantities as in the CMFV
framework and do not involve any flavour violating parameters. Consequently
the CP and flavour violating effects in these models 
are again governed by the CKM matrix.
However, the presence of new operators makes this approach less constraining
than the CMFV framework. We will discuss some other aspects of this approach 
below.

Most general MFV models can also contain FBPs that can have profound 
implications for the phenomenology of weak decays because of the interplay 
of the CKM matrix with these phases. In fact such models became very 
popular recently and we will discuss them below.

In the simplest non-MFV models, the basic operator structure of CMFV models
remains but the functions $F_i$ in addition to real SM contributions can
contain new flavour parameters and new complex phases that can be both 
flavour violating and flavour blind. 

Finally, in the most general non-MFV models, new operators 
(new $B_i$ parameters)
contribute and the functions $F_i$ in addition to real SM contributions can
contain new flavour parameters and new complex phases.

In \cite{Buras:2009if} we have presented a classification of different classes of models
in a form of a  $2\times 2$ flavour matrix which distinguished only between 
models with SM operators and models with new operators on the one hand and 
MFV and non-MFV on the other hand. From the present perspective this matrix 
is insufficient as it does not take into account the possible 
presence of FBPs and moreover does not distinguish sufficiently between 
different Lorentz structures of the operators involved. In particular 
it does not distinguish between right-handed currents that involve gauge 
bosons and scalar currents resulting primarly from Higgs exchanges. Therefore 
at the end of our paper we will attempt to improve on this by proposing 
a flavour code in a form of  
 a bigger matrix: the $3\times 3$ {\it flavour code matrix} (FCM).

 Clearly without a good knowledge of  non-perturbative factors $B_i$ 
no precision studies of 
flavour physics will be possible unless the non-perturbative uncertainties 
can be reduced or even removed by taking suitable ratios of observables. 
In certain rare 
cases it is also possible to measure the relevant hadronic 
matrix elements entering 
rare decays by using leading tree level decays. Examples of such fortunate 
situations are certain mixing induced CP asymmetries and 
the branching ratios for 
$K\to\pi\nu\bar\nu$ decays. Yet, in many cases one has to face the direct
evaluation of
$B_i$. While 
lattice calculations, QCD-sum rules, light-cone sum rules and large-$N$ methods made
significant progress in the last 20 
years, the situation is clearly not satisfactory and one should hope that new 
advances in the calculation of $B_i$ parameters will be made in the LHC era 
in order to 
adequately use improved data. Recently an impressive progress in 
calculating the 
parameter $\hat B_K$, relevant for CP violation in $K^0-\bar K^0$ mixing,
 has been made and we will discuss its implications in Section 6. It should 
be emphasized that also for $B_{d,s}$ and charm systems very significant 
progress has been made in the last years by lattice community. A selection 
of papers can be found in \cite{Lubicz:2008am,Shigemitsu:2009jy,Laiho:2009eu,Kronfeld:2010aw,Bouchard:2010yj}.

Definitely also considerable progress has been made in the study of non-leptonic $B$ decays and radiative exclusive $B$ decays by means of approaches like 
QCD factorization and perturbative QCD but discussing them is not the goal of 
this paper. Nice recent reviews have been published in 
\cite{Buchalla:2008tg,Jager:2010pp} where 
references to original literature can be found.

Concerning the factors $\eta^i_{\rm QCD}$ 
 an
impressive progress has been made during the last 20 years. The 1990s can be 
considered as the era of NLO QCD calculations. Basically,  NLO corrections to all 
relevant decays and transitions have been calculated already 
in the last decade \cite{Buchalla:1995vs},  
with a few exceptions, like the width differences $\Delta\Gamma_{s,d}$ in
the $B^0_{s,d}-\bar B^0_{s,d}$ systems that were completed only in 2003
\cite{Beneke:1998sy,Beneke:2002rj,Ciuchini:2003ww}. The last 
decade can be considered as the era of NNLO calculations. 
In particular one should 
mention here the NNLO calculations of QCD corrections to $B\to X_sl^+l^-$
\cite{Asatryan:2002iy,Asatrian:2002va,Gambino:2003zm,Ghinculov:2003qd,Ghinculov:2002pe,Bobeth:2003at,Beneke:2004dp},
 $\kpn$ \cite{Buras:2005gr,Buras:2006gb,Brod:2008ss},
and 
in particular to $B_s\to X_s\gamma$ \cite{Misiak:2006zs}
  with the latter one being by far 
the most difficult one. Also important steps towards a complete calculation 
of NNLO corrections to non-leptonic decays 
of mesons have been made in \cite{Gorbahn:2004my}. Most recently NNLO QCD 
corrections to the parameter $\eta^{ct}_{\rm QCD}$ in $\varepsilon_K$ 
\cite{Brod:2010mj} and 
two-loop electroweak corrections to $K\to\pi\nu\bar\nu$ 
\cite{Brod:2010hi} have been 
calculated. It should also be stressed that the precision on  $\eta^i_{\rm QCD}$ 
has also been indirectly improved through the more accurate determination 
of $\alpha_s$ for which the most recent result reads \cite{Bethke:2009jm}:
\be
\alpha_s(M_Z)=0.1184\pm0.0007~.
\ee

The final ingredients of our master formula, in addition to $V^i_{\rm CKM}$
factors, are the loop functions $F_i$ resulting from penguin
and
 box diagrams with the exchanges of the
top quark, $W^\pm$, $Z^0$, heavy new gauge bosons, heavy new fermions and
scalars. They are known at one-loop level in 
several extensions of the SM, in particular in the two Higgs doublet model
(2HDM), the
littlest Higgs model without T parity (LH), the ACD model with one universal
extra dimension (UED) \cite{Appelquist:2000nn}, the MSSM with MFV and non-MFV violating interactions,
the flavour blind MSSM (FBMSSM), the littlest Higgs model with T-parity (LHT),
  $Z^\prime$-models, Randall-Sundrum (RS) models,
left-right
symmetric models, the model with the sequential fourth generation of 
quarks and 
leptons. Moreover, in the SM $\ord(\alpha_s)$ corrections
to all relevant one loop functions are known. It should also be stressed
again that
in the loop functions in our master formula one can conveniently absorb tree
level FCNC contributions present in particular in RS models.

\subsection{Local Operators in the SM}

As a preparation for the construction of the new flavour matrix we have 
to make a closer look at the Lorentz structure of the operators involved, 
first in the SM and then beyond it. 
To this end we have to cast our master formula in (\ref{master}) into 
the more familiar formula that results from the relevant 
effective Hamiltonian.

In this more formal picture an amplitude for a decay of a given meson 
$M= K, B,..$ into a final state $F=\pi\nu\bar\nu,~\pi\pi,~DK$,... is then
simply given by
\be\label{amp5}
A(M\to F)=\langle F|{\cal H}_{eff}|M\rangle
=\frac{G_F}{\sqrt{2}}\sum_i V^i_{CKM}C_i(\mu)\langle F|Q_i(\mu)|M\rangle,
\ee
where $\langle F|Q_i(\mu)|M\rangle$ 
are the matrix elements of the local operators 
$Q_i$ between $M$ and $F$, evaluated at the
renormalization scale $\mu$ and $C_i(\mu)$ are the Wilson coefficients 
that collect compactly the effects of physics above the scale $\mu$.

\subsubsection{Nonleptonic Operators}
Of particular interest are the 
operators involving quarks only. In the case of the $\Delta B=1$
transitions the relevant set of operators is given as follows:

{\bf Current--Current:}
\begin{equation}\label{O1} 
Q_1 = (\bar c_{\alpha} b_{\beta})_{V-A}\;(\bar s_{\beta} c_{\alpha})_{V-A}
~~~~~~Q_2 = (\bar c b)_{V-A}\;(\bar s c)_{V-A} 
\end{equation}

{\bf QCD--Penguins:}
\begin{equation}\label{O2}
Q_3 = (\bar s b)_{V-A}\sum_{q=u,d,s,c,b}(\bar qq)_{V-A}~~~~~~   
 Q_4 = (\bar s_{\alpha} b_{\beta})_{V-A}\sum_{q=u,d,s,c,b}(\bar q_{\beta} 
       q_{\alpha})_{V-A} 
\end{equation}
\begin{equation}\label{O3}
 Q_5 = (\bar s b)_{V-A} \sum_{q=u,d,s,c,b}(\bar qq)_{V+A}~~~~~  
 Q_6 = (\bar s_{\alpha} b_{\beta})_{V-A}\sum_{q=u,d,s,c,b}
       (\bar q_{\beta} q_{\alpha})_{V+A} 
\end{equation}

{\bf Electroweak Penguins:}
\begin{equation} 
Q_7 = \frac{3}{2}\;(\bar s b)_{V-A}\sum_{q=u,d,s,c,b}e_q\;(\bar qq)_{V+A} 
\nnb
\ee
\be\label{O4}
Q_8 =  \frac{3}{2}\;(\bar s_{\alpha} b_{\beta})_{V-A}\sum_{q=u,d,s,c,b}e_q
        (\bar q_{\beta} q_{\alpha})_{V+A}
\end{equation}
\begin{equation}
 Q_9 =  \frac{3}{2}\;(\bar s b)_{V-A}\sum_{q=u,d,s,c,b}e_q(\bar q q)_{V-A}
\nnb
\ee
\be\label{O5}
Q_{10} = \frac{3}{2}\;
(\bar s_{\alpha} b_{\beta})_{V-A}\sum_{q=u,d,s,c,b}e_q\;
       (\bar q_{\beta}q_{\alpha})_{V-A} 
\end{equation}
Here, $\alpha,\beta$ denote colours and $e_q$ denotes the electric quark charges reflecting the
electroweak origin of $Q_7,\ldots,Q_{10}$. Finally,
$(\bar c b)_{V-A}\equiv \bar c_\alpha\gamma_\mu(1-\gamma_5) b_\alpha$.

These operators play a crucial role in non-leptonic decays of $B_s$ and $B_d$ 
mesons and have through mixing under renormalization also an impact on other 
processes as is evident from the treatises in 
\cite{Buras:1998raa,Buchalla:1995vs}.
For non-leptonic $K$ decays the quark flavours have 
to be changed appropriately. 
Explicit expressions can be found in \cite{Buras:1998raa,Buchalla:1995vs}.
In particular the analogues of 
$Q_1$ and $Q_2$ govern the $\Delta I=1/2$ rule in $K_L\to\pi\pi$ decays, while 
the corresponding QCD penguins and electroweak penguins enter directly 
the ratio $\epe$.

Before continuing one observation should be made. We have stated before 
that charged current weak interactions are governed by left-handed (LH) 
currents. 
This is indeed the case as seen in (\ref{O1}): only $V-A$ currents are present 
there. This is no longer the case when QCD penguins and Electroweak Penguins 
that govern FCNC processes are considered. Yet, also there the presence of 
only LH charged currents in the SM is signalled by the fact that 
the first currents in each operator have $V-A$ structure. The fact 
that $V+A$ structures appear in (\ref{O3}) is related to the vectorial 
character of gluon interactions that in the process of the renormalization 
 group analysis have to be decomposed into $V-A$ and $V+A$ parts. Similar 
comments apply to (\ref{O4}), where the photon penguins and $Z^0$ 
penguins are involved. 

This discussion implies that in the presence of right-handed (RH) charged currents also 
operators with $V-A$ replaced by $V+A$ in (\ref{O1}) and in the first 
factors in the remaining operators in (\ref{O2})-(\ref{O5}) would 
contribute.

\subsubsection{Magnetic Penguins}
In the case of $B\to X_s\gamma$  and  $B\to X_sl^+l^-$ decays 
and corresponding exclusive decays the crucial role is played by {\it 
magnetic} penguin operators:
\begin{equation}\label{O6}
Q_{7\gamma}  =  \frac{e}{8\pi^2} m_b \bar{s}_\alpha \sigma^{\mu\nu}
          (1+\gamma_5) b_\alpha F_{\mu\nu}\qquad            
Q_{8G}     =  \frac{g}{8\pi^2} m_b \bar{s}_\alpha \sigma^{\mu\nu}
   (1+\gamma_5)T^a_{\alpha\beta} b_\beta G^a_{\mu\nu}  
\end{equation}
The operator $Q_{8G}$ can 
also be relevant in nonleptonic decays. The magnetic operators are often 
called {\it dipole} operators.

\boldmath
\subsubsection{$\Delta S= 2$ and $\Delta B=2$ Operators}
\unboldmath
In the case of  $K^0-\bar K^0$ mixing and  $B_d^0-\bar B^0_d$ mixing
the relevant operators within the SM are
\begin{equation}\label{O7}
Q(\Delta S = 2)  = (\bar s d)_{V-A} (\bar s d)_{V-A}~~~~~
 Q(\Delta B = 2)  = (\bar b d)_{V-A} (\bar b d)_{V-A}~. 
\end{equation}
For $B_s^0-\bar B^0_s$ mixing one has to replace $d$ by $s$ in 
the last operator.

\boldmath
\subsubsection{Semileptonic Operators}
\unboldmath
In the case of $B\to X_sl^+l^-$ also the following operators on top of 
magnetic penguins contribute
\begin{equation}\label{9V}
Q_{9V}  = (\bar s b  )_{V-A} (\bar \mu\mu)_{V}~~~~~
Q_{10A}  = (\bar s b )_{V-A} (\bar \mu\mu)_{A}.
\end{equation}
Changing approprately flavours one obtains the corresponding 
operators relevant for 
$B\to X_dl^+l^-$ and $K_L\to\pi^0l^+l^-$.

The rare decays $B\to X_s\nu\bar\nu$, $B\to K^*\nu\bar\nu$, 
$B\to K\nu\bar\nu$ and $B_s\to\bar\mu\mu$ are governed by
\begin{equation}\label{10V}
Q_{\nu\bar\nu}(B)  = (\bar s b  )_{V-A} (\bar \nu\nu)_{V-A}~~~~~
Q_{\mu\bar\mu}(B)  = (\bar s b )_{V-A} (\bar \mu\mu)_{V-A}~.
\end{equation}

The rare decays $K\to\pi\nu\bar\nu$ and $K_L\to\bar\mu\mu$ are governed on the 
other hand by
\begin{equation}\label{11V}
Q_{\nu\bar\nu}(K)  = (\bar s d  )_{V-A} (\bar \nu\nu)_{V-A}~~~~~
Q_{\mu\bar\mu}(K)  = (\bar s d )_{V-A} (\bar \mu\mu)_{V-A}~.
\end{equation}

\subsection{Local Operators in Extensions of the SM}

NP can generate new operators. Typically new operators 
are generated through the presence of RH currents and 
{\it scalar} currents with the latter strongly suppressed within the SM.
New gauge bosons and scalar exchanges are at the origin of these operators 
that can have important impact on phenomenology. The two-loop anomalous 
dimensions of these operators have been calculated in 
\cite{Ciuchini:1997bw,Buras:2000if}

\boldmath
\subsubsection{$\Delta F=2$ Non-leptonic Operators}
\unboldmath

For definiteness, we shall consider here operators responsible for the
$K^0$--$\bar{K}^0$ mixing and consequently relevant also for $\varepsilon_K$. 
There are 8 such operators of dimension 6.
They can be split into 5 separate sectors, according to the chirality
of the quark fields they contain. The operators belonging to the first
three sectors (VLL, LR and SLL) read \cite{Buras:2000if}

\bea 
Q_1^{\rm VLL} &=& (\bar{s}^{\alpha} \gamma_{\mu}    P_L d^{\alpha})
              (\bar{s}^{ \beta} \gamma^{\mu}    P_L d^{ \beta}),
\nnb\\[4mm] 
Q_1^{\rm LR} &=&  (\bar{s}^{\alpha} \gamma_{\mu}    P_L d^{\alpha})
              (\bar{s}^{ \beta} \gamma^{\mu}    P_R d^{ \beta}),
\nnb\\
Q_2^{\rm LR} &=&  (\bar{s}^{\alpha}                 P_L d^{\alpha})
              (\bar{s}^{ \beta}                 P_R d^{ \beta}),
\nnb\\[4mm]
Q_1^{\rm SLL} &=& (\bar{s}^{\alpha}                 P_L d^{\alpha})
              (\bar{s}^{ \beta}                 P_L d^{ \beta}),
\nnb\\
Q_2^{\rm SLL} &=& (\bar{s}^{\alpha} \sigma_{\mu\nu} P_L d^{\alpha})
              (\bar{s}^{ \beta} \sigma^{\mu\nu} P_L d^{ \beta}),
\label{normal}
\eea
where $\sigma_{\mu\nu} = \f{1}{2} [\gamma_{\mu}, \gamma_{\nu}]$ and
$P_{L,R} =\f{1}{2} (1\mp \gamma_5)$. The operators belonging to the
two remaining sectors (VRR and SRR) are obtained from $Q_1^{\rm VLL}$ and
$Q_i^{\rm SLL}$ by interchanging $P_L$ and $P_R$. For $\Delta B=2$ the 
flavours have to be changed appropriately.

\boldmath
\subsubsection{$\Delta F=1$ Operators}
\unboldmath
The list of $\Delta F=1$ operators in the extensions of the SM is much 
longer and will not be given here. All the dimension six four-quark operators 
are discussed in \cite{Buras:2000if} where also their two-loop anomalous 
dimensions have been calculated. See also \cite{Ciuchini:1997bw}, where a 
different operator basis is used.

Concerning the semileptonic operators in the extensions of the SM the 
typical examples  of operators related to the presence of RH {\it currents} are

\begin{equation}\label{9VR}
\tilde Q_{9V}  = (\bar s b  )_{V+A} (\bar \mu\mu)_{V}~~~~~
\tilde Q_{10A}  = (\bar s b )_{V+A} (\bar \mu\mu)_{A}.
\end{equation}
\begin{equation}\label{10VR}
\tilde Q_{\nu\bar\nu}(B)  = (\bar s b  )_{V+A} (\bar \nu\nu)_{V-A}~~~~~
\tilde Q_{\mu\bar\mu}(B)  = (\bar s b )_{V+A} (\bar \mu\mu)_{V-A}~.
\end{equation}
\begin{equation}\label{11VR}
Q_{\nu\bar\nu}(K)  = (\bar s d  )_{V+A} (\bar \nu\nu)_{V-A}~~~~~
Q_{\mu\bar\mu}(K)  = (\bar s d )_{V+A} (\bar \mu\mu)_{V-A}~.
\end{equation}

If {\it scalar currents} resulting from scalar exchanges like the heavy 
Higgs in the 2HDM models or sparticles in the MSSM are present, scalar operators enter 
the game. The most prominent are the ones that govern the 
$B_s\to \mu^+\mu^-$ decay in 2HDMs and the MSSM at large $\tan\beta$:
\begin{equation}\label{scalarL}
Q_S  = (\bar s P_L b  ) (\bar \mu\mu)~~~~~
Q_P  = (\bar s P_L b ) (\bar \mu\gamma_5 \mu)
\end{equation}

\begin{equation}\label{scalarR}
\tilde Q_S  = (\bar s P_R b  ) (\bar \mu\mu)~~~~~
\tilde Q_P  = (\bar s P_R b ) (\bar \mu\gamma_5 \mu)
\end{equation}

\subsection{Penguin-Box Expansion}
After this rather formal presentation let us just state how our master formula 
(\ref{master}) can be obtained from the expression (\ref{amp5}).
We can start with (\ref{amp5}) but instead of evaluating it at the 
low energy scale we choose for $\mu$ the high energy scale to be called 
$\mu_H$ at which heavy particles are integrated out. Then expressing 
the Wison coefficients  $C_i(\mu_H)$ in terms of the loop functions 
$F_i$ we arrive at (\ref{master}).
As the expansion in (\ref{master}) involves basic one-loop 
functions from penguin and box diagrams it was naturally given the 
name of the  {\it Penguin-Box Expansion} (PBE) \cite{Buchalla:1990qz}.

Originally  PBE was designed to expose the $\mt$-dependence
of FCNC processes \cite{Buchalla:1990qz} which was hidden in 
the Wilson coefficients. 
In particular in the case of $\epe$ where many of these functions enter, this 
turned out to be very useful.
After the top quark mass has been
measured precisely this role of PBE is less important.
On the other hand,
PBE is very well suited for the study of the extensions of the
SM in which new particles are exchanged in the loops and as discussed 
above all these effects are encoded in the functions $F_i$.

If new operators are present it is often useful to work first with 
the coefficients $C_i(\mu_H)$ rather then loop functions.
Then absorbing 
$G_F/\sqrt{2}$ and   $V^i_{CKM}$ in the Wilson coefficients $C_i(\mu_H)$
the amplitude for $M-\overline{M}$ mixing ($M= K, B_d,B_s$) is then
simply given by
\be\label{amp6}
A(M\to \bar M)
=\sum_{i,a} C_i(\mu_H)\langle \overline{M} |Q^a_i(\mu_H)|M\rangle,
\ee
where the sum runs over all the operators in (\ref{normal}), that is 
 $i=1,2$ and $a=VLL,VRR,LR, ...$. The 
matrix elements for $B_d-\bar B_d$ mixing are for instance given as 
follows \cite{Buras:2001ra}
\be\label{eq:matrix}
\langle \bar B^0|Q_i^a|B^0\rangle = \frac{2}{3}M_{B_d}^2 F_{B_d}^2 P_i^a(B_d),
\ee
where the coefficients $P_i^a(B_d)$ 
collect compactly all RG effects from scales below $\mu_H$ as well as
hadronic matrix elements obtained by lattice methods at low energy scales.
Analytic formulae for these coefficients are given in \cite{Buras:2001ra} 
while the recent application of this method can be found in 
\cite{Buras:2010mh,Buras:2010zm,Buras:2010pz}.
As the 
Wilson coefficients $ C_i(\mu_H)$ depend directly on the loop functions 
and fundamental parameters of a given theory, this dependence can be 
exhibited as in PBE or (\ref{master}) if necessary.
Again as in the case of PBE the virtue of using high energy scale rather 
than the low energy scale is that the coefficients $P_i^a(M)$ can be 
evaluated once for all if the hadronic matrix elements are known. 
Other virtues of this approach are discussed 
in \cite{Buras:2010mh,Buras:2010zm,Buras:2010pz}.

\section{A Closer Look at Selected BSM Models}

\subsection{Three Strategies in Waiting for NP in Flavour Physics}
Particle physicists are waiting eagerly for a solid evidence of NP for the
last 30 years. Except for neutrino masses, the BAU and dark matter, no clear 
signal emerged so far.
 While waiting several strategies for finding NP have been 
developed. They can be divided roughly into three classes.
\subsubsection{Precision calculations within the SM}
Here basically the goal is to calculate precisely 
the background to NP coming from
the known dynamics of the SM. At first sight this approach is not very
exciting. Yet, in particular in flavour physics, where the signals of 
NP are generally indirect, this approach is very important. From my 
point of view, having been involved for more than one decade in calculations of 
higher order QCD corrections \cite{Buras:1998raa}, 
I would claim that for most interesting
decays these perturbative and renormalization group improved calculations
reached already the desired level. See references in Section 4.

The main progress is now required from lattice groups. Here the main goals 
for the coming years are more accurate values of weak decay constants 
$F_{B_{d,s}}$ and various $\hat B_i$ parameters relevant for $B_{d,s}$ physics.
For $K^0-\bar K^0$ mixing the relevant parameter $\hat B_K$ is now known 
with an accuracy of $4\%$ \cite{Antonio:2007pb}. An impressive achievement. Let us hope that 
also the parameters $B_6$ and $B_8$, relevant for 
$\varepsilon^\prime/\varepsilon$ will be known with a similar accuracy 
within this decade. Comprehensive reviews on lattice results can be found in 
\cite{Lubicz:2008am,Shigemitsu:2009jy,Laiho:2009eu,Kronfeld:2010aw,Bouchard:2010yj,Colangelo:2010et}. 

Clearly further improvements on the hadronic part of two-body 
non-leptonic decays is mandatory in order to understand more precisely 
the direct CP violation in $B_{s,d}$ decays. 
\subsubsection{The Bottom-Up Approach}
In this approach one constructs effective field theories involving 
only light degrees 
of freedom including the top quark in which the structure of the effective 
Lagrangians is governed by the symmetries of the SM and often other 
hypothetical symmetries. This approach is rather powerful in the case of
electroweak precision 
studies and definitely teaches us something about $\Delta F=2$ 
transitions. In particular lower bounds on NP scales depending on the 
Lorentz structure of operators involved can be derived from the data 
\cite{Bona:2007vi,Isidori:2010kg}.

However, except for the case of  MFV and closely related 
approaches based on flavour symmetries, the bottom-up approach ceases, 
in my view, to be useful in $\Delta F=1$ decays, 
because of very many operators that are allowed to appear
in the effective Lagrangians with coefficients that are basically 
unknown \cite{Buchmuller:1985jz,Grzadkowski:2010es}. In this 
approach then the correlations between various $\Delta F=2$ and $\Delta F=1$ 
observables in $K$, $D$, $B_d$ and $B_s$ systems are either not visible or 
very weak, again except MFV and closely related approaches. Moreover 
the correlations between flavour violation in low energy processes and 
flavour violation in high energy processes to be studied soon at the LHC 
are lost. Again MFV belongs to a few exceptions.
\subsubsection{The Top-Down Approach}
My personal view shared by some of my colleagues is that the top-down 
approach is more useful in flavour physics \footnote{As pointed out by Helmut 
Neufeld this view can only be defended if the number of NP models invented by 
theorists around the world is smaller than the number of possible operators in the bottom-up approach.}. Here one constructs first 
a specific model with heavy degrees of freedom. For high energy processes,
where the energy scales are of the order of the masses of heavy particles 
one can directly use this ``full theory'' to calculate various processes 
in terms of the fundamental parameters of a given theory. For low energy 
processes one again constructs the low energy theory by integrating out 
heavy particles. The advantage over the previous approach is that now the 
coefficients of the resulting local operators are calculable in terms of 
the fundamental parameters of this theory. In this manner correlations between 
various observables belonging to different mesonic systems and correlations 
between low energy and high-energy observables are possible. Such correlations 
are less sensitive to free parameters than separate observables and 
represent patterns of flavour violation characteristic for a given theory. 
These correlations can in some models differ strikingly from the ones of 
the SM and of the MFV approach.

\begin{figure}[hb]
\centerline{\includegraphics[width=0.65\textwidth]{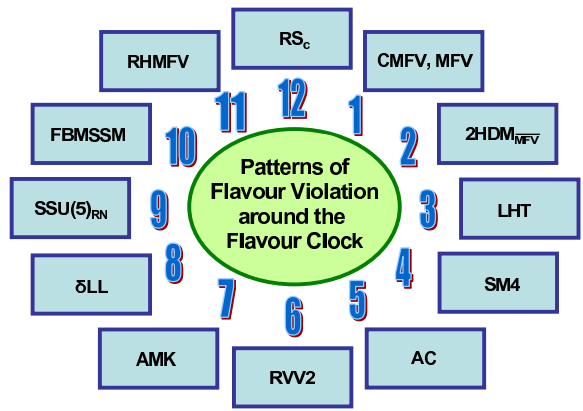}}
\caption{Various patterns of flavour violation around the Flavour Clock.}\label{Fig:2}
\end{figure}

\subsection{Anatomies of explicit models}
Having the last strategy in mind  my group at the Technical University Munich, 
consisting dominantly of diploma students, PhD students and young post--docs 
investigated in the last decade flavour violating and CP-violating processes 
in the following models: CMFV, MFV, MFV-MSSM, 
$Z^\prime$-models, general MSSM, a model with a universal flat 5th dimension, 
the Littlest Higgs model (LH), the Littlest Higgs model with T-parity (LHT),
SUSY-GUTs,  
Randall-Sundrum model with custodial protection (RSc), flavour blind MSSM 
(FBMSSM), four classes of supersymmetric flavour models with the dominance 
of LH currents ($\delta$LL model), the dominance of RH currents  in an abelian flavour model (AC model), non-abelian model with equal 
strength of CKM-like LH and RH currents (RVV2) and 
 the non-abelian AKM model in which the CKM-like RH currents dominate. 
The last comments applying only
to the NP part as the SM part is always there. 
This year we have analyzed the SM4, the 
${\rm 2HDM_{\overline{MFV}}}$ to be defined below, 
quark flavour mixing with RH 
currents in an effective theory approach RHMFV and very recently 
supersymmetric SU(5) GUT enriched through RH neutrinos $SSU(5)_{\rm RN}$.
These analyses where dominated by quark flavour physics, but in the case of the
LHT, FBMSSM, supersymmetric flavour models, $SSU(5)_{\rm RN}$ and the  SM4 also lepton flavour 
violation, EDMs and $(g-2)_\mu$ have been studied in detail. In what follows I will briefly 
describe several of these extensions putting emphasis on flavour violating 
and CP-violating processes. The order of presentation does not correspond 
to the chronological order in which these analyses have been performed. 
The models to be discussed below are summarized in Fig.~\ref{Fig:2}.

\subsection{Constrained Minimal Flavour Violation (CMFV)}
The simplest class of extensions of the SM are  models with 
CMFV
\cite{Buras:2000dm, Buras:2003jf,Blanke:2006ig}.
They are formulated as follows:
\begin{itemize}
\item
All flavour changing transitions are governed by the CKM matrix with the 
CKM phase being the only source of CP violation,
\item
The only relevant operators in the effective Hamiltonian below the weak scale
are those that are also relevant in the SM.
\end{itemize}
This implies that in the master formula (\ref{master}) 
relative to the SM  only the values of $F_i$ are 
modified but their universal character remains intact. In particular they
are real. 

As discussed in detail in \cite{Buras:2003jf}
this class of models can be formulated to a very good approximation 
in terms of 11 parameters: 4 parameters of the CKM matrix and 7 {\it real}
 values 
of the {\it universal} master functions $F_i$ that parametrize 
the short distance 
contributions. In a given CMFV model, $F_i$ can be calculated in 
perturbation theory and are generally correlated with each other but 
in a model independent analysis they must be considered as free 
parameters. 

Generally, several master functions contribute to a given decay,
although decays exist which depend only on a single function.
We have the following correspondence between the most interesting FCNC
processes and the master functions in the CMFV models:
\begin{center}
\begin{tabular}{lcl}
$K^0-\bar K^0$-mixing ($\varepsilon_K$) 
&\qquad\qquad& $S(v)$ \\
$B_{d,s}^0-\bar B_{d,s}^0$-mixing ($\Delta M_{s,d}$) 
&\qquad\qquad& $S(v)$ \\
$K \to \pi \nu \bar\nu$, $B \to X_{d,s} \nu \bar\nu$ 
&\qquad\qquad& $X(v)$ \\
$K_{\rm L}\to \mu \bar\mu$, $B_{d,s} \to l\bar l$ &\qquad\qquad& $Y(v)$ \\
$K_{\rm L} \to \pi^0 e^+ e^-$ &\qquad\qquad& $Y(v)$, $Z(v)$, 
$E(v)$ \\
$\varepsilon'$, Nonleptonic $\Delta B=1$, $\Delta S=1$ &\qquad\qquad& $X(v)$,
$Y(v)$, $Z(v)$,
$E(v)$ \\
$B \to X_s \gamma$ &\qquad\qquad& $D'(v)$, $E'(v)$ \\
$B \to X_s~{\rm gluon}$ &\qquad\qquad& $E'(v)$ \\
$B \to X_s l^+ l^-$ &\qquad\qquad&
$Y(v)$, $Z(v)$, $E(v)$, $D'(v)$, $E'(v)$,
\end{tabular}
\end{center}
where $v$ denotes collectively the arguments of a given function.

This table means that the observables like branching ratios, mass differences
$\Delta M_{d,s}$ in $B_{d,s}^0-\bar B_{d,s}^0$-mixing and the CP violation 
parameters $\varepsilon_K$ und $\varepsilon'$, all can be to a very good 
approximation entirely expressed in
terms of the corresponding master functions, the relevant CKM factors, low 
energy parameters like the $B_i$ and QCD factors $\eta_{\rm QCD}^i$.

All master functions have been defined in \cite{Buras:2003jf,Buchalla:1990qz}.
Phenomenological studies indicate that only 
\be\label{rmasterf}
S(v),\quad X(v), \quad Y(v), \quad Z(v),\quad  D'(v) \quad  E'(v).
\ee
receive significant NP contributions. $S(v)$ represents the 
box diagrams in $\Delta F=2$ processes, $X(v)$ and $Y(v)$ stand for gauge 
invariant combinations of $Z^0$ penguin and $\Delta F=1$ box diagrams, 
$Z(v)$ for a gauge invariant combination of the $Z^0$ penguin and the photon 
penguin. Finally  $D'(v)$ and  $E'(v)$ stand for magnetic photon 
penguin und magnetic gluon penguin, respectively. The NP contributions to
the ordinary gloun penguin $E(v)$ are generally irrelevant in this framework.

In \cite{Buras:2003jf} strategies for the determination of the values of 
these functions have been outlined. 
Moreover, in certain  cases  
these model dependent functions can be eliminated by taking certain combinations of observables. In this manner one obtains
universal correlations between these observables that are characteristic
 for this class of models. The most interesting are the following ones:
\be\label{dmsdmd}
\frac{\Delta M_d}{\Delta M_s}=
\frac{m_{B_d}}{m_{B_s}}
\frac{\hat B_{d}}{\hat B_{s}}\frac{F^2_{B_d}}{F^2_{B_s}}
\left|\frac{V_{td}}{V_{ts}}\right|^2
\end{equation}
\begin{equation}\label{bxnn}
\frac{{\rm Br}(B\to X_d\nu\bar\nu)}{{\rm Br}(B\to X_s\nu\bar\nu)}=
\left|\frac{V_{td}}{V_{ts}}\right|^2
\end{equation}
\begin{equation}\label{bmumu}
\frac{{\rm Br}(B_d\to\mu^+\mu^-)}{{\rm Br}(B_s\to\mu^+\mu^-)}=
\frac{\tau({B_d})}{\tau({B_s})}\frac{m_{B_d}}{m_{B_s}}
\frac{F^2_{B_d}}{F^2_{B_s}}
\left|\frac{V_{td}}{V_{ts}}\right|^2
\end{equation}
that all can be used to determine $|V_{td}/V_{ts}|$ without the knowledge
of $F_i(v)$ \cite{Buras:2000dm}. In particular the relation (\ref{dmsdmd}) 
 offers
a  powerful determination of the 
length of one side of the unitarity triangle, denoted usually by $R_t$. 
Combining this determination with the exparimental value of the CP 
asymmetry $S_{\psi K_S}=\sin2\beta$ allows to determine the unitarity 
triangle that is unversal for CMFV models \cite{Buras:2000dm}.

Out of these three ratios the cleanest 
is (\ref{bxnn}), which is essentially free
of hadronic uncertainties \cite{Buchalla:1998ux}. Next comes (\ref{bmumu}), involving
${\rm SU(3)}$ breaking effects in the ratio of $B$-meson decay constants.
Finally, ${\rm SU(3)}$ breaking in the ratio 
$\hat B_{B_d}/\hat B_{B_s}$ enters in addition in (\ref{dmsdmd}). These 
${\rm SU(3)}$ breaking effects are already known with respectable
precision from lattice QCD.

Eliminating $|V_{td}/V_{ts}|$ from the three relations above allows 
to obtain three relations between observables that are universal within the
MFV models. In particular 
from (\ref{dmsdmd}) and (\ref{bmumu}) one finds \cite{Buras:2003td}  
\be\label{R1}
\frac{{\rm Br}(B_{s}\to\mu\bar\mu)}{{\rm Br}(B_{d}\to\mu\bar\mu)}
=\frac{\hat B_{d}}{\hat B_{s}}
\frac{\tau( B_{s})}{\tau( B_{d})} 
\frac{\Delta M_{s}}{\Delta M_{d}},
\ee
that does not 
involve $F_{B_q}$ and consequently contains 
substantially smaller hadronic uncertainties than the formulae considered 
above. It involves
only measurable quantities except for the ratio $\hat B_{s}/\hat B_{d}$
that is known already now from lattice calculations 
with respectable precision \cite{Shigemitsu:2009jy,Laiho:2009eu}  
\be\label{BBB}
\frac{\hat B_{s}}{\hat B_{d}}=0.95\pm 0.03, \qquad
\hat B_{d}=1.26\pm0.11, \qquad \hat B_{s}=1.33\pm0.06~.
\ee

Moreover, as in the MFV models there are no flavour violating CPV phases 
beyond the CKM phase, 
 we also expect \cite{Buchalla:1994tr,Buras:2001af}
\be\label{R7}
(\sin 2\beta)_{\pi\nu\bar\nu}=(\sin 2\beta)_{\psi K_S}, 
\qquad
(\sin 2\beta)_{\phi K_S}\approx (\sin 2\beta)_{\psi K_S},
\ee
if FBPs are assumed to be negligible. Indeed within the SM the corrections 
to the latter relation amount to roughly $0.02$ \cite{Beneke:2005pu}.

The confirmation of these two relations would be a very important test for the 
MFV idea unless important FBs are present (see below). Indeed, in $K\to\pi\nu\bar\nu$ the phase $\beta$ originates in 
the $Z^0$ penguin diagram, whereas in the case of $S_{\psi K_S}$ in 
the $B^0_d-\bar B^0_d$ box diagram. In the case of the asymmetry 
$S_{\phi K_S}$ it originates also in $B^0_d-\bar B^0_d$ box diagrams 
but the second relation in (\ref{R7}) could be spoiled by new physics 
contributions in the decay amplitude for $B\to \phi K_S$ that is
non-vanishing only at the one loop level.

One can also derive the following relations between 
${\rm Br}(B_{q}\to\mu\bar\mu)$ and $\Delta M_q$ \cite{Buras:2003td}  
\be\label{R2}
{\rm Br}(B_{q}\to\mu\bar\mu)
=4.36\cdot 10^{-10}\frac{\tau(B_{q})}{\hat B_{q}}
\frac{Y^2(v)}{S(v)} 
\Delta M_{q}, \qquad (q=s,d).
\ee
These relations  
allow to predict ${\rm Br}(B_{s,d}\to\mu\bar\mu)$  
in a given CMFV model with substantially smaller hadronic uncertainties 
than found by using directly the formulae 
for the branching ratios in question.
Using the present input parameters we find
\be\label{BRSM1}
{\rm Br}(B_d\to\mu^+\mu^-)_{\rm SM} = (1.0\pm 0.1)\times 10^{-10},
\ee
\be\label{BRSM2}
{\rm Br}(B_s\to\mu^+\mu^-)_{\rm SM} = (3.2\pm 0.2)\times 10^{-9}\,.
\ee

\subsection{Minimal Flavour Violation}
We have already formulated what we mean by CMFV.
 Let us first add here that  the models with CMFV generally contain only
one Higgs doublet and  the top Yukawa coupling dominates.
On the other hand general models with MFV contain more scalar representations, 
in particulat two Higgs doublets. Moreover, the operator structure in these 
models can differ from
the SM one. This is the case when  bottom and top Yukawa 
couplings are of comparable size. A well known example is the MSSM with MFV
and large $\tan\beta$. 

In the more general case of MFV the formulation  
with the help of global symmetries present in the limit of vanishing 
Yukawa couplings \cite{Chivukula:1987py,Hall:1990ac} as formulated in 
\cite{D'Ambrosio:2002ex} is elegant and useful. See also \cite{Feldmann:2006jk}
for a similar formulation that goes beyond the MFV. 
Recent discussions of various aspects of MFV 
can be found in 
\cite{Colangelo:2008qp,Paradisi:2008qh,Mercolli:2009ns,Feldmann:2009dc,Kagan:2009bn,Paradisi:2009ey}.
In order to see what is here involved we follow a compact formulation of 
Gino Isidori \cite{Isidori:2010gz}.

Let us look then at
the Standard Model (SM) Lagrangian which can be divided into two main parts, 
the gauge and the Higgs (or symmetry breaking) sector. The gauge sector
is completely specified by the 
local symmetry ${\mathcal G}^{\rm SM}_{\rm local} =SU(3)_{C}\times SU(2)_{L}\times U(1)_{Y}$
and by the fermion content,
\bea
{\cal L}^{\rm SM}_{\rm gauge} &=& \sum_{i=1\ldots3}\ \sum_{\psi=Q^i_L \ldots E^i_R}
{\bar \psi} i D\sla \psi \no\\
&& -\frac{1}{4} \sum_{a=1\ldots8} G^a_{\mu\nu} G^{a\mu\nu} -\frac{1}{4}
 \sum_{a=1\ldots3} W^a_{\mu\nu} W^{a\mu\nu}  -\frac{1}{4} B_{\mu\nu} B^{\mu\nu}~.
\eea
The fermion content consist of five fields with different quantum numbers 
under the gauge group.
\be
\label{eq:SMfer}
Q^i_{L}(3,2)_{+1/6}~,\ \ U^i_{R}(3,1)_{+2/3}~,\ \
D^i_{R}(3,1)_{-1/3}~,\ \ L^i_{L}(1,2)_{-1/2}~,\ \ E^i_{R}(1,1)_{-1}~,
\ee
each of them appearing in three different flavours ($i=1,2,3$).

As given above ${\cal L}^{\rm SM}_{\rm gauge}$ has 
a large {\em global} flavour symmetry  $U(3)^5$,
corresponding to the independent unitary rotations in flavour space 
of the five fermion fields in (\ref{eq:SMfer}). 
This can be decomposed as follows: 
\be 
{\mathcal G}_{\rm flavour} = U(3)^5 \times  
{\mathcal G}_{q} \times {\mathcal G}_{\ell}~, 
\label{eq:Gtot}
\ee
where 
\be
{\mathcal G}_{q} = {SU}(3)_{Q_L}\times {SU}(3)_{U_R} \times {SU}(3)_{D_R}, \qquad 
{\mathcal G}_{\ell} =  {SU}(3)_{L_L} \otimes {SU}(3)_{E_R}~.
\label{eq:Ggroups}
\ee
Three of the five $U(1)$ subgroups can be identified with the total baryon and 
lepton number and the weak hypercharge. The two remaining $U(1)$ groups can 
be identified with the Peccei-Quinn symmetry of 2HDMs and with a global 
rotation of a single $SU(2)_L$ singlet. 

Both the local and the global symmetries of ${\cal L}^{\rm SM}_{\rm gauge}$
are broken with the introduction of a $SU(2)_L$ Higgs doublet $\phi$.
The local symmetry is {\it spontaneously }
broken by the vacuum expectation value of the Higgs field,
$\langle \phi \rangle  = v = 246$~GeV, while 
the global flavour symmetry is {\em explicitly} broken by 
the Yukawa interaction of $\phi$ with the fermion fields:
\be
\label{eq:SMY}
- {\cal L}^{\rm SM}_{\rm Yukawa}=Y_d^{ij} {\bar Q}^i_{L} \phi D^j_{R}
 +Y_u^{ij} {\bar Q}^i_{L} \tilde\phi U^j_{R} + Y_e^{ij} {\bar L}_{L}^i
\phi E_{R}^j + {\rm h.c.} \qquad  ( \tilde\phi=i\tau_2\phi^\dagger)~.  
\ee
The subgroups controlling flavour-changing dynamics, in particular
flavour non-universality,  are the non-Abelian groups ${\mathcal G}_{q}$ 
and ${\mathcal G}_{\ell}$, which are explicitly broken by $Y_{d,u,e}$ not being 
proportional to the identity matrix. 

The hypothesis of MFV amounts to assuming that the Yukawas are the only sources of the breakdown of flavour and CP-violation.

The phenomenological implications of the MFV hypothesis formulated in this 
more grander manner than the CMFV formulation given above can be found 
model independently 
by using an effective field theory approach (EFT) \cite{D'Ambrosio:2002ex}. 
In this framework the SM lagrangian is supplemented by all higher dimension
 operators consistent with the MFV hypothesis, built using the Yukawa 
couplings as spurion fields. The NP effects in this 
framework
are then parametrized in terms of a few {\it flavour-blind} 
free parameters and SM Yukawa couplings that are solely responsible for 
 flavour violation  and also CP violation if these flavour-blind parameters 
are chosen as {\it real} quantities as done in \cite{D'Ambrosio:2002ex}. 
This approach naturally 
suppresses FCNC processes to the level observed experimentally even in the 
presence of new particles with masses of a few hundreds GeV. It also implies 
specific correlations between various observables, which are not as  stringent 
as in the CMFV but are still very powerful.

Yet, it should be stressed  that the MFV symmetry principle in itself does 
not forbid the presence of
{\it flavour blind} CP violating sources~\cite{Baek:1998yn,Baek:1999qy,Bartl:2001wc,Paradisi:2009ey,Ellis:2007kb,Colangelo:2008qp,Altmannshofer:2008hc,Mercolli:2009ns,Feldmann:2009dc,Kagan:2009bn} that make effectively the flavour blind free parameters 
{\it complex} quantities having flavour-blind phases (FBPs). These phases can in 
turn enhance the electric dipole moments {EDMs} of various particles and 
atoms and in the interplay with the CKM matrix can have also profound 
impact on flavour violating observables, in particular the CP-violating ones. 
In the context of the so-called aligned 2HDM model such effects have also been 
emphasized 
in \cite{Pich :2009sp}.

One concrete example is
MFV MSSM  which in view of these FBPs suffers from the same SUSY CP problem 
as the ordinary MSSM.
Either an extra assumption or a mechanism accounting for a natural suppression
of these CP-violating
phases is then desirable.
Possible solution to this problem is the following \cite{Paradisi:2009ey}: 
the SUSY breaking mechanism is
{\it flavour blind} and CP conserving and the breaking of CP only arises through the MFV compatible
terms breaking the {\it flavour blindness}. That is, CP is preserved by the sector responsible for
SUSY breaking, while it is broken in the flavour sector.
While the generalized MFV ansatz still accounts for a natural solution of the SUSY CP problem, it also
leads to peculiar and testable predictions in low energy CP violating processes~\cite{Paradisi:2009ey}. See more about this below.

Yet, independently of Supersymmetry, that we will discuss later on, the 
introduction of flavour-blind 
CPV phases compatible with the MFV symmetry principle turns out to 
be a very interesting set-up~\cite{Kagan:2009bn,Colangelo:2008qp,Mercolli:2009ns,Paradisi:2009ey,Ellis:2007kb}.
In particular, as noted in \cite{Kagan:2009bn}, 
a large new phase in $B^0_s$--$\bar B^0_s$ mixing could
in principle be obtained in the MFV framework if additional FBPs
are present. This idea cannot be realized in the ordinary MSSM
with MFV, as shown in~\cite{Altmannshofer:2009ne,Blum:2010mj}. The difficulty of realizing this scenario in the MSSM 
is due to the suppression in the MSSM 
of effective operators with several Yukawa insertions. 
Sizable couplings for these operators are necessary both to have an 
effective large CP-violating phase in $B^0_s$--$\bar B^0_s$ mixing and, at the same time,
 to evade bounds from 
other observables, such as $B_s\to \mu^+\mu^-$ and $B \to X_s \gamma$.
However, it could be realized in different underlying models, 
such as the up-lifted MSSM, as recently pointed out in \cite{Dobrescu:2010rh} and in the 2HDM with MFV and FBPs as we will discuss now.  

\subsection{${\rm 2HDM_{\overline{MFV}}}$}
\subsubsection{Preliminaries}
We will next discuss a specific class of 2HDM models, namely  2HDM 
with MFV accompanied by flavour blind CP phases that we will call for 
short ${\rm 2HDM_{\overline{MFV}}}$ \cite{Buras:2010mh} with
 the ''bar'' on MFV indicating the presence of FBPs. 

Before entering the details it will be instructive to recall that
the standard assignment of the $SU(2)_L\times U(1)_Y$ quark charges, 
identified long ago by Glashow, Iliopoulos, and Maiani (GIM)~\cite{Glashow:1970gm},
forbids tree-level flavour-changing couplings of the quarks to the 
SM neutral gauge bosons. 
In the case of only one-Higgs doublet, 
namely within the SM, this structure is effective also 
in eliminating a possible dimension-four flavour-changing neutral-current 
(FCNC) coupling of the quarks to the Higgs field. 
While the $SU(2)_L\times U(1)_Y$ assignment of quarks and leptons 
can be considered as being well established, much less is known 
about the Higgs sector of the theory. In the presence 
of more than one-Higgs field the appearance of tree-level FCNC 
is not automatically forbidden by the standard assignment 
of the $SU(2)_L\times U(1)_Y$ fermion charges: additional conditions have to 
be imposed on the model in order to guarantee a sufficient suppression of 
FCNC processes~\cite{Glashow:1976nt,Paschos:1976ay}. 
The absence of renormalizable couplings contributing 
at the tree-level to FCNC processes, in multi-Higgs 
models, goes under the name of Natural Flavour Conservation (NFC)
hypothesis. Excellent discussions of multi-Higgs models and of NFC 
can be found in \cite{Branco:1999fs,Accomando:2006ga}.

The idea of NFC has been with us for more than 30 years. 
During the last decade the mechanism for the suppression 
of FCNC processes with the help of MFV has been developed and it is 
natural to ask  
how NFC (and GIM) are related to MFV, and vice versa. 
Motivated by a series of recent studies about the strength of 
FCNCs in multi-Higgs doublet 
models~\cite{Pich:2009sp,Joshipura:2007cs,Botella:2009pq,Gupta:2009wn,Ferreira:2010xe}, 
we have presented recently
a detailed analysis of the relation between the NFC and MFV  
hypotheses \cite{Buras:2010mh}.
As we have shown, while the two hypotheses 
are somehow equivalent at the tree-level, important differences 
arise when quantum corrections are included. Beyond the tree
level, or beyond the implementation of these two hypotheses in
their simplest version, some FCNCs are naturally generated
in both cases. In this more general framework, the MFV hypothesis
in its general formulation~\cite{D'Ambrosio:2002ex}
turns out to be more stable in suppressing FCNCs than 
the hypothesis of NFC alone. 

I will not repeat here these arguments as for some readers they could 
appear academic. In short it is probably not surprising that flavour-blind 
symmetries that are used to protect FCNCs in the context of NFC are not 
as powerful as flavour symmetries used in the context of the MFV hypothesis.
A nice summary of our work by one of my collaborators making this point very 
clear appeared recently \cite{Carlucci:2010bj}.
Instead, I would like to summarize the 
phenomenological implications of this framework that were not expected
by us  when we started our analysis. In particular, in our second analysis
the issue of EDMs in this framework has also 
been considered \cite{Buras:2010zm}. 
Other recent interesting 
analyzes of FCNC processes within 2HDMs can be found in 
\cite{Pich:2009sp,Botella:2009pq,Jung:2010ik,Dobrescu:2010rh,Braeuninger:2010td,Ferreira:2010yh,Ferreira:2010xe,Trott:2010iz}.

Let me first list the few important points of the ${\rm 2HDM_{\overline{MFV}}}$
 framework.
\begin{itemize}
\item
The presence of FBPs in this MFV framework modifies through their interplay 
with the standard CKM flavour violation the usual characteristic 
relations of  the MFV framework. In particular the mixing induced CP
asymmetries in $B_d^0\to\psi K_S$ and $B_s^0\to\psi\phi$ take the form known 
from non-MFV frameworks like LHT, RSc and 
SM4:
\begin{equation}
S_{\psi K_S} = \sin(2\beta+2\varphi_{B_d})\,, \qquad
S_{\psi\phi} =  \sin(2|\beta_s|-2\varphi_{B_s})\,,
\label{eq:3.43}
\end{equation}
where $\varphi_{B_q}$ are NP phases in $B^0_q-\bar B^0_q$ mixings.
Thus in the presence of non-vanishing $\varphi_{B_d}$ and $\varphi_{B_s}$, originating here in non-vanishing FBPs, 
these two asymmetries donot measure $\beta$ and $\beta_s$ but $(\beta+\varphi_{B_d})$ and $(|\beta_s|-\varphi_{B_s})$, respectively.
\item
 The FBPs in the ${\rm 2HDM_{\overline{MFV}}}$ can appear  both 
in Yukawa interactions and  in the Higgs potential. While in  
\cite{Buras:2010mh} only the case of FBPs in Yukawa interactions has been 
considered, in \cite{Buras:2010zm} these considerations have been extended
to include also the FBPs in the Higgs potential.
The two flavour-blind CPV mechanisms can be distinguished
through the correlation between $S_{\psi K_S}$ and $S_{\psi\phi}$ that is
strikingly different if only one of them is relevant. We will see this 
explicitely below. See Fig. 5 in \cite{Buras:2010zm}.
\item
Sizable FBPs, necessary to explain possible  sizable 
non-standard CPV effects in $B_{s}$ mixing could, in principle, be 
forbidden
by the upper bounds on 
EDMs of the neutron and the atoms. This question has been adressed 
in \cite{Buras:2010zm} and the answer will be given  below.  
\end{itemize}

Let us then briefly consider these two cases, returning to them in more 
details in 
Section 6 in the context of a general discussion of the anomalies 
observed in the present data.

\subsubsection{FBPs in Yukawa Interactions}
Integrating out the neutral Higgs fields leads to 
tree-level contributions to scalar FCNC operators. 
Working in the decoupling limit for the heavy Higgs doublet,
 the leading Higgs contributions to $\Delta F=1$ and  $\Delta F=2$ 
Hamiltonians thus generated are $(q=d,s)$
\bea
 \Delta\cH_{\rm MFV}^{|\Delta B|=1} &=&
 -\frac{a^*_0+a^*_1}{M_H^2}~ y_\ell y_b y_t^2 V^*_{tb} V_{tq}~
 ({\bar b}_R q_L)({\bar \ell}_L \ell_R) {\rm ~+~h.c.}, 
\label{eq:HeffBll} \\
 \Delta\cH_{\rm MFV}^{|\Delta S|=1} &=&
 -\frac{a^*_0}{M_H^2}~ y_\ell y_s y_t^2 V^*_{ts} V_{td}~
 ({\bar s}_R d_L)({\bar \ell}_L \ell_R) {\rm ~+~h.c.}~,  \\
 \Delta\cH_{\rm MFV}^{|\Delta B|=2} &=&
 - \frac{(a^*_0+a^*_1)(a_0+a_2)}{  M_H^2} ~y_b y_q [y_t^2 V^*_{tb} V_{tq} ]^2~
 ({\bar b}_R q_L) ({\bar b}_L q_R)  {\rm ~+~h.c.} ,  
\label{eq:HeffB}\\
 \Delta\cH_{\rm MFV}^{|\Delta S|=2} &=&
 - \frac{|a_0|^2}{  M_H^2} ~y_s y_d [y_t^2 V^*_{ts} V_{td} ]^2~
 ({\bar s}_R d_L) ({\bar s}_L d_R)  {\rm ~+~h.c.}~,
\label{eq:HeffK}
\eea
where $a_i$ are flavour-blind complex coefficients that are $\ord(1)$.
Their complex phases originate in this framework precisely from FBPs in
the Yukawa interactions. $y_i$ are Yukawa couplings.
Inspecting these formulae we anticipate immediately two key properties 
that can be directly 
deduced by looking at their flavour- and CP-violating structure:

\begin{itemize}
\item
The impact in $K^0$--$\bar{K}^0$,  $B^0_{d}$--$\bar{B}^0_{d}$
and $B^0_{s}$--$\bar{B}^0_{s}$ mixing amplitudes scales, relative to the SM,
with $m_sm_d$, $m_b m_d$ and $m_b m_s$, respectively.
This fact opens the possibility of sizable non-standard 
contributions to the $B_s$ system  without serious constraints 
from  $K^0$--$\bar{K}^0$ and  $B^0_{d}$--$\bar{B}^0_{d}$ mixing.
In particular one has the relation
\be\label{R1a}
\varphi_{B_d}\approx \frac{m_d}{m_s}\varphi_{B_s}
\approx\frac{1}{17}\varphi_{B_s}
\ee

\item
While the possible flavour-blind phases do not contribute to the 
$\Delta S=2$ effective Hamiltonian, they could have an impact in the 
$\Delta B=2$ case, offering the possibility to solve the 
recent experimental anomalies related to the $B_s$ mixing phase. 
However, this happens only if $a_1\not= a_2$.
This requires a  non-trivial underlying dynamics, which does not suppress
effective operators with high powers of Yukawa insertions. 
While in the ${\rm 2HDM_{\overline{MFV}}}$ this is generally possible, 
this is not the case in the MSSM with MFV, where the supersymmetry 
puts these two coefficients to be approximately 
equal \cite{Altmannshofer:2009ne,Blum:2010mj}.
\item
The presence of scalar operators like $Q_{P,S}$ in the $\Delta B=1$ transitions allows 
strong enhancements of  branching ratios for $B_q\to\mu^+\mu^-$ in a correlated 
manner that is characteristic for models with MFV.
\end{itemize}

Let us also note that NP contributions to $\Delta F=2$ transitions 
in this case
 are dominated by the operator $Q_2^{\rm LR}$ whose contributions are strongly 
enhanced by RG effects and in the case of $\varepsilon_K$ by the 
chiral enhancement of its hadronic matrix element. Fortunately the 
suppression of this {\it direct} contrbution to $\varepsilon_K$ by the 
the relevant CKM factor and in particular by $m_dm_s$ does not introduce 
any problems with satisfying the $\varepsilon_K$ constraint and in fact 
this contribution can be neglected.

We are now ready to summarize the main phenomenological results 
obtained in \cite{Buras:2010mh}:

{\bf 1.}
 The pattern of NP effects in this model is
  governed by the quark masses relevant for the particular
 system considered: $m_sm_d$, $m_b m_d$ and $m_b m_s$, 
for the $K$, $B_d$ and $B_s$ systems, respectively. 

  {\bf 2.}
 If we try to accommodate a large CP-violating phase in $B_s^0$--$\bar B_s^0$ mixing
 in this scenario, we find a correlated shift in the relation between $S_{\psi K_S}$ and the CKM phase $\beta$. 
 This shift is determined unambiguously by the relation (\ref{R1a})
and  contains no free parameters. The shift is such that the
 prediction of $S_{\psi K_S}$ {\em decreases} with respect to the SM case 
 at fixed CKM inputs if  a large positive value of $S_{\psi \phi}$ is chosen. 
 This relaxes the existing 
 tension between  $S^{\rm exp}_{\psi K_S}$ and its SM prediction as seen in 
 Figure~2 of \cite{Buras:2010mh}.
 
{\bf 3.}
 The NP contribution to $\varepsilon_K$ is tiny and can be neglected. 
 However, given the modified relation between 
 $S_{\psi K_S}$ and the CKM phase $\beta$ in (\ref{eq:3.43}),
with $\varphi_{B_d}<0$
the true value of $\beta$ extracted in this 
 scenario increases with respect to SM fits in which this phase is absent. 
As a result of this modified value of 
 $\beta$, also the predicted value for $\varepsilon_K$ {\it increases} with respect to 
 the SM case, resulting in a better agreement with data. See Figure~3 in 
\cite{Buras:2010mh}.
 
 {\bf 4.}
The branching ratios ${\rm Br}(B_q\to \mu^+\mu^-)$ can be enhanced by an order 
of magnitude over the SM values in (\ref{BRSM1}) and (\ref{BRSM2}),
thus reaching the upper bounds from CDF and D0. Moreover, the relation 
in (\ref{bmumu}) appears to be basically unaffected by the presence of possible 
FBPs. On the other hand the CMFV relation (\ref{R1}) gets modified as follows
\be\label{eq:r}
\frac{{\rm Br}(B_s\to\mu^+\mu^-)}{{\rm Br}(B_d\to\mu^+\mu^-)}= \frac{\hat
B_{B_d}}{\hat B_{B_s}} \frac{\tau(B_s)}{\tau(B_d)} \frac{\Delta
M_s}{\Delta M_d}\,r\,,\quad r= \frac{M_{B_s}^4}{M_{B_d}^4}
\frac{|S_d|}{|S_s|}\,,
\ee
where for $r=1$ we recover the SM and CMFV relation 
derived in~\cite{Buras:2003td}. $S_q$ are the box functions that 
now are different for $B_s$ and $B_d$ systems.
In the present model
$r$ can deviate from one; however, this deviation is at most of $\ord(10\%)$.
We are looking forward to the test of (\ref{eq:r}) which will be possible 
once ${\rm Br}(B_q\to \mu^+\mu^-)$ will be known.

{\bf 5.} For large $S_{\psi\phi}$, the branching ratios 
${\rm Br}(B_{s,d} \to\mu^+\mu^-)$ are strongly enhanced over SM values as seen in 
Figure~4 of \cite{Buras:2010zm}.

\subsubsection{FBPs in the Higgs Potential}
If the FBPs are present only in the Higgs potential the relation between 
new phases $\varphi_{B_q}$ changes to
\be\label{R2a}
\varphi_{B_d}=\varphi_{B_s}
\ee
modifying certain aspects of the phenomenology. In particular the 
correlation between the CP asymmetries $S_{\psi K_S}$  and
$S_{\psi \phi}$ is now very different as seen in Figure~5 of 
\cite{Buras:2010zm} and in order to 
be consistent with the data on  $S_{\psi K_S}$ the asymmetry $S_{\psi \phi}$ 
cannot exceed 0.3. We will return to this in Section 6.

\subsubsection{Correlation between EDMs and $S_{\psi\phi}$}
In \cite{Buras:2010zm} the correlations
between EDMs and CP
violation in $B_{s,d}$ mixing in ${\rm 2HDM_{\overline{MFV}}}$ including
 FBPs in Yukawa interactions and  in the Higgs potential have been studied 
in detail.
It has been shown that in both cases the upper bounds on 
EDMs of the neutron and the atoms 
do not forbid sizable non-standard CPV effects in $B_{s}$ mixing.
However, if a large CPV phase in $B_s$ mixing will be confirmed, this
will imply hadronic EDMs very close to their present experimental bounds,
within the reach of the next generation of experiments, as well as
$ Br(B_{s,d}\to\mu^+\mu^-)$ typically largely enhanced over its
SM expectation. As demonstrated in Figure~5 of \cite{Buras:2010zm}
 the two flavour-blind CPV mechanisms can be distinguished
through the correlation between $S_{\psi K_S}$ and $S_{\psi\phi}$ that is
strikingly different if only one of them is relevant. Which of these two
CPV mechanisms dominates depends on the precise values of $S_{\psi\phi}$
and $S_{\psi K_S}$. Current data seems to show a mild preference for
 a {\it hybrid}
scenario where both these mechanisms are at work \cite{Buras:2010zm}.

\subsection{Littlest Higgs Model with T-parity}
We will next discuss two models having the operator structure of the 
SM but containing new sources of flavour and CP-violation. This is the Littlest 
Higgs Model with T-parity (LHT) and the SM4, the SM extended by a fourth sequential 
generation of quarks and leptons.

The Littlest Higgs model
without \cite{ArkaniHamed:2002qy} T-parity 
has been invented to solve the problem of the quadratic 
divergences in the Higgs mass without using supersymmetry.
In this approach the cancellation of divergences in $ m_H$ is achieved with 
the help of new particles of the same spin-statistics. 
Basically the SM Higgs is kept light 
because it is a pseudo-Goldstone boson of a spontaneously broken global 
symmetry:
\be 
SU(5)\to SO(5).
\ee
Thus the Higgs is protected from acquiring a large mass by a global symmetry, 
although in order to achieve this the gauge group has to be extended
to
\be
 G_{\rm LHT}=SU(3)_c\times [SU(2)\times U(1)]_1\times[SU(2)\times U(1)]_2
\ee
 and 
the Higgs mass generation properly arranged 
({\it collective symmetry breaking}). The 
dynamical origin of the global symmetry in question and 
the physics behind its 
breakdown are not specified. But in analogy to QCD one could imagine 
a new strong 
force at scales $\ord(10-20\tev)$ between  new very heavy fermions that bind 
together to produce the SM Higgs. In this scenario the SM Higgs is 
analogous to 
the pion. At 
scales well below $ 5\tev$ the Higgs is considered as an elementary particle but at
$20\tev$  its 
composite structure should be seen.  Possibly at these high scales one will
have to cope with non-perturbative strong dynamics and an unknown ultraviolet 
completion with some impact on 
low energy predictions of 
Little Higgs models has to be specified. Concrete perturbative 
completions, albeit very complicated, have been found 
\cite{Batra:2004ah,Csaki:2008se}.
The advantage of these models, relative to supersymmetry, is a 
much smaller number of free parameters but the disadvantage is 
that  
Grand Unification
in this framework is rather unlikely. Excellent reviews can be found in
\cite{Schmaltz:2005ky,Perelstein:2005ka}.

In order to make this model 
consistent with electroweak precision tests and simultaneously having
the new particles of this model in the reach of the LHC, a discrete symmetry,
T-parity, has been introduced \cite{Cheng:2003ju,Cheng:2004yc}. 
Under T-parity all SM particles are {\it even}.
Among the new particles only a heavy $+2/3$ charged T quark belongs to the
even sector. Its role is to cancel the quadratic divergence in the Higgs
mass generated by the ordinary top quark. The even sector and also the 
model without T-parity belong to the CMFV class if only flavour violation 
in the down-quark sector is considered \cite{Buras:2004kq,Buras:2006wk}.

More interesting from the point of view of FCNC processes is the T-odd
sector. It contains three doublets of mirror quarks and three doublets
of mirror leptons that communicate with the SM fermions by means of heavy 
$W^\pm_H$, $Z_H^0$ and $A^0_H$ gauge bosons that are also odd under 
T-parity. These interactions are 
governed by new mixing matrices $V_{Hd}$ and $V_{Hl}$ for down-quarks and
charged leptons, respectively. The corresponding matrices in the up ($V_{Hu}$)
and neutrino $(V_{H\nu})$ sectors are obtained by means of the relations 
\cite{Hubisz:2005bd,Blanke:2006xr}
\be
V_{Hu}^\dagger V_{Hd}=V_{CKM},\qquad V_{H\nu}^\dagger V_{Hl}=V^\dagger_{PMNS}.
\ee
 Thus we have new flavour and CP-violating contributions to decay amplitudes
in this model. These new interactions
 can have a structure very different from the CKM and PMNS matrices.

The difference between the CMFV models and the LHT model can be transparently 
seen in the formulation of FCNC processes in terms of the master functions. 
In the LHT model the real and universal master functions in
(\ref{rmasterf}) become complex quantities and the property of the universality 
is lost. Consequently the usual CMFV relations between $K$, $B_d$ and $B_s$ 
systems can be strongly broken. Explicitly, the new functions are given as 
follows ($i=K,d,s$):
\begin{equation}\label{eq31}
S_i\equiv|S_i|e^{i\theta_S^i},
\quad 
X^\ell_i \equiv \left|X^\ell_i\right| e^{i\theta_X^{i\ell}}, \quad 
Y_i \equiv \left|Y_i\right| e^{i\theta_Y^i}, \quad 
Z_i \equiv \left|Z_i\right| e^{i\theta_Z^i}\,,
\end{equation}

\begin{equation} \label{eq32}
D'_i \equiv \left|D_i'\right| e^{i\theta_{D'}^i}\,, \quad 
E'_i \equiv \left|E_i'\right| e^{i\theta_{E'}^i}\,.
\end{equation}

Let us note also that 
in contrast to the models discussed until now the LHT model 
contains
new heavy gauge bosons $W_H^\pm$, $Z^0_H$ and
 $A^0_H$. The masses of $W_H^\pm$ and
 $Z^0_H$ are typically $\ord(700\gev)$. $A_H$ is significantly lighter with
 a mass of a few hundred GeV and, being the lightest particle 
with odd T-parity, it can play the role of a dark matter candidate. 
The mirror quarks and leptons can have masses typically in the range 
500-1500 GeV and could be discovered at the LHC. 
Their impact on
FCNC processes can be sometimes spectacular. A review on flavour physics in 
the LHT model can be found in 
\cite{Blanke:2007ww} and selected papers containing details
of the pattern of flavour violation in these models can be found in
\cite{Blanke:2006sb,Blanke:2006eb,Blanke:2007db,Goto:2008fj,delAguila:2008zu,Blanke:2009pq,Blanke:2009am}. Critical discussions of the LHT model can be 
found in \cite{Grinstein:2009ex}.

Here we only list the most interesting results from our analyses \cite{Blanke:2006sb,Blanke:2006eb,Blanke:2007db,Blanke:2009pq,Blanke:2009am}.

{\bf 1.} $S_{\psi\phi}$ can be much larger than its SM value but 
 values above 0.3 are rather unlikely.

{\bf 2.} ${\rm Br(\klpn)}$ and ${\rm Br(\kpn)}$ can be enhanced up to factors of
 3 and 2.5, respectively. The allowed points in the $Br(\klpn)$ vs $Br(\kpn)$
 plot cluster around two branches. On one of them $Br(\kpn)$ can reach 
 maximal values while $Br(\klpn)$ is SM-like. Here $Br(\kpn)$ can easily 
reach the central experimental value of E949 Collaboration at Brookhaven 
  \cite{Artamonov:2008qb}.
On the other one $Br(\klpn)$ 
 can reach maximal values but $Br(\kpn)$ can be enhanced by at most  
 a factor of 1.4 and therefore not reaching the central experimental 
 value. 
Some insights on this behaviour have been provided in
\cite{Blanke:2009pq}.

{\bf 3.} Rare $B$-decays turn out to be SM-like but still some enhancements 
are possible. 
In particular $Br(B_{s,d}\to\mu^+\mu^-)$ can be enhanced by $30\%$ and a
significant part of this enhancement comes from the T-even sector.

{\bf 4.} Simultaneous enhancements of $S_{\psi\phi}$ and of
$Br(K\to\pi\nu\bar\nu)$ are rather unlikely.

{\bf 5.} $Br(\mu\to e\gamma)$ can reach the upper bound
of $2\cdot 10^{-11}$ from the MEGA collaboration and in fact some fine
tuning of the parameters is required to satisfy this bound   
\cite{Blanke:2007db,delAguila:2008zu,Blanke:2009am}:
either the
corresponding mixing matrix in the mirror lepton sector has to be at
least as hierarchical as the CKM matrix and/or the masses of mirror
leptons carrying the same electric charge must be quasi-degenerate.
Therefore if the MEG collaboration does not find anything at the level of
$10^{-13}$, significant fine tuning of the LHT parameters 
will be required in order to keep $\mu\to e\gamma$ under control. 

{\bf 6.} It is not possible to distinguish the LHT model from  
         the supersymmetric models discussed below on the basis of 
         $\mu\to e\gamma$ alone. On the other hand as pointed out in 
\cite{Blanke:2007db} such a distinction can be made by measuring any of the 
           ratios $Br(\mu\to 3e)/Br(\mu\to e\gamma)$, 
          $Br(\tau\to 3\mu)/Br(\tau\to \mu\gamma)$, etc. In supersymmetric
          models all these decays are governed by dipole operators so 
         that these ratios are $\ord(\alpha)$ 
\cite{Ellis:2002fe,Arganda:2005ji,Brignole:2004ah,Paradisi:2005tk,Paradisi:2006jp,Paradisi:2005fk}.
        In the
        LHT model the LFV decays with three leptons in the final state are
        not governed by dipole operators but by Z-penguins and box diagrams
        and the ratios in question turn out to be by at least one order of
        magnitude
        larger than in supersymmetric models. The most recent analysis of 
        LFV in the LHT model can be found in \cite{Goto:2010sn}.

 {\bf 7.} CP violation in the $D^0-\bar D^0$ mixing  at a level well 
beyond anything
possible with CKM dynamics has been identified \cite{Bigi:2009df}. 
Comparisons with CP violation
in $K$ and $B$ systems should offer an excellent test of this NP scenario and
reveal the specific pattern of flavour and CP violation in the $D^0-\bar D^0$
system predicted by this model.

\subsection{The SM with Sequential Fourth Generation}
One of the simplest extensions of the SM3
is the addition of a sequential fourth generation (4G)
of quarks and leptons \cite{Frampton:1999xi}
{(hereafter referred to as SM4)}. Therefore it is of interest to study 
its phenomenological implications. Beyond flavour physics possibly the 
most interesting implications of the presence of 4G are the following 
ones:

\begin{enumerate}
 \item While being consistent with electroweak precision data (EWPT) 
\cite{Maltoni:1999ta,He:2001tp,Alwall:2006bx,Kribs:2007nz,Chanowitz:2009mz,Novikov:2009kc}, 
a 4G can remove the tension between the SM3 fit and the lower bound
on the Higgs mass $m_H$ from LEP~II. Indeed, as pointed out in 
\cite{He:2001tp,Kribs:2007nz,Hashimoto:2010at}, a heavy Higgs boson does not contradict 
EWPT as soon as the 4G exists. For additional discussions see 
\cite{Erler:2010sk,Chanowitz:2010bm}

\item Electroweak baryogenesis might be viable \cite{Hou:2008xd,Kikukawa:2009mu,Fok:2008yg}.
\item Dynamical breaking of electroweak symmetry might be triggered by the presence of 
4G quarks
 \cite{Holdom:1986rn,Hill:1990ge,King:1990he,Burdman:2007sx,Hung:2009hy,Hung:2009ia,Holdom:2010za,Hashimoto:2010fp}.
\end{enumerate}

However, the SM4 is also interesting for flavour physics. Several analyses of
flavour physics \cite{Arhrib:2002md,Hou:2005yb,Hou:2006mx,Soni:2008bc,Soni:2010xh,Herrera:2008yf,Bobrowski:2009ng,Eberhardt:2010bm,Eilam:2009hz,Buras:2010pi,Buras:2010nd,Buras:2010cp,Hou:2008yb,Das:2010fh}
have been performed in the last years. 
The SM4
introduces three
new mixing angles $s_{14}$, $s_{24}$, $s_{34}$ and two new phases in the 
quark sector and can still have a significant impact on flavour 
phenomenology. Similarly to the LHT model it does not introduce any 
new operators but brings in new sources of flavour and CP violation 
that originate in the interactions of the four generation fermions 
with the ordinary quarks and leptons that are mediated by the SM 
electroweak gauge bosons. Thus in this model, as opposed to the LHT 
model, the gauge group is the SM one. This implies smaller number of 
 free parameters.

An interesting virtue of the SM4 model is the non-decoupling of new 
particles. Therefore, unless the model has non-perturbative Yukawa 
interactions, the 4G fermions are bound to be observed at the LHC with 
masses below $600\gev$.

Here I will only summarize the results of our analyses of quark flavour 
physics \cite{Buras:2010pi,Buras:2010nd} and lepton flavour violation 
\cite{Buras:2010cp}. Details can be found in these papers, in particular 
many correlations between various observables that are shown there
 in numerous plots. Two nice reviews of our work have been presented by 
Tillmann Heidsieck \cite{Heidsieck:2010ue}.

The most interesting patterns of flavour 
violation in the SM4 are the 
following ones:

{\bf 1.}
All existing tensions in the UT fits can be removed in this NP scenario.

{\bf 2.}
In particular the desire to explain the $S_{\psi\phi}$ anomaly implies
 uniquely the suppressions of 
 the CP asymmetries $S_{\phi K_S}$ and $S_{\eta' K_S}$ in agreement with the data.
 This correlation has been pointed out in \cite{Hou:2005yb,Soni:2008bc} and we confirmed 
 it. However we observed that for $S_{\psi\phi}$ significantly larger
 than 0.6 the values of $S_{\phi K_s}$ and $S_{\eta' K_s}$ are below their central values indicated by the data, although some non-perturbative uncertainties
 are involved here. 

{\bf 3.}
The $S_{\psi\phi}$ anomaly implies a sizable enhancement of ${\rm Br}(B_s\to \mu^+\mu^-)$
over the SM3 prediction although this effect is much more modest than 
in SUSY models where the Higgs penguin  with large $\tan\beta$ is at work. Yet,
values as high as $8\cdot 10^{-9}$ are certainly possible in the SM4, 
which is well beyond those attainable in the LHT model discussed 
previously  and  the RSc model 
discussed below. On the other hand large values of $S_{\psi\phi}$ preclude 
non-SM values of $Br(B_d\to\mu^+\mu^-)$. Consequently the CMFV relations 
in (\ref{bmumu}) and (\ref{R1}) can be strongly violated in this model.

{\bf 4.}
Possible enhancements of ${\rm Br}(\kpn)$ and ${\rm Br}(\klpn)$ over the SM3 values
are much larger than found in the LHT and RSc models and in particular in
SUSY flavour models discussed below, where they are SM3 like. Both branching ratios as high
as several $10^{-10}$ are still possible in the SM4.
 Moreover, in this case, the two branching ratios are strongly correlated
 and close to the Grossman-Nir bound \cite{Grossman:1997sk}.

{\bf 5.}
Interestingly, in contrast to the LHT and RSc models, a high value of 
$S_{\psi\phi}$ does not preclude a sizable enhancements of ${\rm Br}(\kpn)$, 
and ${\rm Br}(\klpn)$.

{\bf 6.}
NP effects in $K_L\to\pi^0\ell^+\ell^-$ and $K_L\to\mu^+\mu^-$ can be visibly
larger than in the LHT and RSc models. In particular 
${\rm Br}(K_L\to\mu^+\mu^-)_{\rm SD}$ can easily violate the existing bound of
$2.5\cdot 10^{-9}$ \cite{Isidori:2003ts}.  Imposition of this bound on top of other constraints 
results in a characteristic shape of the correlation between 
${\rm Br}(\kpn)$ and ${\rm Br}(\klpn)$ that we already encountered in the LHT 
model.

{\bf 7.}
 Even in the presence of SM-like values for $S_{\psi\phi}$ and ${\rm Br}(B_s\to \mu^+\mu^-)$, large effects in the $K$-system are possible.

{\bf 8.}
For large positive values of $S_{\psi\phi}$ the predicted value of $\epe$ is significantly below the data, unless the
 hadronic matrix elements of the electroweak penguins are sufficiently suppressed with respect to the large N result and the ones of QCD penguins enhanced.
We have also reemphasised \cite{Buras:1998ed,Buras:1999da} the important role $\epe$ will play in bounding rare $K$ decay branching ratios once the relevant hadronic matrix elements in $\epe$ will be precisely known.

 {\bf 9.} While simultaneous large 4G effects in the $K$ and $D$ systems are possible, large effects in $B_d$ generally disfavour large NP effects in the $D$ system.
  Moreover, significant enhancement of $S_{\psi\phi}(B_s)$ above the SM3 value will not allow large CP violating effects in the $D$ system within the 4G scenario.
Additional imposition of the $\epe$ constraint significantly diminishes 4G effects in CP violating observables in the $D$ system.

{\bf 10.}
The branching ratios for $\ell_i\to\ell_j\gamma$, $\tau\to\ell\pi$, $\tau\to\ell\eta^{(\prime)}$, $\mu^-\to e^-e^+e^-$, $\tau^-\to e^-e^+e^-$,
 $\tau^-\to \mu^-\mu^+\mu^-$, $\tau^-\to e^-\mu^+\mu^-$ and 
$\tau^-\to \mu^-e^+e^-$ can all still be as large as the present experimental 
upper bounds but not necessarily simultaneously.

{\bf 11.}
The correlations between 
various LFV branching ratios should allow to test this model. This should  be contrasted with the SM3 where all these branching ratios are unmeasurable.

{\bf 12.}
The rate for  $\mu-e$ conversion in nuclei can also reach the corresponding 
upper bound.

{\bf 13.}
The pattern of the LFV branching ratios in the SM4 differs significantly from the one encountered in the MSSM, allowing to distinguish these two models with the help of LFV processes in a transparent manner. 
Also differences from the LHT model are identified.

{\bf 14.}
The  branching ratios for $K_L \to \mu e$, $K_L \to \pi^0\mu e$, $B_{d,s} \to \mu e$, $B_{d,s} \to \tau e$ and $B_{d,s} \to \tau\mu $ turn out to be by several 
orders of magnitude smaller than the present experimental bounds.

In summary, the SM4 offers a very rich pattern of flavour violation which can be tested already in the coming years, in particular through precise measurements
of $S_{\psi\phi}$, ${\rm Br}(B_q\to \mu^+\mu^-)$, ${\rm Br}(K^+\to \pi^+\nu\bar\nu)$
and, later, $S_{\phi K_s}$, $S_{\eta' K_s}$ and ${\rm Br}(K_L\to \pi^0\nu\bar\nu)$. 
Also, precise measurements of the phase $\gamma\approx \delta_{13}$ and 
of LFV decays 
will be important for these investigations.

\subsection{Supersymmetric Flavour Models (SF)}
In supersymmetric models the cancellation of divergences in $m_H$ 
is achieved with the help of new particles of different spin-statistics 
than the SM particles: supersymmetric particles.
 For this approach to work, these new particles should have masses 
 below 1 TeV, otherwise  fine tuning of parameters cannot be avoided. As 
none of the 
supersymmetric particles 
has been seen so far, the MSSM became a rather fine tuned scenario even if much
less than the SM in the presence of the GUT and Planck scales. 
One of the important predictions of 
the 
simplest realization of this scenario, the MSSM with R-parity, 
is  light Higgs with 
$m_H\le 130\gev$ and one of its virtues is its perturbativity up to the GUT
scales.  An excellent introduction
to the MSSM can be found in \cite{Martin:1997ns}.

The unpleasant feature of the MSSM is a large number of parameters residing 
dominantly in the soft sector that has to be introduced in the process of 
supersymmetry breaking. Constrained 
versions of the MSSM can reduce the number of parameters significantly. 
The same is true in the case of the MSSM with MFV. 

Concerning the FCNC processes let us recall that  in addition to a 
light Higgs, 
squarks, sleptons, gluinos, charginos and
 neutralinos, also charged Higgs particles $H^{\pm}$ and additional 
 neutral scalars are present in this framework. 
All these particles can contribute to FCNC transitions through box and 
penguin diagrams. 
New sources of flavour
and CP violation come from the misalignement of quark and squark mass 
matrices
 and similar new flavour and CP-violating effects are present in the lepton
sector. Some of these effects can be strongly enhanced at large $\tan\beta$
and the corresponding observables provide stringent constraints on
the parameters of the MSMM.
In particular $B_s\to \mu^+\mu^-$ can be enhanced up to its experimental
upper bound, branching ratios for $K\to\pi\nu\bar\nu$ can be much larger than
their SM values and the CP asymmetry $S_{\psi\phi}$ can also strongly
deviate from the tiny SM value.

The Higgs sector of the MSSM is at the tree level the same as of the 
2HDM II: only one Higgs doublet couples to a fermion of a given charge 
and there are no FCNCs mediated by Higgs particles.
However, at one-loop level  this is no longer true and the Higgs-penguins, 
in analogy to Z-penguins are born. At large $\tan\beta$ they can be very 
important with their ''smoking gun'' being the order of magnitude enhancements 
of  $B_{s,d}\to \mu^+\mu^-$. When non-MFV sources of flavour and CP violation 
in the squark  sector are present also the asymmetry $S_{\psi\phi}$ 
can be strongly enhanced. However, there is a  striking
 difference between ${\rm 2HDM_{\overline{MFV}}}$ and the MSSM with MFV 
which we already mentioned before:
\begin{itemize}
\item 
While in the ${\rm 2HDM_{\overline{MFV}}}$
 large values of $S_{\psi\phi}$ are possible, this is not the case of the 
MSSM with MFV, even in the presence of FBPs: supersymmetric relations 
between parameters do not allow for such an enhancement 
\cite{Altmannshofer:2009ne,Blum:2010mj}.
\end{itemize}

There is a very rich literature on flavour violation in supersymmetric 
theories. A rather complete collection of references can be found in 
a paper from my group \cite{Altmannshofer:2009ne}, where the supersymmetric 
flavour (SF) models have 
been analyzed in great detail. We will now confine our discussion to these 
models.

The general MSSM framework with very many new flavour parameters in the
soft sector is not terribly predictive and is plagued by flavour and CP problems:
FCNC processes and electric dipole moments are generically well above
the experimental data and upper bounds, respectively. Moreover the MSSM
framework addressing primarily the gauge hierarchy problem and the
quadratic divergences in the Higgs mass does not provide automatically the
hierarchical pattern of quark and lepton masses and of FCNC and CP 
violating interactions.

Much more interesting from this point of view are supersymmetric 
models with flavour symmetries that allow for a simultaneous understanding of 
the flavour structures in the Yukawa couplings and in SUSY soft-breaking 
terms, adequately suppressing FCNC and CP violating phenomena and solving
SUSY flavour and CP problems.

The SF models can be divided into two broad
classes depending on whether they are based on abelian or non-abelian 
flavour symmetries. Moreover, their phenomenological output crucially
depends on whether the flavour and CP violations are governed by 
left-handed (LH)
currents or there is an important new right-handed (RH) current component
 \cite{Altmannshofer:2009ne}.
They can be considered as generalisations of the Froggatt-Nielsen mechanism
for generating hierarchies in fermion masses and their interactions but
are phenomenologically much more successful than the original 
Froggatt-Nielsen model \cite{Froggatt:1978nt}. There is a rich 
literature on SF models and I cannot  refer
to all of them here. Again a  rather complete list of references can be 
found in \cite{Altmannshofer:2009ne}. I will now summarize the 
results obtained in this paper.
See also \cite{Altmannshofer:2009ap}.

In \cite{Altmannshofer:2009ne} we have
performed an extensive study of processes governed by $b\to s$ transitions 
in the SF models and of their correlations with processes governed by 
$b\to d$ transitions, 
$s\to d$ transitions, $D^0-\bar D^0$ mixing, LFV 
decays, electric dipole moments and $(g-2)_{\mu}$. 
Both abelian and non-abelian flavour models have been considered as well as the
flavour blind MSSM (FBMSSM) and the MSSM with MFV. It has been shown how
 the characteristic patterns of correlations among the considered flavour 
observables allow to distinguish between these different SUSY scenarios and 
also to distinguish them from RSc and LHT scenarios of NP.

Of particular importance in our study were the correlations between 
the CP asymmetry $S_{\psi\phi}$ and
$B_s\rightarrow\mu^+\mu^-$, between the observed anomalies in 
$S_{\phi K_s}$ and $S_{\psi\phi}$, between 
$S_{\phi K_s}$ and $d_e$, between $S_{\psi\phi}$ and $(g-2)_{\mu}$ and 
also those involving LFV decays.

In the context of our study of the SF models we have analysed the 
following representative scenarios:
\begin{itemize}
\item [1.] Dominance of RH currents 
(abelian model by Agashe and Carone\cite{Agashe:2003rj}),
\item [2.] Comparable LH and RH currents with CKM-like mixing
  angles represented by the special version (RVV2) 
of the non abelian $SU(3)$ 
model by
Ross, Velasco and Vives \cite{Ross:2004qn} as discussed in 
\cite{Calibbi:2009ja}. 
\item [3.] In the second non abelian $SU(3)$ 
 model by Antusch, King and Malinsky (AKM) \cite{Antusch:2007re} analyzed 
by us the RH contributions are CKM-like but 
 new LH contributions in contrast to the RVV2 model can be 
suppressed arbitrarily at the high scale. Still 
they can be generated by RG effects at lower scales. To first approximation 
the version of this model considered by us can be characterized by NP being
dominated by CKM-like RH currents.
\item [4.] Dominance of CKM-like LH currents in non-abelian 
models~\cite{Hall:1995es}.
\end{itemize}

In the choice of these four classes of flavour models, we were guided by our
model independent analysis in Section 2 of our paper, that I cannot present 
 here because of the lack of 
space. Indeed  these three scenarios
predicting  quite 
different patterns of flavour violation should give a good representation of
most SF models discussed in the literature.
The distinct patterns of flavour violation found in each scenario have
been illustrated with several  plots that can  be found
in figures 11-14 of   \cite{Altmannshofer:2009ne}.

The main messages from our analysis of the models in question are as 
follows:

{\bf 1.}
Supersymmetric models with RH currents (AC, RVV2, AKM) and those with 
exclusively LH currents
can be globally distinguished by the values of the CP-asymmetries 
$S_{\psi\phi}$ and $S_{\phi K_S}$ with the following important result: 
none of
the models considered by us can simultaneously explain the $S_{\psi\phi}$ and
$S_{\phi K_S}$ anomalies observed in the data. 
In the models with RH currents,
$S_{\psi\phi}$ can naturally be much larger than its SM value, while 
$S_{\phi K_S}$ remains either SM-like or its correlation with $S_{\psi\phi}$ 
is inconsistent with the data. 
On the contrary, in the models with LH currents only,
$S_{\psi\phi}$ remains SM-like, while the  $S_{\phi K_S}$  anomaly can be 
easily explained.
Thus already future precise measurements of 
$S_{\psi\phi}$ and $S_{\phi K_S}$ will select one of these two classes of 
models, if any.

{\bf 2.}
The desire to explain the $S_{\psi\phi}$ anomaly within the models with
RH currents unambiguously implies, in the case of the AC and the AKM models,
values of
${\rm Br}(B_s\to\mu^+\mu^-)$ as high as several $10^{-8}$. In the 
RVV2 model such values are also possible but not necessarily implied
by the large value of $S_{\psi\phi}$. Interestingly, in all these models large
values of $S_{\psi\phi}$ can also privide the 
solution to the $(g-2)_\mu$ anomaly. Moreover, the ratio 
${\rm Br}(B_d\to\mu^+\mu^-)/{\rm Br}(B_s\to\mu^+\mu^-)$ in the AC and RVV2  models is 
dominantly below its MFV prediction and can be much smaller than the latter.
In the AKM model this ratio stays much closer to the MFV value of roughly 
$1/33$ \cite{Buras:2003td,Hurth:2008jc} and can be smaller or larger than 
this value with equal probability.
Still, values of $Br(B_d\to\mu^+\mu^-)$ as high as $1\times 10^{-9}$ are
possible in all these models.

{\bf 3.}
In the RVV2 and the AKM models, a large value of $S_{\psi\phi}$  combined with
the desire to explain the $(g-2)_\mu$ anomaly implies 
$Br(\mu\to e\gamma)$ in 
the reach of the MEG experiment.  In the case of the RVV2 model,
$d_e\ge 10^{-29}$ e.cm. is predicted, while in the AKM model it is typically
smaller.
 Moreover, in the case of the RVV2 model, 
$Br(\tau\to\mu\gamma)\ge 10^{-9}$ is then
 in the reach of Super-B machines, while this is not the case in the AKM model.

{\bf 4.}
Next, while the abelian AC model resolves the present UT tensions 
\cite{Lunghi:2008aa,Buras:2008nn,Lunghi:2009sm,Buras:2009pj,Lunghi:2009ke,Lunghi:2010gv,UTfit-web,Lenz:2010gu} to be discussed 
in Section 6
through the modification of the ratio $\Delta M_d/\Delta M_s$, the 
non-abelian flavour models RVV2 and AKM provide the solution through
NP contributions to $\epsilon_K$. Moreover, while the AC model predicts
sizable NP contributions to $D^0-\bar D^0$ mixing, such contributions 
are tiny in the RVV2 and AKM models.

{\bf 5.} The hadronic EDMs represent very sensitive probes of SUSY flavour 
models with RH
currents. In the AC model, large values for the neutron EDM might be easily 
generated by both the
up- and strange-quark (C)EDM. In the former case, visible CPV effects in 
$D^0-\bar D^0$ mixing
are also expected while in the latter case large CPV effects in the
$B_s$ system are unavoidable.
The RVV2 and AKM models predict values for the down-quark (C)EDM and, hence for the neutron
EDM, above the $\approx 10^{-28}e$~cm level 
when a large $S_{\psi\phi}$
is generated. All the above models predict a large strange-quark (C)EDM, hence, a reliable
knowledge of its contribution to the hadronic EDMs, by means of lattice QCD techniques, would
be of the utmost importance to probe or to falsify flavour models embedded in a SUSY framework.

{\bf 6.}
In the supersymmetric models with exclusively LH currents, 
the desire to explain
the $S_{\phi K_S}$ anomaly implies also  the solution to the 
$(g-2)_\mu$ anomaly and the direct CP asymmetry in $b\to s\gamma$ much
larger than its SM value. This is in contrast to the models with RH currents
where this asymmetry remains SM-like.

{\bf 7.}
Interestingly,  in the LH-current-models, the ratio 
$Br(B_d\to\mu^+\mu^-)$ over $Br(B_s\to\mu^+\mu^-)$ can not only deviate
significantly from its MFV value of approximately $1/33$, 
but in contrast to the models with 
RH currents considered by us can also be much larger than the latter value. 
Consequently,
$Br(B_d\to\mu^+\mu^-)$ as high as $(1-2)\times 10^{-9}$ is still 
possible while being consistent with the bounds on all other observables,
in particular the one on $Br(B_s\to\mu^+\mu^-)$. Also interesting
correlations between $S_{\phi K_S}$ and CP asymmetries in 
$B\to K^*\ell^+\ell^-$ are found.

{\bf 8.}
The branching ratios for 
$K\to\pi \nu\bar\nu$ decays in the supersymmetric models considered by us
remain SM-like and can be distinguished from RSc and LHT models where  
they can be significantly enhanced.

{\bf 9.}
In \cite{Altmannshofer:2010ad} a closer look at CP violation in 
$D^0-\bar D^0$ mixing within the SUSY alignment models has been made 
(see also point {\bf 5} above). 
Such models naturally account for large, non-standard
effects in $D^0-\bar D^0$ mixing and within such models
detectable CP violating effects in $D^0-\bar D^0$ mixing would
unambiguously imply a lower bound for the electric dipole moment (EDM)
of hadronic systems, like the neutron EDM and the mercury EDM, in the
reach of future experimental sensitivities. This correlation distinguishes 
the alignment models from 
gauge-mediated SUSY breaking models, SUSY models with MFV and non-Abelian
SUSY flavour models discussed above. 

\subsection{ Supersymmetric $SU(5)$ GUT with RH  neutrinos (RN): 
$SSU(5)_{\rm RN}$}
We will next consider a supersymmetric $SU(5)$ GUT enriched by 
 right-handed neutrinos ($SSU(5)_{\rm RN}$) accounting for the neutrino
 masses and mixing angles by means of a type-I see-saw mechanism.
Since SUSY-GUTS generally predict FCNC and CP violating
processes to occur both in the leptonic and hadronic sectors,
 we have performed in \cite{Buras:2010pm} an
extensive study of FCNC and CP Violation in both sectors, analyzing possible
hadron/lepton correlations among observables. In particular, we have monitored
how in this framework the tensions observed in the UT analysis can be 
resolved. For previous studies of hadron/lepton correlations 
in this framework see \cite{correlations}.

It should be emphasized that in this 
framework, in which flavour symmetries of the type discussed in the previous 
subsection are absent, it is not possible to link model independently 
transitions like $\mu\to e \gamma$ and $b \to s$  or 
$\tau \to \mu\gamma$ and $s\to d$.
On the other hand
the correlations between leptonic and hadronic processes taking place 
between the same
generations like $\mu\to e\,\gamma$ and $s\to d$ or  $\tau\to\mu\,\gamma$ and $b \to s$ transitions exist.

The main results of our study of the  $s\to d$ transitions and of 
their correlations with the $\mu\to e$ transitions are

{\bf 1.}
Sizable SUSY effects in $\varepsilon_{K}$, that might be desirable to solve 
the UT anomalies,
generally imply a lower bound for $Br(\mu\to e\gamma)$ in the reach of the MEG experiment. Furthermore, the simultaneous requirement of an explaination for both the $(g-2)_\mu$ and the UT
anomalies would typically imply $Br(\mu\to e\gamma)\geq 10^{-12}$. 

{\bf 2.} The requirement of sizable non-standard effects in $\varepsilon_{K}$ always implies
large values for the electron and neutron EDMs, in the reach of the planned experimental resolutions.

{\bf 3.}
Finally, as in the SF models,
 $Br(K^{0}_{L}\to \pi^{0}\nu\bar{\nu})$ and $Br(K^{+}\to \pi^{+}\nu\bar{\nu})$ 
 remain SM-like. 

The main results of our study of the $b\to s$ transitions and of
their correlations with the  $\tau\to\mu$ transitions are

{\bf 4.} Non-standard values for $S_{\psi\phi}$ imply a lower bound for 
$ Br(\tau\to\mu\gamma)$
within the reach of Belle II and the SFF. However, the $(g-2)_\mu$ 
anomaly can be solved only for large $\tan\beta$
values where we find $|S_{\psi\phi}|\leq 0.2$ for $\Delta a^{\rm SUSY}_{\mu}\geq 1\times 10^{-9}$
while being still compatible with the constraints from 
$Br(\tau\to\mu\gamma)$.

{\bf 5.}
$S_{\psi K_S}$ remains SM-like to a very good extent and consequently the 
solution of the UT anomalies by means of CPV effects in $b\to d$ mixing 
is not possible. 
However,
the UT anomalies can be solved by means of a negative NP contribution 
to  $\Delta M_d/\Delta M_s$,
implying a lower bound for $Br(\tau\to\mu\gamma)$ within the 
reach of Belle II and the SFF  and large
values for the angle $\gamma$. This scenario will be probed or falsified quite soon at the LHCb through a precise tree level measurement of the latter UT 
angle.

{\bf 6.} Both $Br(B_s\to\mu^+\mu^-)$ and $Br(B_d\to\mu^+\mu^-)$ can reach
large non-standard values. However, sizable departures from the MFV prediction
$Br(B_s\to\mu^+\mu^-)/Br(B_d\to\mu^+\mu^-)\approx |V_{ts}/V_{td}|^2$ would
imply large values for $Br(\tau\to\mu\gamma)$, again well within the 
reach of Belle II and the SFF.

{\bf  7.} The dileptonic asymmetry $A^{b}_{\text{SL}}$ can sizably depart from the SM
expectations but the large value reported by the Tevatron~\cite{Abazov:2010hv}
cannot be accounted for within  the $SSU(5)_{\rm RN}$ model. In particular, we
find that $A^{b}_{\text{SL}} \approx 0.5~A^{s}_{\text{SL}}$ since $A^{d}_{\text{SL}}$
remains SM-like.

{\bf 8.} We also find that
CPV effects in $D^{0}-\bar{D}^{0}$ mixing are negligibly small.

{\bf 9.} The asymmetry $S_{\phi K_S}$ can sizably depart from the SM expectations
and it turns out to be correlated with $S_{\psi\phi}$. 
 In particular, it is possible to simultaneously account for an enhancement of
 $S_{\psi\phi}$ and a suppression of
$S_{\phi K_S}$ (relative to $S_{\psi K_S}$) as required by the data.

This model shares many properties with the RVV2 model but the last 
property on our list provides  an important distinction 
 between $SSU(5)_{\rm RN}$ and the RVV2 model as well as other 
SF models discussed in the previous subsection which we want to emphasize 
here:

{\bf 10.} As stressed in the point {\bf 1} of the previous subsection none of the models discussed in \cite{Altmannshofer:2009ne} was able
to simultaneously account for the current data for 
$S_{\psi\phi}$ and $S_{\phi K_S}$,
in contrast to the $SSU(5)_{\rm RN}$ model discussed here. 
The reason for this can be
traced back recalling that the suppression of
 $S_{\phi K_S}$ relative to $S_{\psi K_S}$ as required by the data 
can be achieved
only at moderate/low $\tan\beta$ values. For such values of $\tan\beta$
 $S_{\psi\phi}$ receives the dominant
contributions from gluino/squark boxes, which in principle could allow 
simultaneously an enhancement of this asymmetry. Yet such effects are strongly constrained in the SF models
either by $D^0-\bar D^0$ mixing (in the case of Abelian flavour models) or by $K^0-\bar K^0$
mixing (in  the case of non-Abelian flavour models). Consequently, in this region 
of parameter space $S_{\psi\phi}$ cannot be large in these models.
In fact  $S_{\psi\phi}$ receives in these models large values
only at large $\tan\beta$ (by means of the double Higgs penguin exchange) where
the sign of the correlation between $S_{\psi\phi}$ and $S_{\phi K_S}$ is 
found to be opposite
to data ~\cite{Altmannshofer:2009ne}, that is $S_{\phi K_S}$ is enhanced rather 
than suppressed.
In constrast, the $SU(5)_{\rm RN}$ model predicts unobservable effects for $D^0-\bar D^0$
mixing while the NP effects in $K^0-\bar K^0$ are generally unrelated to those entering
$B^s-\bar B^s$ mixing and therefore the tight bounds from $\epsilon_K$ can be always avoided. Therefore at moderate/low $\tan\beta$ the suppression of 
$S_{\phi K_S}$ and simultaneous sizable enhancement of $S_{\psi\phi}$ can 
be obtained.

{\bf 11.} Also the comparison with models dominated by LH-currents such as 
$\delta$LL SF models, FBMSSM discussed below or the SM4 is 
interesting.
Indeed in these models the asymmetries $S_{\eta^{\prime} K_S}$ and $S_{\phi K_S}$
 are either both suppressed or both enhanced relative to the $S_{\psi K_S}$.
In the case of  $SSU(5)_{\rm RN}$ they exhibit opposite deviations from 
the latter asymmetry. That is a suppression of $S_{\phi K_S}$ as required 
by the data implies an enhancement of $S_{\eta^{\prime} K_S}$ that is still 
consistent with the data within the experimental uncertainties even if 
a slight suppression is favoured.

Finally, in this context let us recall that within the SM4
the correlation between $S_{\psi\phi}$ and in $S_{\phi K_S}$ is qualitatively similar to the
one found in the $SSU(5)_{RN}$ model, that is with increasing $S_{\psi\phi}$ the asymmetry
$S_{\phi K_S}$ decreases in accordance with the data~\cite{Hou:2005yb,Soni:2008bc,Buras:2010pi}. In this sence the correlation between 
$S_{\eta^{\prime} K_S}$ and $S_{\phi K_S}$ provides an interesting distinction 
between
$SSU(5)_{RN}$ and the SM4 as we just emphasized.

 \subsection{The flavour blind MSSM (FBMSSM)}
The flavour blind MSSM (FBMSSM) scenario \cite{Baek:1998yn,Baek:1999qy,Bartl:2001wc,Ellis:2007kb,Altmannshofer:2008hc} having new FBPs in the soft sector
belongs actually to the class of MFV models but as the functions $F_i$
become complex quantities and it is a supersymmetric framework 
 we mention this model here. In fact our analysis of this scenario 
in \cite{Altmannshofer:2008hc} preceded our detailed analysis of SF models
that we just summarized.

The FBMSSM has fewer parameters than the general MSSM
 implying striking correlations
between various observables that we list below. 
These correlations originate in the fact that 
the SUSY contributions to $S_{\phi K_{S}}$,
$A_{CP}(b\to s\gamma)$ and the EDMs are generated by the same CP
violating invariant $A_{t}\mu$. On the operator level the magnetic 
photon penguin operator in the case of $b\to s\gamma$ and magnetic 
gluon penguin in the case  $S_{\phi K_{S}}$ play here the crucial role 
in the NP sector and as these are dipole operators also  correlations 
with EDMs and under mild assumptions with $(g-2)_\mu$ exist. While 
this framework is of MFV or even better of $\overline{\rm MFV}$ type, 
it does not belong to the CMFV framework as scalar exchanges can 
enhance $B_{s,d}\to\mu^+\mu^-$ by an order of magnitude. 

The main messages from this analysis are as follows:

{\bf 1.}We find that $S_{\phi K_{S}}$ and $S_{\eta^{\prime} K_{S}}$ can both
 differ from $S_{\psi K_{S}}$
 with the effect being typically by a factor of
$1.5$ larger in $S_{\phi K_{S}}$ in agreement with the pattern observed in the data.
Most interestingly, we find that the desire of reproducing the observed low values of $S_{\phi K_{S}}$
and $S_{\eta^{\prime} K_{S}}$ implies uniquely:

{\bf 2.}
Lower bounds on the electron and neutron EDMs $d_{e,n} \gtrsim 10^{-28}\,e\,$cm.

{\bf 3.}
Positive and sizable (non-standard) $A_{CP}(b\to s\gamma)$ asymmetry in the ballpark
of $1\%-5\%$, that is having an  opposite sign to the SM one.

{\bf 4.} The NP effects in $S_{\psi K_{S}}$ and $\Delta M_{d}/\Delta M_{s}$ are very small
so that these observables determine the coupling $V_{td}$, its phase $-\beta$ and its
magnitude $|V_{td}|$, without significant NP pollution. 
In particular we find 
$\gamma=63.5^{\circ}\pm 4.7^\circ$ and $|V_{ub}|\!=\! (3.5\pm 0.2)\cdot 10^{-3}$.
We remark that the latter value differs from the one coming from inclusive 
determinations and consequently from the one favoured by the RHMFV scenario 
discussed soon.

{\bf 5.} $|\epsilon_K|$ turns out to be uniquely enhanced over its SM value up to a level of
$ 15\%$  softening the $\varepsilon_K$ anomaly that we will discuss in 
the next section.

{\bf 6.} Only small effects in $S_{\psi\phi}$ which could, however, be still visible through
the semileptonic asymmetry $A^{s}_{SL}$.

{\bf 7.}  A natural explanation of the $(g-2)_{\mu}$ anomaly (under very mild assumptions).

The FBMSSM belongs to the more general models with MFV discussed in
\cite{Kagan:2009bn} and shares several properties with supersymmetric 
models of the $\delta$LL type. 
The major difference discriminating these two scenarios regards 
their predictions for the
leptonic and hadronic EDMs. The lower bounds on these observables 
are significantly stronger within the FBMSSM if one wants to eliminate the
$S_{\phi K_{S}}$ anomaly. Thus the SUSY
non-MFV models with purely left-handed currents like $\delta$LL
can easier survive the future data. Yet both models will have problems 
if the UT-tensions and the $S_{\psi\phi}$ anomaly will be confirmed 
by more accurate data.

\subsection{The Minimal Effective Model with Right-handed Currents:RHMFV}
One of the main properties of the Standard Model (SM) regarding flavour 
 violating processes is the LH 
 structure of the charged currents that is 
 in accordance with the maximal violation 
 of parity observed in low energy processes. LH charged currents encode
 at the level of the Lagrangian the full information about flavour mixing 
 and CP violation represented compactly by the CKM matrix.
 Due to the GIM
 mechanism this structure has automatically profound implications for
 the pattern of FCNC processes that seems to be remarkably
 in accordance with the 
 present data within theoretical and experimental uncertainties, bearing in 
 mind certain anomalies which will be discussed below and in the next Section.

 Yet, the SM is expected to be only the low-energy limit of a more fundamental
 theory in which in principle parity could be a good symmetry implying the existence of
 RH charged currents. Prominent examples of such fundamental 
 theories are  left-right symmetric models on which a rich literature exists.
 We have seen that some SF models discussed above contained RH currents as well.

 Left-right symmetric models were born 35 years 
 ago~\cite{Pati:1974vw,Pati:1974yy,Mohapatra:1974gc,Mohapatra:1974hk,Senjanovic:1975rk,Senjanovic:1978ev}
 and extensive
 analyses of many observables can be found in the literature~(see 
 e.g.~\cite{Zhang:2007da,Maiezza:2010ic,Guadagnoli:2010sd} and references therein).
 Renewed theoretical interest in models with an underlying $SU(2)_L \times SU(2)_R$ 
 global symmetry has also been motivated by Higgsless 
models~\cite{Csaki:2003zu,Nomura:2003du,Barbieri:2003pr,Georgi:2004iy}. However,
 the recent phenomenological interest in making another look  at the right-handed 
 currents in general, and not necessarily in the context of a given left-right 
 symmetric model, originated in tensions between inclusive and exclusive
 determinations of the elements of the CKM matrix $|V_{ub}|$ and  $|V_{cb}|$. 
 It could be that these tensions are due to the underestimate of theoretical 
 and/or experimental uncertainties. Yet, it is a fact, as pointed out
 and analyzed recently in particular in \cite{Crivellin:2009sd,Chen:2008se}, 
 that the presence of right-handed  currents could either remove 
 or significantly weaken some of these tensions, especially in the 
 case of $|V_{ub}|$.

Assuming that RH currents provide the solution to
 the problem at hand, there is an important question whether the strength
 of RH currents required for this purpose is consistent with
 other observables and whether it implies new effects somewhere else that
 could be used to test this idea more globally.

This question has been addressed in \cite{Buras:2010pz}.
The starting point of our analysis is the assumption that 
the SM is the low-energy limit of a more
fundamental theory. We don't know the exact structure of this
theory, but we assume that in the high-energy limit it is 
left-right symmetric. The difference of LH and 
 RH sectors observed in the SM is 
only a low-energy property, due to appropriate symmetry-breaking 
terms.

In particular, we assume that the theory has 
a $SU(2)_L \times SU(2)_R \times U(1)_{B-L}$ global 
symmetry, explicitly broken only in the Yukawa 
sector and by the $U(1)_Y$ gauge coupling. 
Under this symmetry the SM quark fields can be grouped 
into three sets of LH or RH doublets with $B-L$ charge $1/3$:
\be
\qquad 
Q^i_{L} = \left( \ba{c} u_{L}^i \\ d_{L}^i \ea \right)~, \qquad 
Q^i_{R} = \left( \ba{c} u_{R}^i \\ d_{R}^i \ea \right)~, \qquad 
i=1 \ldots 3~.
\ee
With this assignment the SM hypercharge is given by $Y=T_{3R} + (B-L)/2$.  
In order to recover the SM electroweak gauge group,  
we assume that only the $SU(2)_L$ and $U(1)_Y$ subgroups 
of $SU(2)_L \times SU(2)_R \times U(1)_{B-L}$ are effectively 
gauged below the TeV scale. In close analogy we can 
introduce three sets of LH and RH 
leptons, $L^i_L$ and $L_R^i$ (including three 
RH neutrinos), with $B-L=-1$.

In our effective theory approach for the 
study of RH currents \cite{Buras:2010pz}
 the central role is played by  a left-right symmetric flavour group
 $SU(3)_L \times SU(3)_R$, commuting with 
an underlying $SU(2)_L \times SU(2)_R \times U(1)_{B-L}$ global symmetry and
broken only by two Yukawa couplings. The model contains a new 
unitary matrix $\Vt$  controlling flavour-mixing in the RH sector 
and can be considered as the minimally flavour violating generalization 
to the RH sector. Thus bearing in mind that this model contains non-MFV
interactions from the point of view of the standard MFV hypothesis that
includes only LH charged currents, we decided to  call this model RHMFV.

The new mixing matrix $\Vt$  can be parametrized in terms 
of 3 real mixing angles
and 6 complex phases. Adopting the standard CKM phase convention, 
where the 5 relative phases of the quark  fields are adjusted to 
remove 5 complex phases from the CKM matrix, we have no more
freedom to remove the 6 complex phases from  $\Vt$.  In passing let 
us remark that the same comments apply to the $V_{Hd}$ and $V_{Hl}$ 
matrices in the mirror fermion sector of the LHT model.

In the
standard CKM basis  $\Vt$ can be 
parametrized as follows
\be
\Vt = D_U \tilde V_0 D^\dagger_D~, 
\ee
where $\Vtr$ is a ``CKM-like'' mixing matrix, containing only one non-trivial phase
and $D_{U,D}$ are diagonal matrices containing the remaining CP-violating phases.
It turned out to be useful to choose the following parametrization of 
$\Vtr$ attributing the non-trivial phase of  this matrix to the $2-3$ mixing, 
such that 
\be
 \Vtr = 
\left(\begin{array}{ccc}
\tc_{12}\tc_{13}&\ts_{12}\tc_{13}&\ts_{13}\\ -\ts_{12}\tc_{23}
-\tc_{12}\ts_{23}\ts_{13}e^{-i\phi} &\tc_{12}\tc_{23}-\ts_{12}\ts_{23}\ts_{13}e^{-i\phi}& 
\ts_{23}\tc_{13}e^{-i\phi}\\ 
-\tc_{12}\tc_{23}\ts_{13}+\ts_{12}\ts_{23}e^{i\phi}& -\ts_{12}\tc_{23}\ts_{13}-\ts_{23}\tc_{12}e^{i\phi}&\tc_{23}\tc_{13}
\end{array}\right)~,
\label{eq:Vtrgen}
\ee
and 
\be
D_U={\rm diag}(1, e^{i\phi^u_2}, e^{i\phi^u_3})~,  \qquad 
D_D={\rm diag}(e^{i\phi^d_1}, e^{i\phi^d_2}, e^{i\phi^d_3})~.
\label{eq:Dphases}
\ee

Having this set-up at hand we have performed a detailed phenomenology 
of RH currents, taking all tree level constraints into account and 
solving the $V_{ub}$ problem in this manner. It should be emphasised that 
this solution chooses the {\it inclusive} value of $|V_{ub}|$ as the true 
value of this CKM element and our best value turned out to be in the 
ballpark of $(4.1\pm0.2) \times 10^{-3}$ implying in turn 
\be
\sin 2\beta=0.77\pm0.05~,
\ee
a value much larger than 
the measured value of $S_{\psi K_S}=0.672\pm 0.023$. This has profound 
implications as we will see below.

Returning to the RH mixing matrix we find that a good description of the 
tree level data is provided by the matrix
\be
 \Vtr = 
\left(\begin{array}{ccc}
\pm  \tc_{12} \frac{\sqrt{2}}{2} & \pm \ts_{12} \frac{\sqrt{2}}{2} & - \frac{\sqrt{2}}{2} \\ 
-\ts_{12} &\tc_{12} &  0 \\ 
\tc_{12} \frac{\sqrt{2}}{2} & \ts_{12} \frac{\sqrt{2}}{2} &  \pm \frac{\sqrt{2}}{2}
\end{array}\right)~,
\label{eq:Vtansatz2}
\ee
 where  $\ts_{12}$ is free, provided that ${\rm sgn}(\tc_{13}\ts_{12})=-1$. 
Other still possible but less interesting cases are considered in 
\cite{Buras:2010pz}.

The novel feature of our analysis as compared with~\cite{Crivellin:2009sd}
is the determination of the full right-handed matrix, 
and not only its selected elements, making use of its unitarity.
We also find that while RH currents are very welcome to solve the 
``$|V_{ub}|$ problem'' they do not have a significant impact on the 
determination of $|V_{cb}|$ (as also pointed out in~\cite{Feger:2010qc}). 

The matrix $\Vtr$ in (\ref{eq:Vtansatz2}) has then been used in
our analysis of FCNCs. In fact the particular structure of this matrix is 
a key ingredient to generate a sizable non-standard contribution to $S_{\psi\phi}$.
As a result, after we require a large $S_{\psi\phi}$,  
most of our conclusions listed below do not depend on the choice of this 
ansatz. 
It should also be stressed that, contrary to the CKM case, having a zero in $\Vtr$
does not prevent non-vanishing CP-violating effects thanks to the extra phases in (\ref{eq:Dphases}).  

The mixing structures relevant  to the three down-type $\Delta F=2$ 
and FCNC amplitudes in the SM (LH sector) and in the RH sector are shown 
in Table~\ref{tab:YY}. 
We observe that 
the $\tc_{12}$ and $\ts_{12}$ dependencies in the three systems considered are
non-universal with the observables in the 
$K$-mixing, $B_d$ mixing and $B_s$-mixing dominated by 
 $\tc_{12}\ts_{12}$,  $\tc_{12}$ and $\ts_{12}$, respectively. 
Since both $\Delta S=2$ and $B_d$ mixing are strongly 
constrained, and the data from CDF and D0 give some hints 
for sizable NP contributions in the $B_s$ mixing, 
it is natural to assume  that $\tc_{12} \ll 1$. 
The phenomenological analysis is then rather constrained.

\begin{table}[t] 
\begin{center}
\begin{tabular}{|c||c|c|c|}
\hline
\raisebox{0pt}[10pt][5pt]{Mixing term} &  $s\to d$  & $b\to d$ & $b\to s$ \\
\hline\hline 
\raisebox{0pt}[12pt][5pt]{   $V^*_{ti} V_{tj}$     } 
&  $ V_{ts}^*V_{td} \approx -\lambda^5 e^{-i\beta} $ 
&  $ V_{tb}^*V_{td} \approx \lambda^3 e^{-i\beta} $ 
&  $ V_{tb}^*V_{ts} \approx -\lambda^2 e^{-i\beta_s} $ \\ \hline\hline
\raisebox{0pt}[15pt][10pt]{  $\Vt^*_{ti} \Vt_{tj}$  } 
&  $ \frac{1}{2} \tc_{12} \ts_{12} e^{i(\phi^d_2-\phi^d_1)} $ 
&  $ \pm \frac{1}{2} \tc_{12} e^{i(\phi^d_3-\phi^d_1)} $ 
&  $ \pm \frac{1}{2} \ts_{12} e^{i(\phi^d_3-\phi^d_2)} $ \\  
\hline\hline
\end{tabular}
\caption{Mixing structures relevant to the three down-type $\Delta F=2$ 
and FCNC amplitudes in the SM (LH sector) and in the RH sector.
In the SM case approximate expressions of the CKM factors 
expanded in powers of  $\lambda = |V_{us}|$ are also shown.
In the RH case the parametrization for the RH matrix
is the one in (\ref{eq:Vtansatz2}).} \label{tab:YY}
\end{center}
\end{table}

 Before presenting the main results of our analysis let us emphasise that 
the non-standard contributions to $\Delta S=2$ amplitudes 
are exceedingly large compared to the SM term (and compared 
with data) unless the Wilson coefficients of the relevant operators 

\be
Q_1^{\rm VRR} = (\bar{s}^{\alpha} \gamma_{\mu}    P_R d^{\alpha})
              (\bar{s}^{ \beta} \gamma^{\mu}    P_R d^{ \beta}),
\quad
Q_1^{\rm LR} =  (\bar{s}^{\alpha} \gamma_{\mu}    P_L d^{\alpha})
              (\bar{s}^{ \beta} \gamma^{\mu}    P_R d^{ \beta}),
\ee
are very small. This can be for instance achieved if
one of the two mixing terms $\tc_{12}$ or $\ts_{12}$ is very 
small. The most problematic is the second operator for which the 
$P_i$ parameter in (\ref{eq:matrix}) takes the value $P_1^{LR}(K)\approx -52$ to be compared 
with $P_1^{VRR}\approx 0.5$.
Due to
the  $(V-A)\times(V+A)$ structure of $Q_1^{\rm LR}$, its 
 contributions  are 
known to be strongly enhanced at low energies through renormalization 
group effects and in the case of $\varepsilon_K$ and $\Delta M_K$ through its
chirally enhanced hadronic matrix elements. 
Consequently these observables put severe constraints on the model 
parameters  as  known from various studies 
in explicit left-right symmetric models~\cite{Zhang:2007da} and also RS 
models.

Having determined the size and the flavour structure of 
RH currents that is consistent with the present data on tree 
level processes and which removes the ``$|V_{ub}|$-problem'', 
we have investigated how this NP would manifest itself in 
neutral current processes, including particle-antiparticle mixing, 
$Z\to b \bar b$, $B_{s,d}\to \mu^+\mu^-$,
 $B \to \{X_s,K, K^*\}\nu\bar \nu$ and $K\to \pi\nu\bar\nu$
decays. Most importantly, we have also addressed the possibility to explain a 
non-standard CP-violating phase 
in $B_s$ mixing in this context and the issue of the anomalies in 
the UT-triangle.

The main messages from our analysis of these processes are as follows:

{\bf 1.} 
The desire to generate large CP-violating effects in $B_s$-mixing, 
hinted for by the enhanced value of $S_{\psi\phi}$ observed by the 
CDF and D0 collaborations, in conjunction with the $\varepsilon_K$-constraint,
implies additional constraints on the shape of $\tilde V$. In particular 
$\tc_{12}\ll 1$ and consequently $\tilde s_{12}\approx 1$. 
The pattern of deviations from the SM in this model is then as follows.

{\bf 2.}
The $S_{\psi\phi}$ and $\varepsilon_K$ anomalies can be understood.

{\bf 3.}
As a consequence of the large value of $\tilde s_{12}$, 
it should be possible  
to resolve the presence of RH currents also in $s\to u$ 
charged-current transitions. Here RH currents imply 
a  $\ord(10^{-3})$ deviation in 
the determination of $|V_{us}|$ from $K\to \pi\ell\nu$
and $K\to \ell\nu$ decays.

{\bf 4.}
The ``true value'' of $\sin 2\beta$ determined in our framework, 
namely the determination of the CKM phase $\beta$ on the basis
of the tree-level processes only, and in particular of $|V_{ub}|$,
is   $\sin 2\beta=0.77\pm 0.05$. This result is roughly $2\sigma$ 
larger than the measured value $S_{\psi K_S}=0.672\pm0.023$. 
This is 
a property of any explanation of the ``$|V_{ub}|$-problem'' by means 
of RH currents, unless the value of $|V_{ub}|$ from inclusive 
decays will turn out to be much lower than determined presently.

{\bf 5.}
In general, such a  discrepancy could be removed by a negative new 
CP-violating phase $\varphi_{B_d}$ in $B_d^0-\bar B_d^0$ mixing. However, we 
have demonstrated that this is not possible in the present framework once the 
$\varepsilon_K$ constraint is imposed and large $S_{\psi\phi}$ is 
required. Indeed as seen in Table~\ref{tab:YY} for $\tc_{12}\ll 1$ the NP 
effects in $S_{\psi K_S}$ are tiny. 
Thus we pointed out that simultaneous explanation of 
the  ``$|V_{ub}|$-problem'',
of $S_{\psi K_S}=0.672\pm0.023$ and large $S_{\psi\phi}$ is problematic through RH
currents alone.

{\bf 6.}
The present constraints from $B_{s,d}\to\mu^+\mu^-$ eliminate the possibility 
of removing the known anomaly in the $Z\to b\bar b$ decay with the help
of RH currents.
On top of it, the constraint from $B\to X_s l^+l^-$ precludes 
$B_{s}\to \mu^+\mu^-$ to be close to its present experimental bound. 
However, still
values as high as $1\cdot 10^{-8}$ are possible.
Moreover NP effects in $B_{d} \to \ell^+\ell^-$ are found generally 
smaller than in $B_{s} \to \ell^+\ell^-$.

{\bf 7.}
Contributions from RH currents to 
$B \to \{X_s,K, K^*\} \nu\bar \nu$ and
$K\to\pi\nu\bar\nu$ decays can still be significant. 
Most important, the deviations from the SM in these decays 
would exhibit a well-defined pattern of correlations.

Thus our analysis casts a shadow on the explanation of the $|V_{ub}|$-problem 
with the help of RH currents alone unless the $S_{\psi\phi}$ anomaly goes 
away and $\tc_{12}$ can be large solving the problem with $S_{\psi K_S}$ 
naturally.

Particularly interesting is the comparison with the 
${\rm 2HDM_{\overline{MFV}}}$ model,
where the $S_{\psi\phi}$ and  $\varepsilon_K$ anomalies 
can also be accommodated~\cite{Buras:2010mh} as seen in our 
presentation of this model. 
What clearly distinguishes these two models at low-energies
is how they face the ``$|V_{ub}|$-problem"
(which can be solved only in the RHMFV case) and the 
``$\sin2\beta$--$S_{\psi K}$ tension'' (which can be   
softened only in the 2HDM case). 
But also the future results on rare $B$ and $K$ decays listed 
above could in principle help to distinguish 
these two general NP frameworks.

 Restricting the discussion to these two NP frameworks, it appears 
 that a model with an extended scalar sector and RH currents
 could provide solutions to all the existing tensions in flavour
 physics simultaneously. This possibility can certainly be realized
 in explicit left-right symmetric models, where an extended Higgs
 sector  is also required to break the extended gauge symmetry.
 However, these
 extensions contain many free parameters and clear cut conclusions on
 the pattern of flavour violation cannot be as easily reached as it was
 possible in the simple frameworks like RHMFV ~\cite{Buras:2010pz} and 
${\rm 2HDM_{\overline{MFV}}}$ ~\cite{Buras:2010mh}. A nice short summary 
of our results on RHMFV has been presented in \cite{Gemmler:2010fv}.

\subsection{A Randall-Sundrum Model with Custodial Protection}
When the number of space-time dimensions is increased,
 new solutions to the hierarchy 
problems are possible. Most 
ambitious proposals are models with a warped extra dimension first proposed
by Randall and Sandrum (RS)  \cite{Randall:1999ee} which provide a geometrical
explanation of the
hierarchy  between the Planck scale and the EW scale. Moreover, when the SM
fields, except for the Higgs field, are
allowed to propagate in the bulk 
\cite{Gherghetta:2000qt,Chang:1999nh,Grossman:1999ra}, 
these models naturally generate the
hierarchies in the fermion masses and mixing angles 
\cite{Grossman:1999ra,Gherghetta:2000qt} through different localisations 
of the fermions in the bulk. Yet, this way of explaining the hierarchies in masses 
and mixings necessarly   
implies FCNC transitions
at the tree level 
\cite{Burdman:2003nt,Huber:2003tu,Agashe:2004cp,Csaki:2008zd}.
Most problematic is the parameter 
$\varepsilon_K$ which receives tree level KK gluon contributions
 and some fine-tuning
of parameters in the flavour sector is necessary in order to achieve 
consistency with the data for KK scales in the reach of the LHC 
\cite{Csaki:2008zd,Blanke:2008zb}.

Once this fine-tuning is made,
the RS-GIM mechanism 
\cite{Huber:2003tu,Agashe:2004cp}, combined with an additional custodial
protection of  flavour violating $Z$ couplings 
\cite{Blanke:2008zb,Blanke:2008yr,Buras:2009ka}, 
 allows yet to achieve 
the agreement with
existing data for other observables 
without an additional fine tuning of parameters\footnote{See however comments 
at the end of this subsection.}.
New theoretical ideas addressing the issue of large FCNC transitions in the
RS framework and proposing new protection mechanisms occasionally leading
to MFV can be found in 
\cite{Csaki:2008eh,Cacciapaglia:2007fw,Cheung:2007bu,Santiago:2008vq,Csaki:2009bb,Csaki:2009wc}.

Entering some details it should be emphasised that to avoid problems
with electroweak precision tests (EWPT) and FCNC processes, the gauge group 
is
generally larger than the SM gauge group \cite{Agashe:2003zs,Csaki:2003zu,Agashe:2006at}:
\be
G_{RSc}=SU(3)_c\times SU(2)_L\times SU(2)_R\times U(1)_X
\ee
 and similarly to the LHT model
new heavy gauge bosons are present. The increased symmetry provides 
a custodial protection.  We will denote such framework by RSc.

The lightest new gauge bosons are the KK--gluons, the KK-photon and the 
electroweak KK gauge bosons $W^\pm_H$, $W^{\prime\pm}$, $Z_H$ and $Z^\prime$,
all with masses $M_{KK}$ around $2-3\tev$ as required by the consistency 
with the EWPT \cite{Agashe:2003zs,Csaki:2003zu,Agashe:2006at}.
 The fermion sector is 
enriched through heavy KK-fermions (some of them with exotic electric charges)
 that could in principle be discovered at 
the LHC. The fermion content
of this model is explicitly given in \cite{Albrecht:2009xr}, where also 
 a complete set of 
Feynman rules has been worked out. Detailed analyses of electroweak precision
tests and of the parameter $\varepsilon_K$ in a RS model without and with 
custodial 
protection can also be found in \cite{Casagrande:2008hr,Bauer:2008xb}. These authors 
analyzed also rare and non-leptonic decays in \cite{Bauer:2009cf}.
Possible flavour protections
in warped Higgsless models have been presented in  \cite{Csaki:2009bb}.

Here we summarize the main results obtained in Munich:

{\bf 1.} The CP asymmetry $S_{\psi\phi}$
can reach values as high as  0.8 
to be compared with its SM value 0.04.

{\bf 2.} The branching ratios for $\kpn$, $\klpn$, $K_L\to\pi^0l^+l^-$
         can be enhanced relative to the SM expectations up to factors of
         1.6, 2.5 and 1.4, respectively, when only moderate fine tuning 
in $\varepsilon_K$ is required. Otherwise the enhancements can be larger.
          $Br(\kpn)$ and $Br(\klpn)$ can
         be simultaneously enhanced but this is not necessary as the
         correlation between these two branching ratios is not evident
         in this model. On the other hand $Br(\klpn)$ and 
         $Br(K_L\to\pi^0l^+l^-)$ ($l=e,\mu)$ are strongly correlated and the
         enhancement of one of these three branching ratios implies the
         enhancement of the remaining two.

{\bf 3.} A large enhancement of the short distance part of 
        $Br(K_L\to \mu^+\mu^-)$ is possible, up to a factor of 2-3, but
         not simultaneously with $Br(\kpn)$.

 {\bf 4.} More importantly simultaneous large NP effects in $S_{\psi\phi}$
          and $K\to\pi\nu\bar\nu$ channels are very unlikely and this 
          feature is even more pronounced than in the LHT model.

{\bf 5.} The branching ratios for $B_{s,d}\to \mu^+\mu^-$ and 
             $B\to X_{s,d}\nu\bar\nu$ remain SM-like: the maximal enhancements
            of these branching ratios amount to $15\%$.

{\bf 6.} The relations \cite{Buras:2003jf,Buras:2009us}
         between various observables present in models with CMFV
         can be strongly
         violated.
 
In particular this pattern of flavour violation 
  implies that in the case of the confirmation of large
 values of $S_{\psi\phi}$ by future experiments  significant
 deviations of $Br(\klpn)$ and $Br(\kpn)$ from their SM values in
 this framework similarly to the LHT model are very unlikely. 
On the other hand  SM-like value of $S_{\psi\phi}$
 will open the road for large enhancements of these branching ratios that
 could be tested by KOTO at J-Parc and NA62 at CERN, respectively.

 Next, let me just mention  that large NP contributions in the RS framework
 that require some tunings of parameters
in order to be in agreement with the experimental data have been found in 
$Br(B\rightarrow X_s\gamma)$ \cite{Agashe:2008uz}, $Br(\mu\rightarrow
 e\gamma)$ \cite{Agashe:2006iy,Davidson:2007si,Agashe:2009tu} and EDM's
 \cite{Agashe:2004cp,Iltan:2007sc}, that are all dominated by dipole operators. Also the new 
contributions to $\varepsilon^{\prime}/{\varepsilon}$ can
be large \cite{Gedalia:2009ws}. Moreover it appears that the fine tunings 
in this ratio are not consistent necessarily with the ones required in the 
case of $\varepsilon_K$.

Finally a personal comment. My excursion into the fifth dimension was very 
interesting and I learned a lot about the structure of weak interactions 
in this NP scenario. Yet after this study I am sceptical that the nature 
is so violent 
to provide this physics already at the LHC scales. Therefore I am 
delighted to be back in $D=4$.

Having said this let me stress that many of the ideas and
concepts that characterize most of the physics discussed above 
do not rely on the assumption
of additional dimensions.\footnote{I would like to thank Andreas 
Weiler for discussions related to this point.} 
As indicated by AdS/CFT correspondence
 we can regard RS models
as a mere computational tool for certain 
strongly coupled theories. This had already an important impact 
on the study of dynamical electroweak symmetry breaking and 
can be used to make 4D techni-colour or composite Higgs models 
more realistic~\cite{Csaki:2003zu,Agashe:2004rs}. 
The particular mechanism for example that is employed
in RS to give masses to the SM fields is a realization
of an old 4D idea which is known as partial compositeness~\cite{Kaplan:1991dc}
(a simple 4D effective theory using a two site model has been proposed recently~\cite{Contino:2006nn}).
The fact that the 4D description of RS looks like a theory
with a conformal window allows us to explain the hierarchies
in the flavour sector as a result of conformal dynamics.
Thus afterall this physics could turn out to be useful also for 
strong dynamics in 4D and if such physics will show up at 
LHC, the techniques developed in the last decade will 
certainly play an important role in the phenomenology to which I hope 
to be able to contribute again.

\newcommand{\three}{{\color{red}$\bigstar\bigstar\bigstar$}}
\newcommand{\two}{{\color{blue}$\bigstar\bigstar$}}
\newcommand{\one}{{\color{black}$\bigstar$}}

\begin{table}[t]
\addtolength{\arraycolsep}{4pt}
\renewcommand{\arraystretch}{1.5}
\centering
\begin{tabular}{|l|c|c|c|c|c|c|}
\hline
&  AC & RVV2 & AKM  & $\delta$LL & FBMSSM & $SSU(5)_{\rm RN}$
\\
\hline\hline
$D^0-\bar D^0$& \three & \one & \one & \one & \one & \one
\\ \hline
$\epsilon_K$& \one & \three & \three & \one & \one & \three
\\ \hline
$ S_{\psi\phi}$ & \three & \three & \three & \one & \one & \three
\\ \hline\hline
$S_{\phi K_S}$ & \three & \two & \one & \three & \three & \two
\\ \hline
$A_{\rm CP}\left(B\rightarrow X_s\gamma\right)$ & \one & \one & \one & \three & \three & \one
\\ \hline
$A_{7,8}(K^*\mu^+\mu^-)$ & \one & \one & \one & \three & \three & \one
\\ \hline
$B_s\rightarrow\mu^+\mu^-$ & \three & \three & \three & \three & \three & \three
\\ \hline
$K^+\rightarrow\pi^+\nu\bar\nu$ & \one & \one & \one & \one & \one & \one
\\ \hline
$K_L\rightarrow\pi^0\nu\bar\nu$ & \one & \one & \one & \one & \one & \one
\\ \hline
$\mu\rightarrow e\gamma$& \three & \three & \three & \three & \three & \three
\\ \hline
$\tau\rightarrow \mu\gamma$ & \three & \three & \one & \three & \three & \three 
\\ \hline
$\mu + N\rightarrow e + N$& \three & \three & \three & \three & \three & \three
\\ \hline\hline
$d_n$& \three & \three & \three & \two & \three & \three
\\ \hline
$d_e$& \three & \three & \two & \one & \three & \three
\\ \hline
$\left(g-2\right)_\mu$& \three & \three & \two & \three & \three & \three
\\ \hline
\end{tabular}
\renewcommand{\arraystretch}{1}
\caption{\small
``DNA'' of flavour physics effects for the most interesting observables in a selection of SUSY
models. \three\ signals large NP effects, \two\ moderate to small NP effects and \one\
implies that the given model does not predict visible NP effects in that observable. From \cite{Altmannshofer:2009ne} and \cite{Buras:2010pm}.}
\label{tab:DNA}
\end{table}

\begin{table}[t]
\addtolength{\arraycolsep}{4pt}
\renewcommand{\arraystretch}{1.5}
\centering
\begin{tabular}{|l|c|c|c|c|c|}
\hline
& LHT & RSc & 4G & 2HDM  & RHMFV
\\
\hline\hline
$D^0-\bar D^0$ (CPV)&  \three & \three & \two & \two & 
\\ \hline
$\epsilon_K$& \two & \three & \two &\two & \two
\\ \hline
$ S_{\psi\phi}$ & \three & \three & \three & \three  &  \three
\\ \hline\hline
$S_{\phi K_S}$ &  \one & \one  & \two & &
\\ \hline
$A_{\rm CP}\left(B\rightarrow X_s\gamma\right)$ & \one &  & \one & &
\\ \hline
$A_{7,8}(K^*\mu^+\mu^-)$ & \two &\one  & \two & &
\\ \hline
$B_s\rightarrow\mu^+\mu^-$ & \one & \one & \three & \three & \two
\\ \hline
$K^+\rightarrow\pi^+\nu\bar\nu$ & \three & \three & \three & & \two
\\ \hline
$K_L\rightarrow\pi^0\nu\bar\nu$ &  \three & \three & \three & & \two
\\ \hline
$\mu\rightarrow e\gamma$& \three & \three & \three & &
\\ \hline
$\tau\rightarrow \mu\gamma$ & \three & \three & \three & &
\\ \hline
$\mu + N\rightarrow e + N$& \three & \three & \three & &
\\ \hline\hline
$d_n$& \one & \three & \one & \three &
\\ \hline
$d_e$& \one & \three & \one & \three&
\\ \hline
$\left(g-2\right)_\mu$& \one & \two & \one & &
\\ \hline

\end{tabular}
\renewcommand{\arraystretch}{1}
\caption{\small
``DNA'' of flavour physics effects for the most interesting observables in a selection of non-SUSY models. \three\ signals large NP effects, \two\ moderate to small NP effects and \one\
implies that the given model does not predict visible NP effects in that observable. Empty spaces reflect my present ignorance about the given entry.}
\label{tab:DNA2}
\end{table}

\subsection{''DNAs'' of Flavour Models}
The ``DNA's'' of flavour physics effects for the most interesting observables 
constructed in \cite{Altmannshofer:2009ne} and extended by the recent results obtained in the 
${\rm 2HDM_{\overline{MFV}}}$, SM4, RHMFV and $SSU(5)_{\rm RN}$ are
presented in Tables~\ref{tab:DNA} and \ref{tab:DNA2}. These tables only 
indicate whether large, moderate 
or small NP effects in a given observable are still allowed in a given model 
but do not exhibit correlations between various observables characteristic 
for a given model. 
Still they could turn out to be useful in eliminating models in which large 
NP effects for certain observables, hopefully seen soon in the data, are 
not possible.

Concerning correlations
we have discussed them above and they will also enter the discussion below. 
In Table~\ref{tab:corr} the references to papers from my group that analyzed 
various correlations in all models discussed above have been collected for 
convenience.

\begin{table}[thb]
\begin{center}
\begin{tabular}{|c|c|}
\hline
{\bf Model} & {\bf Reference}   \\
\hline
CMFV &   \cite{Buras:2000dm,Buras:2003jf,Blanke:2006ig,Buras:2003td}\\
${\rm 2HDM_{\overline{MFV}}}$ &  \cite{Buras:2010mh,Buras:2010zm} \\
ACD Model & \cite{Buras:2002ej,Buras:2003mk} \\
LH        &  \cite{Buras:2004kq,Buras:2006wk}\\
LHT &  \cite{Blanke:2009am,Bigi:2009df}\\
SM4 &  \cite{Buras:2010pi,Buras:2010nd,Buras:2010cp}\\
AC,~RVV2,~AMK,~$\delta$LL &  \cite{Altmannshofer:2009ne,Altmannshofer:2010ad}
 \\ 
$SSU(5)_{RN}$ & \cite{Buras:2010pm}\\
FBMSSM & \cite{Altmannshofer:2008hc}\\
RHMFV & \cite{Buras:2010pz}\\
RSc & \cite{Blanke:2008yr,Blanke:2008zb}\\
\hline
\end{tabular}
\end{center}
\vspace{0.1cm}
\caption[]{References to correlations in various models.
\label{tab:corr}}
\end{table}

\section{Facing Anomalies and Distinguishing between Various BSM Models through
 Correlations}
\subsection{Preliminaries}
Armed with the information on patterns of flavour violation in a large 
number of 
concrete models and NP scenarios, we will now look from a different angle 
at various anomalies observed in the data. The readers, who followed the 
previous sections will notice certain repetitions of statements made before 
but now these statements are in a different context and could be useful anyway. 
On the other hand the readers who skept all previous sections should be able, because 
of these repetitions, to follow this section, only occassionally looking 
up the summaries of specific models presented before.

\subsection{The $\varepsilon_K-S_{\psi K_S}$ Anomaly}
It has been pointed out in \cite{Buras:2008nn,Buras:2009pj} that the SM prediction for $\varepsilon_K$ 
implied by the measured value of $S_{\psi K_S}=\sin 2\beta$, the ratio 
$\Delta M_d/\Delta M_s$ and the value of $|V_{cb}|$ turns out to be too
small to agree well with experiment. This tension between $\varepsilon_K$ and
 $S_{\psi K_S}$ has been pointed out from a different perspective in
\cite{Lunghi:2008aa,Lunghi:2009sm,Lunghi:2009ke,Lunghi:2010gv}.
These findings have been confirmed by a 
UTfitters  analysis \cite{UTfit-web}. 
The CKMfitters having a different treatment of uncertainties find less significant effects \cite{Lenz:2010gu}.

 The main reasons 
for this tension are on the one hand a decreased value of the relevant non-perturbative 
parameter $\hat B_K=0.724\pm0.008\pm0.028$ \cite{Antonio:2007pb} 
\footnote{Interestingly this value is very close to $\hat B_K=0.75$ obtained 
in the large-N limit of QCD \cite{Gaiser:1980gx,Buras:1985yx}. Including 1/N corrections Bardeen, Gerard and myself
\cite{Bardeen:1987vg,Gerard:1990dx} found some indications for
$\hat B_K\le 0.75$. A very recent more precise 
analysis of Gerard \cite{Gerard:2010jt} puts this result on firm footing. On the other hand the 
most recent message from RBC and UKQCD collaborations \cite{Aoki:2010pe} 
reads  $\hat B_K=0.749\pm0.027$. Not only 
an impressive accuracy but also an indication for small 
1/N corrections in accordance with the results in \cite{Bardeen:1987vg,Gerard:1990dx} obtained more than 20 years ago. In this context also the paper by 
Bijnens and Prades \cite{Bijnens:1995br} should be mentioned. See \cite{Gerard:2010jt} for more details.
}
resulting from unquenched lattice 
calculations and on the other hand the decreased value of $\varepsilon_K$ in 
the SM arising from a multiplicative factor, estimated first to be 
$\kappa_\varepsilon=0.92\pm0.02$ \cite{Buras:2008nn}. This factor took into account the departure 
of $\phi_\varepsilon$ from $\pi/4$ and the long distance (LD) effects in 
${\rm Im}\Gamma_{12}$ in the $K^0-\bar K^0$ mixing. The recent inclusion of LD effects 
in ${\rm Im}M_{12}$ modified this estimate to $\kappa_\varepsilon=0.94\pm0.02$ 
\cite{Buras:2010pza}. 
Very recently also NNLO-QCD corrections to the QCD factor $\eta_{ct}$ in 
$\varepsilon_K$ \cite{Brod:2010mj} have 
been calculated enhancing the value of $\varepsilon_K$ by $3\%$. Thus while 
in \cite{Buras:2008nn} the value $|\varepsilon_K|_{\rm SM}=(1.78\pm0.25)\cdot 10^{-3}$ has 
been quoted and with the new estimate of LD effects and new input one 
finds $|\varepsilon_K|_{\rm SM}=(1.85\pm0.22)\cdot 10^{-3}$, including NNLO
corrections gives the new value \cite{Brod:2010mj}
\begin{equation}\label{epnew}
|\varepsilon_K|_{\rm SM}=(1.90\pm0.26)\cdot 10^{-3},
\end{equation}
significantly closer to the experimental value 
$|\varepsilon_K|_{\rm exp}=(2.23\pm0.01)\cdot 10^{-3}$. 

Consequently, the $\varepsilon_K$-anomaly softened considerably but it is still 
alive. Indeed, the $\sin 2\beta=0.74\pm 0.02$ from UT fits is visibly larger
than the experimental value $S_{\psi K_S}=0.672\pm 0.023$. The difference is 
even larger if one wants to fit $\varepsilon_K$ exactly: 
$\sin 2\beta\approx 0.85$  \cite{Lunghi:2008aa,Buras:2008nn,Buras:2009pj,Lunghi:2009sm,Lunghi:2009ke,Lunghi:2010gv}.

One should also recall the tension between inclusive and exclusive determinations of $|V_{ub}|$ with the exclusive ones in the ballpark of $3.5\cdot 10^{-3}$ and the 
inclusive ones typically above $4.0\cdot 10^{-3}$. 
As discussed in detail in the previous Section, an interesting solution to 
this problem is the presence of RH charged currents, which selects the inclusive value as the true value, implying again $\sin 2\beta\approx 0.80$ 
\cite{Buras:2010pz}.

As discussed in \cite{Lunghi:2008aa,Buras:2008nn} and subsequent papers of these authors a small negative NP phase $\varphi_{B_d}$ in 
$B^0_d-\bar B^0_d$ mixing would solve both problems, provided such a phase 
is allowed by other constraints.  Indeed we have 
then 
\begin{equation}
S_{\psi K_S}(B_d) = \sin(2\beta+2\varphi_{B_d})\,, \qquad
S_{\psi\phi}(B_s) =  \sin(2|\beta_s|-2\varphi_{B_s})\,,
\label{eq:basic}
\end{equation}
where the corresponding formula for $S_{\psi\phi}$ in the presence of 
a NP phase $\varphi_{B_s}$ in 
$B^0_s-\bar B^0_s$ mixing has also been given. With a negative $\varphi_{B_d}$ 
the true $\sin 2\beta$ is larger than $S_{\psi K_S}$, implying a higher value
on $|\varepsilon_K|$, in reasonable agreement with data and a better UT-fit. This 
solution would favour the inclusive value of $|V_{ub}|$ as chosen 
e.g. by RH currents but as pointed out in \cite{Buras:2010pz} this 
particular solution of the ''$V_{ub}-problem''$ does not allow for a good 
fit to $S_{\psi K_S}$ if large $S_{\psi\phi}$ is required.

Now making a  universality hypothesis of $\varphi_{B_s}=\varphi_{B_d}$ 
\cite{Ball:2006xx,Buras:2008nn},
a negative $\varphi_{B_d}$ would automatically imply an enhanced value of
$S_{\psi\phi}$ which in 
view of $|\beta_s|\approx 1^\circ$ amounts to roughly 0.04 in the SM. However, 
in order to be in agreement with the experimental value of $S_{\psi K_S}$ this 
type of NP would imply $S_{\psi\phi}\le 0.25$. All this shows that correlations 
between various observables are very important in this game.

The  universality hypothesis of $\varphi_{B_s}=\varphi_{B_d}$ in 
\cite{Ball:2006xx,Buras:2008nn} was clearly
ad hoc. Recently, in view of the enhanced value of $S_{\psi\phi}$ at CDF and 
D0 a more dynamical origin of this relation has been discussed by other 
authors and different relations between these two phases corresponding still
to a different dynamics have been discussed in the literature. 
 Let us elaborate
on this topic in more detail.
\subsection{Facing an enhanced CPV in the $B_s$ mixing}
Possibly the most important highlight in flavour physics in 2008, 2009 
\cite{Aaltonen:2007he} and even 
more in 2010 was the enhanced value of $S_{\psi\phi}$ measured by the CDF and 
D0 collaborations, seen either directly or indirectly through the 
correlations with various semi-leptonic asymmetries. While in 2009 and in 
the Spring of 2010  \cite{Abazov:2010hv}, the messages from Fermilab indicated good prospects 
for $S_{\psi\phi}$ above 0.5, the recent messages from ICHEP 2010 in Paris, 
softened such hopes significantly \cite{Aaltonen:2010xx}.
Both CDF and D0 find the enhancement 
by only one $\sigma$. Yet, this does not yet preclude $S_{\psi\phi}$ above 0.5, 
which would really be a fantastic signal of NP. But $S_{\psi\phi}$ below 
0.5 appears more likely at present. Still even a value of 0.2 would 
be exciting. Let us hope that the future data from Tevatron and in 
particular from the LHCb, will measure this asymmetry with sufficient 
precision so that we will know to which extent NP is at work here. 
One should also hope that the large CPV in dimuon CP asymmetry from D0,
  that triggered 
new activities, will be better understood. I have nothing to add here
at present and can only refer to numerous papers  
\cite{Dobrescu:2010rh,Ligeti:2010ia,Blum:2010mj,Lenz:2010gu,Bauer:2010dg}.

Leaving the possibility of $S_{\psi\phi}\ge 0.5$ still open but keeping in 
mind that also $S_{\psi\phi}\le 0.25$ could turn out to be the final value, 
let us investigate how different models described in Section 5 would face these two different 
results and what kind of dynamics would be behind these two scenarios.
\boldmath
\subsubsection{$S_{\psi\phi}\ge 0.5$}
\unboldmath
Such large values can be obtained in the RSc model due to KK gluon 
exchanges and also heavy neutral KK electroweak gauge boson exchanges. 
In the  supersymmetric flavour model with the dominance of RH
currents  like the AC model, double Higgs penguins constitute the dominant NP 
contributions responsible for $S_{\psi\phi}\ge 0.5$, while in the RVV2 model where NP LH current 
contributions are equally important, also gluino boxes are relevant.
On the operator level, it is $Q_2^{LR}$ operator in (\ref{normal}) with 
properly 
changed quark flavours,
 which is primarly responsible 
for this enhancement.

Interestingly the SM4 having only $(V-A)\times (V-A)$ operator $Q_1^{VLL}$ is also capable 
in obtaining 
high values of $S_{\psi\phi}$ \cite{Hou:2005yb,Soni:2008bc,Buras:2010pi} but not as easily as the RSc, AC and RVV2 models.
The lower scales of NP in the SM4 relative to the latter models and the
non-decoupling effects of $t^\prime$ compensate to some extent the absence 
of LR scalar operators. In the LHT model where only $(V-A)\times (V-A)$ operators 
are present 
and the NP enters at higher scales than in the SM4, $S_{\psi\phi}$ above 
0.5 is out of reach \cite{Blanke:2009am}. Similar comment applies to the 
AKM model even if it has RH currents.

All these models contain new sources of flavour  and CP violation 
and it is not surprising that in view of many parameters involved large 
values of $S_{\psi\phi}$ can be obtained. The question then arises whether 
strongly
enhanced values of this asymmetry would uniquely imply new sources of 
flavour violation beyond the MFV hypothesis. The answer to this question is as 
follows:
\begin{itemize}
\item
In models with MFV and FBPs set to zero, $S_{\psi\phi}$
remains indeed SM-like.
\item
In supersymmetric models with MFV even in the presence of 
 non-vanishing FBPs, at both 
small and 
large $\tan\beta$, the supersymmetry constraints do not allow values 
of $S_{\psi\phi}$ visibly different from the SM value 
\cite{Altmannshofer:2009ne,Blum:2010mj,Altmannshofer:2008hc}.
\item
In the ${\rm 2HDM_{\overline{MFV}}}$ in which at one-loop both Higgs
doublets couple to up- and down-quarks, the interplay  of FBPs with 
the CKM matrix allows to obtain $S_{\psi\phi}\ge 0.5$ while satisfying 
all existing constraints \cite{Buras:2010mh}.
\end{itemize}

In the presence of a large $S_{\psi\phi}$
the latter model allows also for a simple 
and unique softening of the $\varepsilon_K$-anomaly and of the tensions in the 
UT analysis if the FBPs in the Yukawa interactions are the dominant source 
of new CPV. In this case the NP phases 
 $\varphi_{B_s}$ and $\varphi_{B_d}$ are related through 
\begin{equation}\label{BCGI}
 \varphi_{B_d}\approx\frac{m_d}{m_s}\varphi_{B_s}\approx \frac{1}{17} \varphi_{B_s},
\end{equation}
in visible contrast to the hypothesis $\varphi_{B_s}=\varphi_{B_d}$ of 
\cite{Ball:2006xx,Buras:2008nn}. 
Thus in this scenario large $\varphi_{B_s}$ required to obtain values of 
$S_{\psi\phi}$ above 0.5 imply a unique small shift in $S_{\psi K_S}$ 
that allows to lower $S_{\psi K_S}$ from 0.74 down to 0.70, 
that is closer to the experimental value $0.672\pm0.023$. This in turn 
implies that it is $\sin 2\beta=0.74$ and not  $S_{\psi K_S}=0.67$ that 
should be used in calculating $\varepsilon_K$ resulting in a value of 
 $\varepsilon_K\approx 2.0\cdot 10^{-3}$ within one $\sigma$ from the 
experimental value. The direct Higgs contribution to $\varepsilon_K$ 
is negligible because of small masses $m_{d,s}$. We should emphasize that 
once  $\varphi_{B_s}$ is determined from the data on $S_{\psi\phi}$ by means 
of (\ref{eq:basic}), the implications for $\varepsilon_K$ and  $S_{\psi K_S}$ are 
unique. 
All these correlations are explicitely seen in (\ref{eq:HeffB}) and 
(\ref{eq:HeffK}).

It is remarkable that such a simple set-up 
allows basically to solve all these tensions provided $S_{\psi\phi}$ is 
sufficiently above 0.5. The plots of $\varepsilon_K$ and  $S_{\psi K_S}$ 
versus $S_{\psi\phi}$ in \cite{Buras:2010mh} show this very transparently.
On the other hand this scenario does not provide any clue for the difference between inclusive and exclusive determinations of $|V_{ub}|$.
\boldmath
\subsubsection{$S_{\psi\phi}\approx 0.25$}
\unboldmath
Now, as signalled recently by  CDF and D0 data \cite{Aaltonen:2010xx},  $S_{\psi\phi}$ could be 
smaller. In this case all non-MFV models listed above can reproduce 
such values and in particular this time also the LHT model \cite{Blanke:2009am} and another 
supersymmetric flavour model (AKM) analysed by us stay alive \cite{Altmannshofer:2009ne}.

Again MSSM-MFV cannot reproduce such values. On the other hand the
${\rm 2HDM_{\overline{MFV}}}$ can still provide interesting results. Yet 
as evident from the plots in \cite{Buras:2010mh} the FBPs in Yukawa interactions
cannot now solve the UT tensions. Indeed the relation in (\ref{BCGI}) 
precludes now any interesting effects in $\varepsilon_K$ and $S_{\psi K_S}$:
$S_{\psi\phi}$ and the NP phase  $\varphi_{B_s}$ are simply too small. 
Evidently, this time the relation
\begin{equation}\label{BG}
 \varphi_{B_d}=\varphi_{B_s}
\end{equation}
would be more appropriate.

Now, the analyses in \cite{Ligeti:2010ia,Blum:2010mj} indicate how such a relation could be 
obtained within the ${\rm 2HDM_{\overline{MFV}}}$. This time the FBPs in the 
Higgs potential are at work, the relation in (\ref{BG}) follows and 
the plots of $\varepsilon_K$ and  $S_{\psi K_S}$
versus $S_{\psi\phi}$ are strikingly modified \cite{Buras:2010zm}: the dependence is much 
stronger and even moderate values of $S_{\psi\phi}$ can solve all tensions.
This time not  $Q_2^{LR}$  but  $Q_1^{SLL}$ in (\ref{normal}) is responsible 
for this behaviour.

Presently it is not clear which relation between $\varphi_{B_s}$ and 
$\varphi_{B_d}$ fits best the data but the model independent analysis 
of \cite{Ligeti:2010ia} indicates that $\varphi_{B_s}$ should be significantly larger than 
$\varphi_{B_d}$, but this hierarchy appears to be smaller than in (\ref{BCGI}).
Therefore as pointed out in \cite{Buras:2010zm} in the ${\rm 2HDM_{\overline{MFV}}}$ the best agreement 
with the data is obtained by having these phases both in Yukawa interactions 
and the Higgs potential, which is to be expected  in any case.
Which of the two flavour-blind CPV mechanisms dominates depends on the value of
$S_{\psi\phi}$, which is still affected by a sizable experimental error, and
 also by the precise amount of NP allowed in $S_{\psi K_S}$.

Let us summarize the dynamical picture behind an enhanced value of 
$S_{\psi\phi}$ within ${\rm 2HDM_{\overline{MFV}}}$. For $S_{\phi\phi}\ge 0.7$ the 
FBPs in Yukawa interactions are expected to dominate. On the other hand 
for $S_{\phi\phi}\le 0.25$ the FBPs in the Higgs potential are expected to
dominate the scene. If $S_{\psi\phi}$ will eventually be found somewhere between 
0.3 and 0.6, a hybrid scenario analyzed in \cite{Buras:2010zm} would be most efficient 
although not as predictive as the cases in which only one of these 
two mechanism is at
work.

\subsection{Implications of an enhanced  $S_{\psi\phi}$}
\subsubsection{Preliminaries}
Let us then assume that indeed $S_{\psi\phi}$ will be found to be significantly 
enhanced over the SM value. The studies of different observables in different 
models summarized in the previous Section 
allow then immediately to make some concrete predictions for a number
of observables which makes it possible to distinguish different models. This 
is important as  $S_{\psi\phi}$ alone is insufficient for this purpose.

In view of space limitations I will discuss here only the implications for
$B_{s,d}\to \mu^+\mu^-$ and $K\to\pi\nu\bar\nu$ decays, which we declared to be
the superstars of the coming years. Subsequently I will make brief comments 
on a number of other superstars: EDMs, $(g-2)_\mu$, lepton flavour violation, a number of rare $B$ and $K$ decays 
and $\varepsilon'/\varepsilon$.
\boldmath
\subsubsection{$S_{\psi\phi}\ge 0.5$ Scenario}
\unboldmath
The detailed studies of several models in which such high values of 
$S_{\psi\phi}$ can be attained imply the following pattern:
\begin{itemize}
\item
In the AC model and the ${\rm 2HDM_{\overline{MFV}}}$, $Br(B_{s,d}\to \mu^+\mu^-)$ will be 
automatically enhanced up to the present upper limit of roughly 
$3\cdot 10^{-8}$ from CDF and D0. The double Higgs penguins are responsible 
for this correlation 
\cite{Buras:2010mh,Buras:2010zm,Altmannshofer:2009ne}.
\item
In the SM4 this enhancement will be more moderate: up to $(6-9)\cdot 10^{-9}$, 
that is a factor of 2-3 above the SM value \cite{Soni:2008bc,Buras:2010pi}.
\item
In the non-abelian supersymmetric flavour model RVV2, 
$Br(B_{s,d}\to \mu^+\mu^-)$ can be enhanced up to a few $10^{-8}$ but it is not
uniquely implied due to the pollution of double Higgs penguin contributions 
to $B_s^0-\bar B^0_s$ mixing through 
gluino boxes, that disturbs the correlation between $S_{\psi\phi}$ 
and $Br(B_{s,d}\to \mu^+\mu^-)$
 present in the AC model 
\cite{Altmannshofer:2009ne} and ${\rm 2HDM_{\overline{MFV}}}$.
\item
In the RSc, $Br(B_{s,d}\to \mu^+\mu^-)$ is SM-like independently of the value of 
 $S_{\psi\phi}$ \cite{Blanke:2008yr}. If the custodial protection for $Z^0$ flavour violating 
couplings is removed values of $10^{-8}$ are possible \cite{Blanke:2008yr,Bauer:2009cf}.
\end{itemize}

The question then arises what kind of implications does one have for 
$Br(B_{d}\to \mu^+\mu^-)$. Our studies show that
\begin{itemize}
\item
The ${\rm 2HDM_{\overline{MFV}}}$ implies automatically an enhancement of $Br(B_{d}\to \mu^+\mu^-)$ 
with the ratio of these two branching ratios governed solely 
by $|V_{td}/V_{ts}|^2$ and weak decay constants.
\item
This familiar MFV relation between the two branching ratios $Br(B_{s,d}\to \mu^+\mu^-)$ is strongly violated in non-MFV scenarios like AC and RVV2 models 
and as seen in Fig.~5 of \cite{Buras:2009if} 
for a given $Br(B_{s}\to \mu^+\mu^-)$ 
the range for $Br(B_{d}\to \mu^+\mu^-)$ can be large with the values of the
latter branching ratios being as high as $5\cdot 10^{-10}$.
\item
Interestingly, in the SM4, large $S_{\psi\phi}$ accompanied by 
large $Br(B_{s}\to \mu^+\mu^-)$  precludes a large departure of 
$Br(B_{d}\to \mu^+\mu^-)$ from the SM value $1\cdot 10^{-10}$ 
\cite{Buras:2010pi}.
\end{itemize}

We observe that simultaneous consideration of $S_{\psi\phi}$ and 
 $Br(B_{s,d}\to \mu^+\mu^-)$ can already help us in eliminating
some NP scenarios. Even more insight will be gained when 
$Br(K^+\to\pi^+\nu\bar\nu)$ and $Br(K_L\to\pi^0\nu\bar\nu)$ will be 
measured:
\begin{itemize}
\item
First of all the supersymmetric flavour models mentioned above predict by 
construction tiny NP contributions to $K\to\pi\nu\bar\nu$ decays. 
However, it 
does not mean that in supersymmetric models large effects in these 
decays are not possible. Examples of large enhancements of the 
rates for $K\to\pi\nu\bar\nu$ decays in supersymmetric theories can be found 
in \cite{Nir:1997tf,Buras:1997ij,Colangelo:1998pm,Buras:1999da,Buras:2004qb,Isidori:2006jh} and are reviewed in \cite{Buras:2004uu}.
\item
In the RSc model significant enhancements of both branching ratios are
generally possible \cite{Blanke:2008yr,Bauer:2009cf}
 but not if $S_{\psi\phi}$ is large. Similar comments would
apply to the LHT model where the NP effects in $K\to\pi\nu\bar\nu$ can 
be larger than in the RSc \cite{Blanke:2009am}. However, the LHT model has difficulties to
reproduce a very large $S_{\psi\phi}$ and does not belong to this scenario.
\item
Interestingly, in the SM4 large enhancements of $S_{\psi\phi}$, 
$Br(K^+\to\pi^+\nu\bar\nu)$ and $Br(K_L\to\pi^0\nu\bar\nu)$ can coexist 
with each other \cite{Buras:2010pi}.
\end{itemize}
\boldmath
\subsubsection{$S_{\psi\phi}\approx 0.25$ Scenario}
\unboldmath
In this scenario many effects found in the large $S_{\psi\phi}$ scenario
 are significantly weakend. Prominent exceptions are 
\begin{itemize}
\item
In the SM4, $Br(B_{s}\to \mu^+\mu^-)$ is not longer enhanced and can even 
be suppressed, while $Br(B_{d}\to \mu^+\mu^-)$ can be significantly enhanced
\cite{Buras:2010pi}.
\item
The branching ratios $Br(K^+\to\pi^+\nu\bar\nu)$ and $Br(K_L\to\pi^0\nu\bar\nu)$
can now be strongly enhanced in the LHT model \cite{Blanke:2009am} and the
RSc model \cite{Blanke:2008yr,Bauer:2009cf} with  respect to 
the SM but this is not guaranteed.
\end{itemize}

These patterns of flavour violations demonstrate very clearly the power of 
flavour physics in distinguishing different NP scenarios.
\subsection{EDMs, $(g-2)_\mu$ and $Br(\mu\to e\gamma)$}
These observables are governed by dipole operators but describe 
different physics as far as CP violation and flavour violation is concerned. 
EDMs are flavour conserving but CP-violating, $\mu\to e \gamma$ is CP-conserving but lepton flavour violating and finally $(g-2)_\mu$ is lepton flavour conserving and CP-conserving. A nice paper discussing all these observables 
simultaneously is \cite{Hisano:2009ae}.

In concrete models there exist correlations between these three observables 
of which EDMs and $\mu\to e\gamma$ are very strongly suppressed within the 
SM and have not been seen to date. $(g-2)_\mu$ on the other hand has been very precisely measured and exhibits a $3.2\sigma$ departure
 from the very precise SM value 
(see \cite{Prades:2009qp} and references therein).
Examples of these correlations can be found in 
\cite{Altmannshofer:2009ne,Altmannshofer:2008hc}. 
In certain supersymmetric 
flavour models with non-MFV interactions the solution of the $(g-2)_\mu$ 
anomaly implies simultaneously $d_e$ and $Br(\mu\to e \gamma)$ in the reach of experiments in this decade. In these two papers several correlations 
of this type have been presented. We have listed them in the previous 
Section.

The significant FBPs required to reproduce the enhanced value of $S_{\psi\phi}$ 
in the  ${\rm 2HDM_{\overline{MFV}}}$ model, necessarily  imply large EDMs 
 of the neutron, Thallium and Mercury atoms.
Yet, as a detailed
analysis in  \cite{Buras:2010zm} shows the present upper bounds on the
 EDMs do not forbid sizable non-standard CPV effects in $B_{s}$ mixing.
However, if a large CPV phase in $B_s$ mixing will be confirmed, this
will imply hadronic EDMs very close to their present experimental bounds,
within the reach of the next generation of experiments. For a recent model 
independent analysis of EDMs see \cite{Batell:2010qw}.

\subsection{Waiting for precise predictions of $\varepsilon'/\varepsilon$}
The flavour studies of the last decade have shown that provided the hadronic 
matrix elements of QCD-penguin and electroweak penguin operators will be 
known with sufficient precision, $\varepsilon'/\varepsilon$ will play a very 
important role in constraining NP models. We have witnessed recently an 
impressive progress in the lattice evaluation of $\hat B_K$ that elevated 
$\varepsilon_K$ to the group of observables relevant for precision studies 
of flavour physics. Hopefully  this could also 
be the case of  $\varepsilon'/\varepsilon$ already in this decade.

\subsection{$B\to K^*l^+l^-$}
Let us next mention briefly these decays that will be superstars at the LHCb. 
While  the branching ratios for $B\to X_sl^+l^-$ and  $B\to K^*l^+l^-$
put already significant constraints on NP, the angular observables, 
CP-conserving ones like the well known forward-backward asymmetry 
and CP-violating ones will definitely be very useful for distinguishing
various extensions of the SM. A number of detailed analyses 
of various CP averaged symmetries and CP asymmetries provided by the 
angular distributions in the exclusive decay $B\to K^*(\to K\pi)l^+l^-$
have been performed in 
\cite{Bobeth:2008ij,Egede:2008uy,Altmannshofer:2008dz}. A nice summary of 
the last paper can be found in \cite{Straub:2010ih}.
In particular the zeroes of some of these 
observables can be accurately predicted. Belle and BaBar provided already
interesting results for the best known forward-backward asymmetry but 
the data have to be improved in order to see whether some signs of NP 
are seen in this asymmetry. Future studies by the LHCb, Belle II and SFF in Rome
will be able to contribute here in a significant manner.

\subsection{$B^+\to \tau^+\nu$ and  $B^+\to D^0\tau^+\nu$}
Another prominent anomaly in the data not discussed by us sofar is found in
the tree-level decay $B^+ \to \tau^+ \nu$. The relevant branching ratio
 is given by
\begin{equation} \label{eq:Btaunu}
{Br}(B^+ \to \tau^+ \nu)_{\rm SM} = \frac{G_F^2 m_{B^+} m_\tau^2}{8\pi} \left(1-\frac{m_\tau^2}{m^2_{B^+}} \right)^2 F_{B^+}^2 |V_{ub}|^2 \tau_{B^+}~.
\end{equation}
In view of the parametric uncertainties induced in (\ref{eq:Btaunu}) by $F_{B^+}$ and $V_{ub}$,
in order to find the SM prediction for this branching ratio one can 
rewrite it as follows \cite{Altmannshofer:2009ne}:
\begin{eqnarray} \label{eq:Btaunu_DMd}
{Br}(B^+ \to \tau^+ \nu)_{\rm SM} &=& \frac{3 \pi\Delta M_d}{4
    \, \eta_B \, S_0(x_t) \, \hat B_{B_d}} \frac{m_\tau^2}{M_W^2} \left(1 -
  \frac{m_\tau^2}{m_{B^+}^2} \right)^2 \left\vert \frac{V_{ub}}{V_{td}}
  \right\vert^2 \tau_{B^+}.
\end{eqnarray}
 Here $\Delta M_d$ is supposed to be taken from experiment and
\begin{equation}
\left\vert \frac{V_{ub}}{V_{td}}\right\vert^2 =
\left( \frac{1}{1-\lambda^2/2} \right)^2
~\frac{1+R_t^2-2 R_t\cos\beta}{R^2_t} ~,
\end{equation}
with $R_t$ and $\beta$ determined by means of $\Delta M_d/\Delta M_s$ and 
$S_{\psi K_S}$, respectively.
In writing (\ref{eq:Btaunu_DMd}),
we used $F_B \simeq F_{B^+}$ and $m_{B_d}\simeq m_{B^+}$. We then find
\cite{Altmannshofer:2009ne}
\begin{equation}\label{eq:BtaunuSM1}
{Br}(B^+ \to \tau^+ \nu)_{\rm SM}= (0.80 \pm 0.12)\times 10^{-4}.
\end{equation}
 This result
agrees well with the result presented by the UTfit collaboration
\cite{Bona:2009cj}. 

On the other hand, the present experimental world avarage based 
on results by BaBar and Belle
reads \cite{Bona:2009cj}
\begin{equation} \label{eq:Btaunu_exp}
{Br}(B^+ \to \tau^+ \nu)_{\rm exp} = (1.73 \pm 0.35) \times 10^{-4}~,
\end{equation}
which is roughly by a factor of 2 higher than the SM value.
We can talk about a tension at the $2.5\sigma$
level.

With a higher value of $|V_{ub}|$ as obtained through inclusive determination 
this discrepancy can be decreased significantly. For instance with a value 
of $4.4\times 10^{-3}$, the central value predicted for this branching 
ratio would be more like $1.25\times 10^{-4}$. Yet, this would then require 
NP phases in $B_d^0-\bar B_d^0$ mixing to agree with the data on $S_{\psi K_S}$. 
In any case values of ${Br}(B^+ \to \tau^+ \nu)_{\rm exp}$ significantly 
above $1\times 10^{-4}$ will signal NP contributions either in this decay 
or somewhere else. For a very recent discussion of such correlations see 
\cite{Lunghi:2010gv}.

While the final data from BaBar and Belle will lower the exparimental
error on  $Br(B^+\to\tau^+\nu)$, the full clarification of a possible
discrepancy between the SM and the data will have to wait for the
data from Belle II and SFF in Rome. Also improved values for $F_B$ from lattice 
and $\vub$ from tree level decays will be important if some NP like
charged Higgs is at work here. The decay $B^+\to D^0\tau^+\nu$ being 
sensitive to different couplings of $H^\pm$ can contribute significantly 
to this discussion but formfactor uncertainties make this decay less
theoretically clean. A thorough analysis of this decay is presented 
in \cite{Nierste:2008qe} where further references can be found.

Interestingly, the tension between  theory and experiment in the 
case of $Br(B^+\to\tau^+\nu)$
 increases in the presence of a tree level $H^\pm$
exchange which interferes destructively with the $W^\pm$ contribution. 
As addressed long time ago by 
Hou \cite{Hou:1992sy} and in 
modern times calculated first by Akeroyd and Recksiegel 
\cite{Akeroyd:2003zr}, and later by
Isidori and Paradisi \cite{Isidori:2006pk}, 
one has in the MSSM with MFV and large $\tan\beta$
 \be\label{BP-MSSM}
\frac{Br(B^+\to\tau^+\nu)_{\rm MSSM}}
{Br(B^+\to\tau^+\nu)_{\rm SM}}=
\left[1-\frac{m_B^2}{m^2_{H^\pm}}\frac{\tan^2\beta}{1+\epsilon\tan\beta}
\right]^2,
\ee
with $\epsilon$ collecting the dependence on supersymmetric parameters.
This means that in the MSSM this decay can be strongly suppressed unless 
the choice of model parameters is such that the second term in the parenthesis
is larger than 2. 
Such a
possibility that would necessarily imply a light charged Higgs and large $\tan\beta$ values
seems to be very unlikely in view of the constraints from other 
observables~\cite{Antonelli:2008jg}.
Recent summaries of  $H^\pm$ physics can be found in
\cite{Barenboim:2007sk,Ellis:2009qx}.

\subsection{Rare B Decays $B\to X_s\nu\bar\nu$, $B\to K^*\nu\bar\nu$ and 
        $B\to K\nu\bar\nu$}
Finally we discuss these three superstars that 
provide a very
good test of modified $Z$ penguin contributions 
\cite{Colangelo:1996ay,Buchalla:2000sk}, 
but their measurements appear to be
even harder than those of the rare $K\to\pi\nu\bar\nu$ decays discussed 
previously. Recent analyses
of these decays within the SM and several NP scenarios can be found in 
\cite{Altmannshofer:2009ma,Bartsch:2009qp}.

The inclusive decay $B\to X_s\nu\bar\nu$ is theoretically as clean as 
$K\to\pi\nu\bar\nu$ decays but the parametric uncertainties are a bit
larger. The two exclusive channels are affected by  formfactor uncertainties 
but last year in the case of $B\to K^*\nu\bar\nu$ \cite{Altmannshofer:2009ma} 
and $B\to K\nu\bar\nu$ 
\cite{Bartsch:2009qp}
significant progress has been made in improving this situation. 
In the latter paper this has been achieved 
by considering simultaneously also $B\to K l^+l^-$. 
Last year also non-perturbative tree level contributions from $B^+\to
\tau^+\nu$ to $B^+\to K^+\nu\bar\nu$ and $B^+\to K^{*+}\nu\bar\nu$ at the 
level of roughly $10\%$ 
have been pointed out \cite{Kamenik:2009kc}.

The interesting feature of these three $b\to s\nu\bar\nu$ transitions, in particular when 
taken together, is their sensitivity to right-handed currents 
\cite{Altmannshofer:2009ma}. Belle II and SFF in Rome
machines should be able to measure them at a satisfactory level and various 
ideas put forward in the latter paper will be tested.
 Nice summaries of the potential of these decays in searching for new physics 
have been presented by Straub and Kamenik
\cite{Straub:2010ih,Kamenik:2010na}.

\boldmath
\section{The $3\times 3$ Flavour Code Matrix (FCM)}
\unboldmath

\begin{table}[thb]
\addtolength{\arraycolsep}{4pt}
\renewcommand{\arraystretch}{1.5}
\centering
\begin{tabular}{|l|c|c|c|}
\hline
Model &  LH & RH  & SH
\\
\hline
MFV  & $F_{11}$  &   $F_{12}$   & {$F_{13}$}  
\\ \hline
BMFV & $F_{21}$     & $F_{22}$    & $F_{23}$ 
\\ \hline
 FBPs & $F_{31}$  & $F_{32}$  & $F_{33}$ 
\\ \hline
\end{tabular}
\renewcommand{\arraystretch}{1}
\vskip0.2truecm
\caption{
The Flavour Code Matrix for a given model. FBPs denotes important flavour blind phases. 
BMFV  denotes 
new flavour violating interactions. LH  denotes the left-handed currents,
RH denotes right-handed currents  and SH
denotes scalar currents.}
\label{tab:F44}
\end{table}

\subsection{Basic Idea}
In Section 5 we have reviewed a large number of BSM models with rather 
different patterns of flavour and CP violation. There are other models 
in the literature that I did not discuss here but it appears to me that
already on the basis of the models considered by us 
 a rough picture is emeraging. 
The question then is, how to draw a grand picture of all these NP effects 
and to summarize them in a transparent manner. The problem with this goal 
is the multitude of free parameters present basically in all extensions 
of the SM, making any transparent classification a real challenge. 

In this context let me give one example. At first sight it is evident 
that the presence of the $Q_2^{LR}$ operator in a NP scenario promises 
interesting and often dangerous effects due to large RG effects 
accompaning this operator (large anomalous dimension) and 
the chiral enhancements of its matrix elements in the $K$ system. 
Such effects are typically  much smaller in models dominated by 
LH currents and the study of the LHT model confirms this picture:
the CP-asymmetry $S_{\psi\phi}$ in this model cannot be as large 
as in models containing RH currents which in collaboration with LH
currents produce the $Q_2^{LR}$ operator. Yet, this clear picture 
is polluted to some extent by the results obtained in the SM4. 
This model has only LH currents but still is capable of obtaining 
values for $S_{\psi\phi}$ above 0.5, even if this is not as easy 
as in the case of models with RH currents. 

There are other challenges as we will see below. In spite of this let 
us make an attempt to classify  systematically the knowledge gained in 
previous sections. The first question which arises is the choice
of the right ''degrees of freedom'' or 
''coordinates'' for such a grand picture. We have made the first steps in 
this direction in \cite{Buras:2009if}, where a $2\times 2$ Flavour Matrix has been 
proposed. This matrix distinguishes between models with  SM and non-SM operators on the one hand and 
 between MFV models and models with non-MFV sources on the other hand. 
The previous 
Sections demonstrate clearly that this matrix is too small to transparently uncover all possibilities so that 
a proper distinction between models belonging to a given element 
of this matrix cannot be made.  

There are three aspects which are missing in the $2\times 2$ Flavour Matrix:
\begin{itemize}
\item
 The distinction between models having significant FBPs and those 
 not having them.
\item
 The distinction between BSM models with dominant LH currents, RH 
 currents and scalar (SH) currents.
\item
Moreover, there are models in which LH, RH and SH currents play comparable
role. 
 \end{itemize}

This situation indicates that it is probably a better idea to invent 
a new classification of various NP effects by means of a
 coding system in a form of a  $3\times 3$ {\it flavour code matrix} (FCM)
which instead of attaching a given model to a specific entry of a 
flavour matrix describes each model separately. Thus each NP model
is characterized by a special code in which only some entries of 
the matrix in question are occupied. MFV, non-MFV sources and
FBPs  on the one hand and LH currents, 
RH currents
and SH currents on the other hand are the fundamental coordinates 
in this 
code. They allow often to distinguish compactly the models 
discussed in Section 5.  The basic structure of a FCM 
is shown schematically  in Table~\ref{tab:F44}. 

In other words the main goal is to identify the main
fundamental "ingredients" of NP models that lead to specific/distinct
features in flavour observables.
Once these ingredients are identified one has a classification of a NP
model simply based on the ingredients that are present in that model.
The FCM presented here should be a good starting point
for such a classification scheme but more elaborate schemes could  also be
constructed. We will return to them in the future.

Ultimately a given {\it flavour code} should correspond to a unique flavour DNA
pattern representing a given model and just having this code should be sufficient to determine the interesting flavour observables that can probe this model.

 \subsection{Arguments for the Choice of Coordinates of FCM}
Let us still elaborate on our choice of the coordinates. 
 
First, MFV restricts both flavour and CP violation  to
the CKM matrix.
Next, FBPs only provide additional sources of CP violation.
Finally, BMFV allows for non-trivial complex flavour structures, i.e. both new
sources of flavour and CP violation are present.
This  choice of 
{\it flavour} coordinates MFV, non-MFV and FBPs is then rather convincing
 even if the 
non-MFV interactions can vary from model to model by a significant 
amount. However, some differences between models with non-MFV sources 
arise precisely from the different structure of contributing operators 
and this difference is taken care of at least partially by the 
remaining three coordinates LH, RH and SH.

The clasification of Lorentz structures in only LH, RH and SH oversimplifies 
of course the general situation but we think that it is sufficient as the 
leading order approximation. There would be no point in choosing 
as coordinates all different operators involved as one would end 
with very large code matrices.

Let us note that for $\Delta F=1$ semi-leptonic processes the 
division in LH, RH and SH structures is rather transparent as one can convince 
oneself by looking at the operators involved that we listed in Section 4. 

The case of $\Delta F=2$ processes is still simple if one only 
considers diagrams with exclusively LH currents and separately diagrams with 
only RH 
currents \footnote{I would like to thank Wolfgang Altmannshofer and Paride 
Paradisi for 
interesting and inlightning discussions related to the points presented here and  other parts of this Section.}. Then operators $Q_1^{\rm VLL}$ and $Q_1^{\rm VRR}$ result, 
respectively. This is also the case of box diagrams with internal 
squarks and gluinos 
in SUSY models. $d_{\rm LL}$ and $d_{\rm RR}$ mass insertions when considered 
separately generate $Q_1^{\rm VLL}$ and $Q_1^{\rm VRR}$, respectively. 
The necessary $\gamma_\mu$ is provided by the gluino propagator. These 
two cases are also easily distinguishable from the third coordinate 
related to scalar interactions.

The situation becomes more complicated when LH and RH currents or 
$d_{\rm LL}$ and $d_{\rm RR}$ mass insertions are considered simultaneously. 
Then the operators $Q_1^{\rm LR}$ and $Q_2^{\rm LR}$ enter and 
have to be considered 
simultaneously as they mix under QCD renormalization. Moreover up to 
the colour indices these two operators are related by Fierz transformation. 
But this is not really a problem for our classification as simply the 
presence of LH and RH currents just generates such operators with 
related phenomenological implications. Yet, one should stress that
the operator $Q_2^{\rm LR}$ can also 
be generated by scalar exchanges and the renormalization under QCD 
generates then $Q_1^{\rm LR}$ as well. For large $\tan\beta$ the scalar 
currents can however dominate and as our studies of RHMFV and 
${\rm 2HDM_{\overline{MFV}}}$ indicate the phenomenology of RH currents 
and LH currents and of SH currents is quite different in each case. 
One should also not forget that scalar exchanges can 
generate the operators $Q_{1,2}^{\rm SLL}$ and $Q_{1,2}^{\rm SRR}$.

In summary, the coordinates chosen here appear as a good starting 
point towards some optimal classification of flavour dynamics at very 
short distance scales.

\newcommand{\UU}{{\color{red}$\bigstar$}}
\newcommand{\AC}{{\color{blue}$\bigstar$}}
\newcommand{\BA}{{\color{red}$\blacksquare$}}
\newcommand{\BB}{{\color{black}$\blacksquare$}}
\newcommand{\BC}{{\color{blue}$\blacksquare$}}
\newcommand{\CA}{{\color{red}$\blacktriangle$}}
\newcommand{\CC}{{\color{blue}$\blacktriangle$}}

\begin{table}[htb]
\addtolength{\arraycolsep}{4pt}
\renewcommand{\arraystretch}{1.5}
\centering
\begin{tabular}{|l|c|c|c|}
\hline
CMFV &  LH & RH  & SH
\\
\hline
MFV  & \UU  &      &   
\\ \hline
BMFV &     &     &  
\\ \hline
 FBPs &   &   &  
\\ \hline
\end{tabular}
\quad\quad
\begin{tabular}{|l|c|c|c|}\hline
    ${\rm 2HDM_{\overline{MFV}}}$   &  LH & RH  & SH
\\
\hline
MFV  & \UU  &      & \AC  
\\ \hline
BMFV &     &     &  
\\ \hline
 FBPs &\CA   &   &  \CC
\\ \hline
\end{tabular}
\renewcommand{\arraystretch}{1}
\vskip0.2truecm
\caption{
The FCM for the CMFV models (left) and 
${\rm 2HDM_{\overline{MFV}}}$  (right).}
\label{tab:CMFV}
\end{table}

\begin{table}[htb]
\addtolength{\arraycolsep}{4pt}
\renewcommand{\arraystretch}{1.5}
\centering
\begin{tabular}{|l|c|c|c|}
\hline
LHT &  LH & RH  & SH
\\
\hline
MFV  & \UU  &      &   
\\ \hline
BMFV &  \BA   &     &  
\\ \hline
 FBPs &   &   &  
\\ \hline
\end{tabular}
\quad\quad
\begin{tabular}{|l|c|c|c|}\hline
SM4 &  LH & RH  & SH
\\
\hline
MFV  & \UU  &      &  
\\ \hline
BMFV & \BA    &     &  
\\ \hline
 FBPs &   &   &  
\\ \hline
\end{tabular}
\renewcommand{\arraystretch}{1}
\vskip0.2truecm
\caption{
The FCM for the LHT model (left) and 
SM4  (right).}
\label{tab:LHT}
\end{table}

\begin{table}[htb]
\addtolength{\arraycolsep}{4pt}
\renewcommand{\arraystretch}{1.5}
\centering
\begin{tabular}{|l|c|c|c|}
\hline
FBMSSM &  LH & RH  & SH
\\
\hline
MFV  & \UU  &      &  \AC
\\ \hline
BMFV &     &     &  
\\ \hline
 FBPs & \CA  &   &  
\\ \hline
\end{tabular}
\quad\quad
\begin{tabular}{|l|c|c|c|}\hline
$\delta$LL &  LH & RH  & SH
\\
\hline
MFV  & \UU  &      & \AC 
\\ \hline
BMFV & \BA    &     & \BC 
\\ \hline
 FBPs &   &   &  
\\ \hline
\end{tabular}
\renewcommand{\arraystretch}{1}
\vskip0.2truecm
\caption{
The FCM for the FBMSSM model (left) and 
$\delta$LL  (right).}
\label{tab:FBMSSM}
\end{table}

\begin{table}[htb]
\addtolength{\arraycolsep}{4pt}
\renewcommand{\arraystretch}{1.5}
\centering
\begin{tabular}{|l|c|c|c|}
\hline
RHMFV &  LH & RH  & SH
\\
\hline
MFV  & \UU  &      &   
\\ \hline
BMFV &     & \BB    &  
\\ \hline
 FBPs &   &   &  
\\ \hline
\end{tabular}
\quad\quad
\begin{tabular}{|l|c|c|c|}\hline
RSc &  LH & RH  & SH
\\
\hline
MFV  & \UU  &      &  
\\ \hline
BMFV & \BA    & \BB    &  
\\ \hline
 FBPs &   &   &  
\\ \hline
\end{tabular}
\renewcommand{\arraystretch}{1}
\vskip0.2truecm
\caption{
The FCM for the RHMFV model (left) and 
RSc  (right).}
\label{tab:RHMFV}
\end{table}

\begin{table}[htb]
\addtolength{\arraycolsep}{4pt}
\renewcommand{\arraystretch}{1.5}
\centering
\begin{tabular}{|l|c|c|c|}
\hline
AMK &  LH & RH  & SH
\\
\hline
MFV  & \UU  &      &   \AC
\\ \hline
BMFV & {\color{red}$\rule[1mm]{1mm}{1mm}$}   & \BB    &  \BC
\\ \hline
 FBPs &   &   &  
\\ \hline
\end{tabular}
\quad\quad
\begin{tabular}{|l|c|c|c|}\hline
AC &  LH & RH  & SH
\\
\hline
MFV  & \UU  &      &  \AC
\\ \hline
BMFV &     & \BB    &  \BC
\\ \hline
 FBPs &   &   &  
\\ \hline
\end{tabular}
\renewcommand{\arraystretch}{1}
\vskip0.2truecm
\caption{
The FCM for the  AMK model (left) and 
AC model  (right).}
\label{tab:AMK}
\end{table}

\begin{table}[thb]
\addtolength{\arraycolsep}{4pt}
\renewcommand{\arraystretch}{1.5}
\centering
\begin{tabular}{|l|c|c|c|}
\hline
RVV2 &  LH & RH  & SH
\\
\hline
MFV  & \UU  &      &   \AC
\\ \hline
BMFV & {\BA}    & \BB    &  \BC
\\ \hline
 FBPs &   &   &  
\\ \hline
\end{tabular}
\quad\quad
\begin{tabular}{|l|c|c|c|}
\hline
 $SSU(5)_{RN}$    &  LH & RH  & SH
\\
\hline
MFV  & \UU  &      &   \AC
\\ \hline
BMFV & {\BA}    & \BB    &  \BC
\\ \hline
 FBPs &   &   &  
\\ \hline
\end{tabular}
\renewcommand{\arraystretch}{1}
\vskip0.2truecm
\caption{
The FCM for the RVV2 model and the $SSU(5)_{RN}$.}
\label{tab:RVV2}
\end{table}

\subsection{Examples of Flavour Code Matrices (FCM)}
 We will now present  FCMs for models discussed in the text. 
 While the 
 patterns of flavour violation  in a given model are in details specific 
 for this model, models having similar codes will have similar predictions 
 for various observables. We distinguish MFV, BMFV and FBPs by different 
 shapes and LH, RH and SH by different colours. The latter colour coding is 
 evident: red for LH, black for RH and Bavarian blue for SH.

 It should be emphasized that although in each model considered by us 
 almost all entries of the related FCM could be filled, we indicate only 
 those which play a significant role in the phenomenology of a given model. 

 In Tables~\ref{tab:CMFV}-\ref{tab:RVV2} we show the FMC's for models discussed 
in Section 5.

In Table~\ref{tab:CMFV} we summarize CMFV and ${\rm 2HDM_{\overline{MFV}}}$. The 
NP effects in CMFV are generally small and CPV is SM-like. 
In ${\rm 2HDM_{\overline{MFV}}}$ the presence of scalar currents accompanied by 
FBPs allows to obtain, in spite of MFV, interesting CPV effects in FCNC 
processes and EDMs. Also significantly enhanced $Br(B_{s,d}\to\mu^+\mu^-)$ are 
possible in the latter case.

In Table~\ref{tab:LHT} we show FCMs for the LHT model and the SM4. These 
FCMs look identical and in fact the patterns of flavour violation in these
two models show certain similarities although NP effects in the SM4 can 
be larger. A prominent difference is noticed in the case of  
$Br(B_{s,d}\to\mu^+\mu^-)$. They can be enhanced up to factors of 2-3 in 
the SM4, while this is not possibly in the LHT.

In Table~\ref{tab:FBMSSM} we show the FBMSSM and the FS model 
with the dominance of 
LH currents ($\delta$LL). The pattern of flavour violations in these two 
models is similar but the presence of BMFV sources makes the $\delta$LL 
model less constrained. In particular in the FBMSSM the ratio of 
$Br(B_d\to \mu^+\mu^-)$ to $Br(B_s\to\mu^+\mu^-)$ is MFV-like as in 
(\ref{bmumu}), while in the ($\delta$LL) model due to BMFV sources this relation can 
be strongly violated. This different pattern could be used to distinguish 
these two models. Characteristic for these two models are sizable effects 
in $S_{\phi K_S}$ and SM-like $S_{\psi\phi}$. 
 Finally, let us note that within ($\delta$LL) model and all other 
SF models discussed by us, one assumes that
 CP is a symmetry of the theory (hence no FBPs) that
 is broken only by flavour effects after the breaking of the flavour
 symmetry.

In Table~\ref{tab:RHMFV} we show FCMs for RHMFV and RSc. The presence of 
sizable contributions from BMFV sources in LH and RH currents in RSc makes 
the effects in this model generally larger than in the RHMFV and some 
fine tuning of parameters is required in order to be in accordance with 
the data. In turn some observables, in particular $Br(B_{s,d}\to\mu^+\mu^-)$, 
turn out to be SM-like in the RSc. The RHMFV has a much simpler structure than RSc 
and the pattern of flavour violation is more transparent.

In Table~\ref{tab:AMK} we show the FCMs of AMK and AC models. It should 
be emphasized that the AMK model is based on the SU(3) flavour  symmetry, 
while the AC model is an abelian model. Moreover whereas the RH currents 
in the AMK model are CKM-like, they are $\ord(1)$ in the case of the AC model.
The small red box in the FCM of the AMK model indicates that the LH currents 
in this model as analyzed by us are much weaker than the RH currents. 
Still many implications of these two models are similar although the 
effects in the AC model are generally larger. The smoking gun of both 
models is the strong correlation between $S_{\psi\phi}$ and the lower 
bounds on $Br(B_{s,d}\to\mu^+\mu^-)$ that can be for large $S_{\psi\phi}$ 
by an order of magnitude larger than the SM values. The neutral Higgs 
exchanges with large $\tan\beta$ are responsible for this strong 
correlation.

Finally, in Table~\ref{tab:RVV2} we show FCM of the RVV2 model and of 
 the $SSU(5)_{RN}$ model. RVV2 
having SU(3) flavour symmetry has some similarities with the AMK model 
but the presence of larger LH currents in the RVV2 model leads to 
differences as summarized in Section 5.
The patterns of flavour violation in the RVV2 model and in  the $SSU(5)_{RN}$ 
has some similarities but as we emphasized in Section 5 also few striking 
difference have been identified in \cite{Buras:2010pm}. They are related to 
$S_{\phi K_S}$ and its correlations with $S_{\psi\phi}$ and $S_{\eta^\prime K_S}$
which our FCMs still do not incorporate.

We should stress that we have left in several models the entries related to FBPs empty as 
this issue requires further study. Moreover, in certain models, flavour 
diagonal but not flavour blind phases like in the RSc model are present.

This discussion shows that the idea of FCMs, although more powerful than 
the $2\times 2$ matrix discussed by us previously, cannot yet fully  depict 
all properties of a given model and the differences between various  models. 
However, in conjunction with the correlations between various observables it 
could turn out to be a useful step towards a grand picture of various 
patterns of flavour and CP violation. The references to such correlations 
have been collected in Table~\ref{tab:corr}.

Let us also note that the entry (MFV,RH) is always empty as the 
MFV flavour structure implies automatically the absence of RH currents
or their suppression by mass ratios $m_s/m_b$, $m_d/m_b$ and $m_d/m_s$.

\section{Grand Summary}
Our presentation of various BSM models is approaching the end. I hope I 
convinced the readers that flavour physics is a very rich field which 
necessarily will be a prominent part of a future theory of fundamental 
interactions both at large and short distance scales. While MFV could 
work to the first approximation, various studies show that models 
attempting the explanation of the hierarchies of fermion masses and of 
its hierarchical flavour violating and CP violating interactions in 
most cases imply non-MFV interactions. This is evident from the study 
of supersymmetric flavour models \cite{Altmannshofer:2009ne} and more general recent studies \cite{Lalak:2010bk,Dudas:2010yh}.

What role will be played by flavour blind phases in future phenomenology 
depends on the future experimental data on EDMs. Similar comment applies 
to LFV. A discovery of $\mu\to e\gamma$ rate at the level of $10^{-13}$ 
would be a true mile stone in flavour physics. Also the discovery of 
$S_{\psi\phi}$ at the level of 0.3 or higher would have a very important 
impact on quark flavour physics. The measurements of 
$Br(B_{s,d}\to \mu^+\mu^-)$ in conjunction with $S_{\psi\phi}$, $\kpn$ 
and at later stage $\klpn$ will allow to distinguish between various 
models as explicitely shown in Section 6. Here the correlations between 
various observables will be crucial. It is clearly important to clarify 
the origin of the tensions between $\varepsilon_K$, $S_{\psi K_S}$, $|V_{ub}|$ 
and $Br(B^+\to \tau^+\nu_\tau)$ but this possibly has to wait until Belle II 
and later SFF will enter their operation.

At the end of our presentation we made a new attempt to classify 
various extensions of the SM with the help of a $3\times 3$ 
flavour code matrix. Whether this classification and earlier proposed 
DNA tests of flavour physics will turn out to be a step forward should be 
clear in the next five years when new measurements will be available 
hopefully showing clear patterns of deviations from the SM. In 
any case I have no doubts that we will have a lot of fun with flavour 
physics in this decade and that this field will offer very important 
insights into the short distance dynamics.

{\bf Acknowledgements}\\
I would like to thank the organizers of the Cracow School 2010, in 
particular Michal Praszalowicz, for inviting me to such a pleasant 
school.  I wish my Cracow colleagues next fruitful 50 years in 
Zakopane. I also thank all my collaborators for exciting time we spent together 
exploring the short distance scales with the help of flavour violating 
processes and in particular Wolfgang Altmannshofer and Paride Paradisi for 
comments on the manuscript. 
I would also like to thank Walter Grimus and Helmut Neufeld for the 
great hospitality extended to me during my stay at the University of 
Vienna as the Schr\"odinger guest professor where the final part of these lectures 
has been written.
This research was partially supported by the Cluster of Excellence `Origin and Structure
of the Universe' and  by the German `Bundesministerium f\"ur Bildung und Forschung'under contract 05H09WOE.



\end{document}